\documentclass[aps,prl,groupedaddress]{revtex4-1}  
\usepackage{graphicx}  
\usepackage{bm}        
\usepackage{amssymb,amsmath}   
\usepackage{verbatim}
\usepackage[T1]{fontenc}
\usepackage{amsmath,amssymb,amsfonts}
\usepackage{cleveref}               

\usepackage{ifpdf}
\usepackage{color}

\usepackage{amsmath}
\usepackage{graphics}
\usepackage{mathtools}
\usepackage[usenames,dvipsnames]{xcolor}
\usepackage{epsfig}
\usepackage{epstopdf}
\usepackage{dcolumn}
\usepackage[reset]{geometry} 
\usepackage{float}
\makeatletter
\Gm@restore@org
\makeatother
\usepackage{tikz}
\usetikzlibrary{shapes.geometric, arrows}
\usepackage{upgreek}
\usepackage{setspace}
\usepackage{enumitem}
\usepackage{array,multirow,bigdelim}

\def\be{\begin{equation}}
\def\ee{\end{equation}}
\def\ba{\begin{eqnarray}}
\def\ea{\end{eqnarray}}

\def\CP1{\mathbb{CP}^1}
\def\SL2C{\mathrm{SL}(2,\mathbb{C})}

\def\Z2{\mathbb{Z}_2}

\def\su2{{SU(2)}}

\def\[{\left[}
\def\]{\right]}

\newcommand{\dint}[1]{\! d#1\,}

\def\({\left(}
\def\){\right)}
\def\[{\left[}
\def\]{\right]}

\def\i2{\frac{i}{2}}

\def\2F1{\,_2{\rm F}_1}

\newcommand{\beq}{\begin{equation}}
\newcommand{\eeq}{\end{equation}}
\newcommand{\beqq}{\begin{equation*}}
\newcommand{\eeqq}{\end{equation*}}
\newcommand\beqa{\begin{eqnarray}}
\newcommand\eeqa{\end{eqnarray}}
\newcommand\beqaa{\begin{eqnarray*}}
\newcommand\eeqaa{\end{eqnarray*}}
\newcommand\bea{\begin{array}}
\newcommand\eea{\end{array}}

\begin{document}

\preprint{}

\title{Charged test-particle scattering and effective one-body metrics with spin}

\author{Jitze Hoogeveen$^1$\\}
\affiliation{\smallskip {$^1$}Niels Bohr International Academy\\ Niels Bohr Institute, University of Copenhagen\\
Blegdamsvej 17, DK-2100 Copenhagen \O, Denmark}
\date{\today}

\begin{abstract}
    Using recently developed techniques, we consider weak-field test-particle scattering angle calculations in two distinct settings: Charged test-particles in spacetimes of charged sources and Effective One-Body theory with spin. We present scattering angle calculations up to $\mathcal O(G^4)$ of charged particles in the Kerr-Newman metric, including electromagnetic interactions up to second order in charge. Coulomb scattering is also discussed, and the well-known Darwin scattering formula is rederived by resummation. An Effective One-Body metric for a Kerr-Schwarzschild binary is constructed in a post-Minkowskian framework up to $\mathcal O(G^2)$ and first order in spin. Facilitated by explicit scattering calculations, our approach is equivalent with existing literature through gauge-like transformations. Finally, we investigate if the Newman Janis Algorithm applied to an Effective One-Body metric of non-spinning binaries represents a binary system with spin.
\end{abstract}
\pacs{04.60.-m, 04.62.+v, 04.80.Cc}
\maketitle
\section{Introduction}\label{sec:introduction}
The breakthrough observation of Gravitational Waves \cite{Abbott.etal.2016_GW150914} opens a new window to the universe, allowing for the first time detailed testing of General Relativity. Among these initial observations, signals from binary black holes intricately encode both dynamics of the binary, and single black hole properties. The prospect of gaining insight into these previously unprobed areas has in recent years catalysed a great theoretical effort, developing new analytical and numerical techniques solving the highly non-linear dynamics of General Relativity. One such area is the investigation of test-particle trajectories. This historically well-established subject provides both insight to single black hole properties, and facilitates a simple setting for developing new calculational tools, some of which have proven useful even to the full binary problem.\\
\\
Analytical expressions for geodesics of test-particles have been found for (off-)equatorial trajectories, expressed with elliptical functions (for a review, see ref. \cite{LH2016}). Calculations with a non-spinning Schwarzschild black hole were carried out in \cite{Hagihara1931_therel,Darwin1959,Darwin1961} followed by a charged non-spinning Reissner Nordström black hole \cite{Gackstatter1983_orbray,GK2011}. Similar geodesics for a spinning Kerr black hole \cite{Carter1968,Mino2003_perapp,CFC1998_linemi,FH2009_anasol} including plunging orbits \cite{DM2023}, notably introduced the Carter constant \cite{Carter1968}, and Mino time which is essential for integration \cite{Mino2003_perapp}. A collection of results is presented e.g. in refs. \cite{Sharp1979_geobla,Kraniotis2005_fradra}. Kerr-Newman geodesics are similarly expressible in terms of elliptical functions \cite{HX2013}, considering even charge on the test-particle.

Recently, test-particle scattering trajectories have seen increased attention. The associated calculations of the scattering angle are closely related to binary dynamics. For the purposes of this article, we will restrict ourselves to planar scattering. Examples of non-planar paths are given in e.g. refs. \cite{Hackman2010_thesis,HX2013}. Although planar scattering angles are easily encapsulated within a Hamilton-Jacobi formalism, difficulties dealing with integration limits make actual calculations a non-trivial matter. Few closed form scattering angle calculations are possible (see e.g. ref. \cite{KOT2022}).

In a weak-field limit, where the scattering angle $\chi$ is expanded in Newtons gravitational constant $G$, these difficulties may be overcome. Hadamard regularization is the traditional approach to scattering angle evaluations in this limit \cite{Damour2016_grasca,Damour2020_claqua}. It substitutes the difficult lower integration limit with something simple and manually removes emergent divergencies. A different method was recently developed in ref. \cite{DHLV2022}. Building on work in isotropic metrics \cite{Wallace1973,BCD2020}, this technique provides a simple formula applicable to very general situations. It considers the scattering angle integral
\begin{equation}
    \chi/2+\pi/2=\int_{r_m}^\infty \!dr\, \frac{d\phi}{dr}.
\end{equation}
The lower integration limit, $r_m$, is the distance of minimum approach, which may not always be explicitly obtained. This complicates integration. Ref. \cite{DHLV2022} showed, by writing the integral in a very general form and assuming a weak-field limit, that one may explicitly render the scattering angle a sum of easily calculable integrals independent of $r_m$. Notably, this applies to test-particles in very general metrics (not necessarily restricting to black holes), for both scalar and spinning particles. In ref. \cite{DHLV2022}, the formalism was specifically employed for scattering in the equatorial plane of a Kerr metric for spinning test-particles up to second order in spin.

Test-particle scattering in a weak-field regime is linked to binary (black hole) dynamics in numerous ways. The extreme mass ratio limit of a two-body calculation naturally retrieves the test-particle regime. As such test-particle scattering in itself provides a useful tool for cross checking two-body calculations. However beyond this connection, dynamics of test-particles may encode dynamics of a two-body system through Effective One-Body (EOB) theory. Crucial for the EOB approach, to be introduced shortly, are the calculations of full binary Hamiltonians and scattering angles.\\
\\
Outside the test-particle limit, the work on binary dynamics in GR is rapidly evolving, due to its connection with Gravitational Wave observations.
Precise knowledge of binary trajectories is crucial for constructing the waveform. Both numerical and analytical approaches have proven fruitful. On the analytical side, various classical methods yield the two-body Hamiltonian in an expanded form; a post-Newtonian (PN) approach expands around weak-field Newtonian gravity in velocities $v^2/c^2\ll1$ and Newton's gravitational constant $GM/c^2r\ll1$, whereas a post-Minkowskian (PM) expansion considers weak-field interactions with arbitrary velocity as perturbations of Minkowski space in $GM/c^2r\ll1$. This latter regime is equivalent to the weak-field approach discussed above in the test-particle limit. Partial expressions for the non-spinning binary black hole Hamiltonian are available up to 6PN, ie. $\mathcal O[(v/c)^{12}]$, in the post-Newtonian expansion \cite{BMMS2020,BDG2020,BGD2020_secondpaper,BMMS2021,FRT2021,BMM2021} and 4PM, ie. $\mathcal O(G^4)$, in the post-Minkowskian expansion \cite{KLP2020,KLP2020_contid,PKLP2022}. For spinning binaries, results up to 5PN including both spin-orbit \cite{LMH2021,KLY2022,MMPS2022} and spin-spin couplings \cite{Levi2012_bindyn,LS2014,LS2016,LS2021,KLY2022_quaspi,LMH2021_graqua,CPY2022,KLY2022_secondpaper,LY2022} have been found. For the post-Minkowskian expansion all-order in spin expressions are available at 1PM \cite{Vines2018_scaspi}, whereas second order in spin results are available at 2PM and 3PM \cite{BD2017,BD2018,FKLRZ2022,JM2022_linres}. Starting at $\mathcal O(G^4)$, radiative processes contribute to conservative dynamics. Dealing with these subtleties is still an open problem \cite{BKORS2022}.

The post-Minkowskian (PM) expansion naturally lends itself to scattering trajectories. For reviews see e.g. refs. \cite{BDPV2022,KMO2022,BKORS2022}. Importantly, the two-body Hamiltonian may be recovered from this regime \cite{VSB2019,JM2022_linres,Damour2016_grasca,Damour2017_higgra}.
Binary scattering angles, computed order by order in $G$, encode information about the Hamiltonian. These may be found by a plethora of methods. Early landmark calculations were performed by Westpfahl \cite{Westpfahl1985}. A linearised form of Einstein's equations have yielded exact results for aligned spinning binaries at first order in $G$ \cite{Vines2018_scaspi}.
Other manifestly classical approaches such as the world-line formalism \cite{MPS2021,JMPS2021,JMPS2022,JMPS2022_SUSY,JM2022,JMPS2022_allthi,SV2022,JM2022_linres,BCC2022} and effective field theory \cite{Foffa2014,KP2020_posmin,KLP2020,KP2020,KP2020_II,KLP2020_contid,LPY2021,CKP2022,DKLP2022,PKLP2022,KNP2022} employ techniques lent from quantum field theory. Using these, scattering angles with both non-spinning and spinning binaries are available up to $\mathcal O(G^3)$ and second order in spin. Quantum calculations, e.g. amplitudes of massive particle scattering mediated by gravitons, provide further results up to $\mathcal O(G^5)$ \cite{ZCRC2019,ABSMV2019,PRZ2020,VHRV2020,Damour2020_radcon,MRV2021,HPRZ2021,VHRV2021_radrea,VHRV2021,BDPV2021,DBPV2021_ampcla,DPV2021,BCTW2021,BPRRSSZ2021,BPRSS2022,BPV2022} and various orders in spin \cite{Guevara2019_holcla,DHH2019,VSB2019,GOV2019,GOV2019_blasca,BG2019,MOV2019,CHKL2019,AHO2020,SV2020,CHKL2020,CHK2020,CHK2020_kernew,AHH2020,GMOV2021,CGMORSW2021,CCHK2022,CCHK2022_lenthi,Kim2022_quacor,HH2020,Haddad2022_explea,AHH2022,AHH2022_clagra,BLRSZ2021,KL2021,BKLRT2022,Cruz2022_kinthe,FKLRZ2022,AV2022,AAT2022,MS2022,RVW2022, CJP2022,CCJOPS2022,ACT2023,BCS2023,KS2023}. Remarkably, the classical limit of such amplitude calculations is in correspondence with macroscopic black hole scattering \cite{CG2017,GOV2019,GOV2019_blasca}, reproducing classical results at least up to 3PM and second order in spin \cite{GOV2019,GOV2019_blasca,JM2022_linres}.
Especially relevant in our current treatment is the scattering angle of aligned spinning particles from amplitudes, up to 2PM and fourth order in spin \cite{GOV2019},
\begin{subequations}\label{eq:GOV2PMalignedspin}
\begin{gather}
\begin{aligned}
    \begin{split}
    \chi^{(a_1,a_2)}=& \frac{2G E \left(-2 (a_1+a_2) v+b v^2+b\right)}{v^2 \left(b^2-(a_1+a_2)^2\right)}
    -\pi G^2E\frac{\partial}{\partial b}[m_2f(a_1,a_2)+m_1f(a_2,a_1)]+\mathcal O(G^3)
    \end{split}
\end{aligned}\\
    f(\sigma,a)=\frac{1}{2a^2}\left(-b+\frac{(j+\varkappa-2a)^5}{4v\varkappa[(j+\varkappa)^2-(2va)^2]^{3/2}}\right)+\mathcal O(\sigma^5), \quad\quad j=vb+\sigma+a,\quad\varkappa=\sqrt{j^2-4va(b-v\sigma)}
\end{gather}
\end{subequations}
evaluated in center of mass coordinates. This matches classical computations of aligned Kerr black holes at all orders in spin for $\mathcal O(G)$, up to linear order in spin at $\mathcal O(G^2)$, and conjecturally up to $\mathcal O(spin^4)$. Parameters $(a_1,a_2)$ are the binary spins with $m_1$ and $m_2$ their masses, $E$ is the total (center of mass) energy of the system, $v$ is the relative asymptotical (center of mass) velocity between objects, $b$ is their impact parameter and $\gamma=1/\sqrt{1-v^2}$.\\
\\
As suggested in refs. \cite{Damour2016_grasca,BDFPV2018,CRS2018,CBDV2019,KP2020,KP2020_II,CKP2022}, post-Minkowskian angles encode binary dynamics even for bound orbits. This is crucial for the study of gravitational waves, typically emitted by inspiraling bound systems. One way to recover bound orbit dynamics from scattering data is with Effective One-Body (EOB) theory. Originally formulated in a Post Newtonian expansion of velocities \cite{BD1999,BD2000,DJS2000,Damour2001}, in 2016 it was naturally adopted to a post-Minkowskian, scattering-based approach \cite{Damour2016_grasca,ABSMV2019}. The EOB formalism translates full binary motion (outside the scattering regime), to an effective test-particle moving in an EOB metric. This metric may be constructed by matching scattering angles of the full binary with those of the effective test-particle. This formalism is thus heavily reliant on scattering angle calculations of test-particles in complicated (EOB) metrics. The construction of such a metric has three major benefits, i) calculations of test-particle motion are much easier than directly solving Einstein's equations, ii) geodesics of the metric readily include bound orbits, and iii) the metric in effect resums post-Minkowskian data, widening the regime of applicability. 

EOB formalisms have been constructed up to $\mathcal O(G^2)$ without spin in refs. \cite{Damour2016_grasca,Damour2017_higgra,Damour2020_claqua}. Including spin, current literature goes up to $\mathcal O(G^2)$ and $\mathcal O(spin^1)$ \cite{BD2017,BD2018}. EOB mappings at higher orders in spin have been considered in ref. \cite{VSB2019} at $\mathcal O(G^2)$, and an all order in spin result at $\mathcal O(G)$ was published by Justin Vines \cite{Vines2018_scaspi}. For recent developments in the post-Newtonian approach, see refs. \cite{BDG2020,BGD2020_secondpaper}. Ref. \cite{BD2012} includes a discussion on non-conservative contributions. For older results see \cite{BD1999,DJS2000,Damour2001,BD2000} of which a review is given in \cite{Damour2008}. Notably, ref. \cite{DJS2000} presents a post-geodesic $\mathcal Q$ term to the Hamiltonian which is reintroduced in post-Minkowskian theory in \cite{Damour2017_higgra}. See also ref. \cite{Damour2020_claqua} for discussions hereof, especially the rewriting to an effective potential $W$.
\\
\\
Test-particle scattering calculations provide a way to both probe single black hole properties, and encode full black hole binary dynamics. The method of ref. \cite{DHLV2022} provides a novel tool for evaluating angles, enabling new streamlined analysis of these areas. In this paper, demonstrating the versatility of the method, we consider scattering in two distinct settings. First, the formalism of ref. \cite{DHLV2022} is applied to scattering of charged test-particles in the Kerr-Newman metric. Applications to other spacetimes and electric potentials is discussed. Among these is the treatment of relativistic Coulomb scattering.

Second, the method of ref. \cite{DHLV2022}, given its broad applicability, is considered with the EOB approach. A post-Minkowskian framework, such as that from refs. \cite{Damour2016_grasca,DV2021} is used. Particularly, an EOB metric for a Kerr-Schwarzschild black hole binary is constructed, up to second order in $G$ and fourth order in spin. Comparisons with earlier approaches \cite{Vines2018_scaspi,DV2021,Damour2020_claqua} are made. The formalism is restricted to orbits in the equatorial plane. As an accompanying study, the Newman-Janis Algorithm (NJA) is explored in context of the EOB formalism. Does the application of the NJA algorithm to non-spinning EOB metrics produce an EOB metric with spin? The success of this approach, based on a Schwarzschild-Schwarzschild binary EOB metric from ref. \cite{DV2021}, will be explicitly checked by comparing post-Minkowskian scattering angles of the NJA-transformed metric with those of aligned Kerr black holes from amplitude methods \cite{GOV2019}.\\
\\
Section \ref{sec:scatteringPM} introduces the general formalism of ref. \cite{DHLV2022} with a view towards Kerr-Newman, but emphasises its general applicability. Sections \ref{sec:nonchargedKerrNewman} and \ref{sec:chargedKerrNewman} then compute Kerr-Newman scattering angles of scalar and charged test-particles respectively. Sections \ref{sec:EOBdeformedKerr} and \ref{sec:EOBNJA} turn to the EOB formalism based on a post-Minkowskian approach. In section \ref{sec:EOBdeformedKerr}, a 2PM EOB metric describing Kerr-Schwarzschild binaries is constructed based on a deformed Kerr metric. Section \ref{sec:EOBNJA} treats the application of the NJA to the EOB metric from ref. \cite{DV2021}.

Throughout we adopt natural units for the speed of light $c=1$ and the Coulomb constant $1/(4\pi\epsilon_0)=1$. Newtons gravitational constant is denoted $G$, and the mostly plus sign convention ($-+++$) is used.
\section{Scattering angles in post-Minkowskian expansion}\label{sec:scatteringPM}
We introduce test-particle scattering in this section. Let us first consider a scalar (uncharged, non-spinning) test-particle. Adding electrodynamic behavior is covered in section \ref{sec:chargedKerrNewman}. We start by reviewing the work of ref. \cite{DHLV2022}, establishing a general method of evaluating test-particle scattering angles. The assumptions listed below are used implicitly throughout the article. 

We concern ourselves only with planar scattering in asymptotically flat metrics $g_{\mu\nu}=g_{\mu\nu}(r)$ parametrised by polar coordinates $\{t,r,\phi\}$. The test-particle, given mass $m$, follows a geodesic scattering trajectory $r=\{\infty\rightarrow r_m\rightarrow\infty\}$ with angular deflection $\phi=\{0\rightarrow \pi/2+\chi/2\rightarrow \pi+\chi\}$. The incident direction is chosen as $\phi=0$ without loss of generality and $r_m$ denotes the distance of minimum approach. Test-particle impact parameter is denoted $b$, asymptotical velocity $v$, asymptotical momentum $p_\infty$, energy $E$, and orbital angular momentum $L$. Energy and momenta $E$, $L$ and $p_\infty$ may be expressed in terms of $b$ and $v$ as
\begin{equation}
    E=\gamma m, \quad p_\infty = \gamma mv, \quad L=bp_\infty, \quad \text{where}\quad \gamma=1/\sqrt{1-v^2}.
\end{equation}
Hamilton-Jacobi theory readily determines the test-particle trajectory, and therefore also the scattering angle. For a scalar test-particle, the Hamiltonian and associated Hamilton-Jacobi equations which normalise canonical momentum are
\begin{equation}\label{eq:hamscalar}
    H=\frac{1}{2}g^{\mu\nu}p_{\mu}p_{\nu}, \quad\quad p_\mu=g_{\mu\nu}\frac{dx^\nu}{d\lambda}, \quad
        -m^2= g^{\mu\nu}p_\mu p_\nu \quad\Rightarrow \quad \dot x^\mu = \frac{\partial H}{\partial p_\mu},
\end{equation}
with affine parameter $d\lambda=ds/m$, given in terms of line element $ds$, parameterizing the particle path such that canonical momentum $p_\mu$ matches test-particle four-momentum. Dots denote differentiation with respect to $\lambda$. The equations of motion for $\dot x^\mu$ may be readily found from the Hamiltonian. Translational symmetry of $H$ in $t$ and $\phi$ yields conservation of energy and orbital angular momentum, $p_t\equiv -E$ and $p_\phi\equiv L$. The radial component $p_r$ may be determined from the Hamilton-Jacobi equation as a function only of coordinate $r$.

The scattering angle may readily be found from the Hamiltonian. It may generically be written as
\begin{equation}\label{eq:chigenbase}
    \chi/2+\pi/2=\int_{r_m}^\infty\dint{r} \frac{d\phi}{dr}=-\int_{r_m}^\infty \dint{r} \frac{h(r)}{p_r}, \quad\quad h(r)\equiv -p_r\frac{d\phi}{dr}
\end{equation}
simply by integrating angular deflection over half a scattering trajectory. A conventional $\pi/2$ has been added to yield $\chi=0$ for straight-line motion. The lower integration limit $r_m$ may be found for Hamiltonians quadratic in $p_r$ by the requirement 
\begin{equation}
    \left.\dot r=0\right|_{r=r_m}\quad\Rightarrow\quad p_r(r_m)=0.
\end{equation}
Following ref. \cite{DHLV2022}, we have rewritten the scattering angle integral simply by defining the function $h(r)$ in terms of $\frac{d\phi}{dr}$ and $p_r$. However this form is suggestive: Note that the integral is naturally divergent in the lower integration limit $p_r(r_m)=0$. Factoring out this divergence, $h(r)$ often takes very simple non-divergent forms. In fact from eq. \eqref{eq:hamscalar} one identifies
\begin{equation}\label{eq:hrscalar}
    h(r)=-\frac{Lg^{\phi\phi}-Eg^{\phi t}}{g^{rr}}
\end{equation}
for a scalar test-particle in the general metric discussed above. This identification of $h(r)$ is useful beyond scalar particles, and was also shown applicable to spinning test-particles in ref. \cite{DHLV2022}. Below, the same will be shown true also for charged test-particles.

Exact calculation of the scattering angle, often does not yield a closed expression, instead returning an elliptical integral. However, in a weak-field expansion in $G$ closed expressions may be found order by order in $G$. In ref. \cite{DHLV2022} a general calculation of the scattering angle in the weak-field regime was provided. It considers a scattering angle written in the form of eq. \eqref{eq:chigenbase} with $h(r)$ obeying the requirements
\begin{equation}\label{eq:hrreq}\text{
    $h(r)$ is analytical on $r\in[r_m,\infty[$ and falls off at least like $1/r^2$ as $r\rightarrow\infty$}.
\end{equation}
One may readily confirm that eq. \eqref{eq:hrscalar} indeed obeys these requirements. Furthermore, the metric is assumed written in what ref. \cite{DHLV2022} defines as a normal form. A metric is of normal form when it has the property
\begin{equation}\label{eq:normaldef}
    g_{\mu\nu}(r)\rightarrow \text{(Minkowski)} = \begin{pmatrix}
        -1 & 0 & 0\\
        0 & 1 & 0\\
        0 & 0 & r^2
    \end{pmatrix}\quad\quad \text{as $G\rightarrow0$}
\end{equation}
in the scattering plane. Metrics may readily be written in normal form by setting $G\rightarrow0$, and performing a coordinate transformation to recover the above structure. For asymptotically flat metrics, $p_r$ then takes the form
\begin{equation}\label{eq:prnormalform}
    p_r^2=p_\infty^2-\frac{J^2}{r^2}-U(r)
\end{equation}
with some function $U$ specified by the metric reminiscent of potentials in classical and isotropic amplitude calculations (see e.g. \cite{BCD2020}). This potential may safely be assumed to drop off to zero as $r\rightarrow\infty$, and may depend on any metric and test-particle quantities, e.g. angular momentum, energy etc.\\

The above requirements are satisfied throughout this article. Under such requirements, ref. \cite{DHLV2022} provides the scattering angle in a summed form which readily yields arbitrary high orders in the weak-field expansion in $G$,
\begin{equation}\label{eq:chigen}
    \chi+\pi=-2\sum_{n=0}^\infty \int_0^\infty \dint{u} \left(\frac{d}{du^2}\right)^nh(r)\frac{r^{2n}U(r,b)^n}{n!p_\infty^{2n+1}}, \quad\quad r^2=u^2+b^2.
\end{equation}
A very similar formula for isotropic metrics was previously obtained in ref. \cite{BCD2020}, and may readily be recovered by restricting $U(r,b)\rightarrow V_{eff}(r)$ independent of $b$ and $h(r)=-bp_\infty/r^2$. In this special case $V_{eff}$ is independent of impact parameter $b$ or, equivalently, angular momentum $L$.

Restricting to scalar test-particles, eq. \eqref{eq:chigen} obtains another useful form. Inspired by a derivative form of eq. \eqref{eq:chigenbase},
\begin{equation}\label{eq:chibasescalar}
    \chi/2+\pi/2 = -\frac{d}{dL}\int_{r_m}^\infty \dint{r} p_r, \quad\quad \text{(scalar test-particle)}
\end{equation}
the $h(r)$-dependence may be converted to a derivative of the impact parameter,
\begin{equation}\label{eq:chiscalar}
    \chi=\sum_{n=0}^\infty \int_0^\infty \dint{u}\left(\frac{d}{du^2}\right)^n\frac{d}{db}\left[\frac{r^{2n}U(r,b)^{n+1}}{(n+1)!p_\infty^{2(n+1)}}\right],\quad\quad r^2=u^2+b^2 \quad\text{(scalar test-particle)},
\end{equation}
removing the need to specify $h(r)$ for each individual metric.

In the next section, we demonstrate the use of eqs. \eqref{eq:chigen} and \eqref{eq:chiscalar} by explicitly calculating the scattering angle in the Kerr-Newman metric of a scalar test-particle, up to $\mathcal O(G^{4})$.
\section{Non-charged test-particles in Kerr-Newman spacetime}\label{sec:nonchargedKerrNewman}
The Kerr-Newman metric describes a charged, spinning black hole, its mass denoted by $M$, charge $Q$ and spin $a$. The $Q\rightarrow0$ and $a\rightarrow0$ limits are Kerr and Reissner-Nordström black holes respectively. Written in Boyer-Lindquist coordinates, it is rotationally symmetric only in the $\theta=\pi/2$ plane. In this plane the metric reads
\begin{equation}\label{eq:kerrnewman}
    g_{\mu\nu}=
    \left(
\begin{array}{ccc}
 -\left(1-2GM/r+\frac{r_Q^2}{r^2}\right) & 0 & -a\left(2GM/r-\frac{r_Q^2}{r^2}\right) \\
 0 & \frac{r^2}{a^2-2GMr+r_Q^2+r^2} & 0 \\
 -a\left(2GM/r-\frac{r_Q^2}{r^2}\right) & 0 & a^2 (1+2GM/r-\frac{r_Q^2}{r^2})+r^2 \\
\end{array}
\right),
\end{equation}
which is effectively a Kerr metric with $2GM/r$ replaced by $2GM/r-r_Q^2/r^2$. Parameter $r_Q=GQ^2$ encodes the black hole electric charge.

Notice that eq. \eqref{eq:kerrnewman} is not in normal form, which setting $G\rightarrow0$ confirms. One recovers a result identical to that of Kerr,
\begin{equation}\label{eq:normalcoordinateident}
    g_{\mu\nu}\rightarrow \begin{pmatrix}
        -1 & 0 & 0\\
        0 & \frac{r^2}{r^2+a^2} & 0 \\
        0 & 0 & r^2+a^2
    \end{pmatrix} \quad \text{as $G\rightarrow0$},
\end{equation}
which may be brought to normal form by a coordinate transformation $r^2\rightarrow \rho^2 = r^2+a^2$. In these coordinates, the potential $U$ may readily be found from eqs. \eqref{eq:hamscalar} and \eqref{eq:prnormalform},
\begin{equation}
    U(\rho,b)=
    \frac{G \left(Q^2-2 M \sqrt{\rho ^2-a^2}\right) \left(2 a^2 b^2 p_{\infty }^2-a^2 \rho ^2 p_{\infty }^2-2 a b E \rho ^2 p_{\infty }-b^2 \rho ^2 p_{\infty }^2+2 \rho ^4 p_{\infty }^2+m^2 \rho ^4\right)}{\rho ^4 \left(\rho ^2-a^2\right)}+\mathcal O(G^2),
\end{equation}
given here up to $\mathcal O(G)$ for brevity. Of course, equation \eqref{eq:hamscalar} readily yields $U$ to all orders in $G$. With this information, equation \eqref{eq:chiscalar} straight forwardly gives the scalar test-particle scattering angle in such a spacetime. No explicit computation of $h(r)$ is needed. Take the $\mathcal O(G^1)$ calculation as an example. The results at higher orders are listed in table \ref{tab:kerrnewmanscalar}. Extracting only $G^1$ terms from eq. \eqref{eq:chiscalar}, one finds
\begin{equation}
\begin{split}
    \chi_1=&
    \int_0^\infty\dint{u}\frac{d}{db}\left[\frac{U(\rho,b)}{p_\infty^{2}}\right], \quad \rho^2 = u^2+b^2\\
    =&
    \frac{1}{p_\infty^2}\int_0^\infty \dint{u}\frac{d}{db}\left[    \frac{G \left(Q^2-2 M \sqrt{\rho ^2-a^2}\right) \left(2 a^2 b^2 p_{\infty }^2-a^2 \rho ^2 p_{\infty }^2-2 a b E \rho ^2 p_{\infty }-b^2 \rho ^2 p_{\infty }^2+2 \rho ^4 p_{\infty }^2+m^2 \rho ^4\right)}{\rho ^4 \left(\rho ^2-a^2\right)}\right]\\
    =&
    \frac{2 G M \left(-2 a v+b v^2+b\right)}{v^2 \left(b^2-a^2\right)}+\frac{\pi  G Q^2 \left(2 a^3 v-a^2 \left(-v^2 \sqrt{b^2-a^2}+2 b v^2+b\right)+b^2 v^2 \left(b-\sqrt{b^2-a^2}\right)\right)}{2 a^2 v^2 \left(b^2-a^2\right)^{3/2}}.
\end{split}
\end{equation}
The $Q\rightarrow0$ limit is just the Kerr result. At any order $\mathcal O(G^n)$, $Q$ only appears in even orders no greater than $Q^{2n}$. The scattering angle $\chi$ may thus be decomposed into orders of $G$ and $Q$,
\begin{equation}
        \chi = \sum_{n=1}^\infty \sum_{k=0}^n \chi^{(k)}_n, \quad\quad \chi^{(k)}_n\sim \pi^{(n+k+1) \text{ mod } 2}\frac{Q^{2k}G^n M^{n-k}}{v^{2n}(b^2-a^2)^{(3n+k-1)/2}} f_{n,k}(v).
\end{equation}
These expressions are structurally similar to those found in ref. \cite{DHLV2022} for Schwarzschild and Kerr metrics. A prefactor $\pi$ appears in $\chi_n^{(k)}$ only when $n+k$ is even, much like the factors of $\pi$ appearing at even order in ref. \cite{DHLV2022} for Kerr and Schwarzschild. Indeed, the Kerr-Newman potential $U\sim 1/r^{n+k}$ at order $\mathcal O(G^n)$ and $\mathcal O(Q^{2k})$ resembles in its $r$ dependence a Kerr potential at order $n+k$ in $G$. Additional similarities may be found. Polynomials $f_{n,k}(v)$ of velocity $v$ depend on fractional powers of $b$ and $a$ for even $n+k$, and integer powers of $b$ and $a$ for odd $n+k$. This mimics the identical behavior in Kerr, for $n$ respectively even and odd.
Interestingly, the $v$ and $(a,b)$ dependence of $f_{n,k}$ is partially factorized at orders $(n,k)=(1,0),(2,1),(3,2),(4,3)$, ie. when $n-k=1$. Here we observe
\begin{equation}\label{eq:fnkstructn-k=1}
    f_{n,k}= [f^{(v)}_-f^{(ab)}_{-+}+ f^{(v)}_+ f^{(ab)}_{+-}]_{n,k}
\end{equation}
where $f^{(v)}$ and $f^{(ab)}$ are polynomials exclusively dependent on either $v$ or $(a,b)$ respectively. Subscript $+,-$ signs indicate whether $v$, $a$ and $b$ appear in even or odd powers, ie. $f^{(v)}_+$ is even in $v$ and $f^{(ab)}_{+-}$ is even in $a$ but odd in $b$. All polynomials differ in structure with varying $n,k$. We expect the structure of eq. \eqref{eq:fnkstructn-k=1} to continue to higher orders.
\section{Charged test-particles in Kerr-Newman spacetime}\label{sec:chargedKerrNewman}
Continuing our treatment of the Kerr-Newman metric, we next consider test-particles with charge $e$ and mass $m$ in a spacetime of a charged source. For completeness, as before, our initial analysis pertains to a general metric $g_{\mu\nu}$ as defined in the previous section. Hereafter, the specific case of the Kerr-Newman metric is treated. A short study of Coulomb scattering will also be discussed, as results are easily compared to the all-order exact Coulomb scattering angle \cite{KOT2022} obtained by direct integration of eq. \eqref{eq:chigenbase}.\\
\\
Apart from gravitational effects, electromagnetic interactions with coupling constant $e$ (charge of test-particle) and electromagnetic 4-potential $A_\mu$ need to be accounted for. Charged bodies in curved spacetime may be treated with Einstein-Maxwell theory (for a review, see e.g. ref. \cite{PPV2011}, including \cite{Westpfahl1985} for early scattering calculations). Charged test-particle orbits around black holes are covered in ref. \cite{Young1976_cappar,JR1974,Sharp1979_geobla,HX2013} and fully characterized for Kerr-Newman in ref. \cite{HX2013}. The treatment presented in this article is restricted to test-particle limits of $e$ and $m$, neglecting self-force effects \cite{Westpfahl1985,PPV2011,BP2019}. $A_\mu$ is then entirely produced by the gravitational source.

Subjected to an external potential $A_\mu$, the Hamiltonian and associated equations of motion of a test-particle with charge $e$ and mass $m$ in metric $g_{\mu\nu}$ are
\begin{equation}\label{eq:hamcharged}
    H=\frac{1}{2}g^{\mu\nu}(p_\mu-eA_\mu)(p_\nu-eA_\nu), \quad p_\mu = g_{\mu\nu} \frac{dx^\nu}{d\lambda}+eA_\mu \quad\Rightarrow \quad\dot x^\mu = \frac{\partial H}{\partial p_\mu}
\end{equation}
where $p_\mu$ denotes canonical momentum. Affine parameter $d\lambda=ds/m$ is defined in terms of the line element $ds$ as in eq. \eqref{eq:hamscalar}, and we continue to denote $\dot x^\mu= \frac{dx^\mu}{d\lambda}$. For generality, the discussion below will not assume any specific form of $A_\mu$, save require
\begin{equation}\label{eq:Amu_requirements}
    \text{$A_\mu\rightarrow0$ as $r\rightarrow\infty$},\quad\quad A_\mu=(A_t(r),0,0,A_\phi(r))_{\theta=\pi/2}
\end{equation}
with $\theta=\pi/2$ the equatorial plane of orbit. The scattering angle may still be calculated from eq. \eqref{eq:chigenbase}. Equation \eqref{eq:hamjac_charged} yields
\begin{equation}\label{eq:dpihdrcharge}
    \frac{d\phi}{dr}=\frac{(L-eA_\phi)g^{\phi\phi}+(-E-eA_t)g^{\phi t}}{g^{rr}p_r},
\end{equation}
and per identification
\begin{eqnarray}
    h(r)=-\frac{(L-eA_\phi)g^{\phi\phi}+(-E-eA_t)g^{\phi t}}{g^{rr}}.
\end{eqnarray}
Asymptotically, $h(r)\sim 1/r^2$ as required, provided $A_\mu$ obeys eq. \eqref{eq:Amu_requirements}. The Hamilton-Jacobi equation from which $p_r$ may be determined, is found by normalizing $\dot x^2=-m^2$ with eq. \eqref{eq:hamcharged}
\begin{equation}\label{eq:hamjac_charged}
    -m^2=g^{\mu\nu}(p_\mu-eA_\mu)(p_\nu-eA_\nu).
\end{equation}
Having found the radial momentum with $g_{\mu\nu}$ implicitly in normal form, the corresponding potential $U$ is identified from eq. \eqref{eq:prnormalform}
\begin{equation}
p_r^2=T-U.
\end{equation}
Crucially, $T$ is now taken independent of both $G$ and $e$,
\begin{equation}
    T\equiv\left.p_r^2\right|_{G=e=0}
\end{equation}
as we are dealing with two interactions.\\
\\
We now treat the Kerr-Newman metric, and subsequently relativistic Coulomb scattering in flat space. Both conform to the requirements set by eq. \eqref{eq:chigen}. The respective electrodynamic potentials read
\begin{equation}\label{eq:Amu_kerrnewman}
    A_\mu=(-\frac{Q}{r},0,0,\frac{aQ}{r})\quad \text{(Kerr-Newman)}, \quad\quad  A_\mu=(-\frac{Q}{r},0,0,0) \quad\text{(Coulomb)}.
\end{equation}
where the Kerr-Newman potential, see ref. \cite{GP2009}, has been evaluated at $\theta=\pi/2$ in Boyer-Lindquist coordinates. The Coulomb potential may simply be considered the $G=a=0$ limit of the Kerr-Newman solution. \\
Let us therefore focus on Kerr-Newman. The normal form of the metric is recovered by the coordinate transformation of eq. \eqref{eq:normalcoordinateident}, $\rho^2=r^2+a^2$ where $r$ is the Boyer-Lindquist radial coordinate appearing in eq. \eqref{eq:kerrnewman}. Equations \eqref{eq:prnormalform} and \eqref{eq:hamjac_charged} then yield
\begin{equation}\label{eq:Ukerrnewman}
\begin{split}
    U(\rho)=&\frac{  Q e  \left(-2 a b \sqrt{\rho ^2-a^2} p_{\infty }+\rho ^2 \left(2 E \sqrt{\rho ^2-a^2}-  Q e \right)+a^2  Q e \right)}{\rho ^4-a^2 \rho ^2}\\
    &\begin{aligned}
    +G\frac{\left(Q^2-2 M \sqrt{\rho ^2-a^2}\right)}{\rho ^6-a^2 \rho ^4} \Big[&p_{\infty } \left(p_{\infty} \left(-\rho ^2 \left(a^2+b^2\right)+2 a^2 b^2+2 \rho ^4\right)+4 a b  Q e  \sqrt{\rho ^2-a^2}-2 a b E \rho ^2\right)\\
    &+2   Q \rho ^2 e  \left(  Q e -2 E \sqrt{\rho ^2-a^2}\right)-2 a^2 Q^2 e ^2+m^2 \rho ^4\Big]
    \end{aligned}\\
    &+\mathcal O(G^2),
\end{split}
\end{equation}
here truncated at $\mathcal O(G)$ for simplicity. The scattering angle is now found readily from eq. \eqref{eq:chigen}.  It may be expanded simultaneously in $G$ and $e$, ie. as a collective expansion in gravitational and electromagnetic interactions,
\begin{equation}
    \chi= \sum_{n=0}^\infty \sum_{j=0}^\infty \chi_{n,j}, \quad \chi_{n,j}\sim G^n e^j.
\end{equation}
As before, owing to the $Q^2$-dependence of the Kerr-Newman metric, $\chi_{n,j}$ may be decomposed as
\begin{equation}
    \chi_{n,j}=\sum_{k=0}^n \chi_{n,j}^{(k)}, \quad \chi_{n,j}^{(k)} \sim Q^{2k},
\end{equation}
where $k$ denotes the order of $Q^2$ coming from the metric. $\chi_{n,j}^{(k)}$ are tabulated in table \ref{tab:kerrnewmancharged}, B) up to $\mathcal O(G^3)$ and $\mathcal O(e^3)$. We again find definite structure
\begin{equation}\label{eq:chichargestructure}
    \chi_{n,j}^{(k)}\sim \pi^{(n+j+k+1) \text{mod }2} \frac{G^{n} e^j Q^{2k+j} M^{n -k }}{m^j \gamma^{j}v^{2 (j+n) } \left(b^2-a^2\right)^{(3 (n+j) +k -1)/2}} \sum_{p=0}^{2(n+j)-1} d_{n,j,k}a^pb^{2(n+j)-p} f_{n,j,k}(v),
\end{equation}
involving {\it only} whole powers of $a$ and $b$, contrary to fractional powers encountered with non-charged test-particles (table \ref{tab:kerrnewmanscalar}). This behavior is structurally equivalent to odd powers of $n+k$ for non-charged test-particles. $Q$ appears exclusively as a prefactor $Q^{2k+j}$, containing a contribution from the metric ($Q^{2k}$) and a contribution from the electromagnetic potential ($Q^j$). $f_{n,j,k}(v)$ are polynomials in $v$ and $d_{n,j,k}$ are numerical constants. As observed with scalar test-particles in eq. \eqref{eq:fnkstructn-k=1}, some angles factorise $v$ and $(a,b)$ dependence. Specifically, for $(n,j,k)=(0,1,0), (1,1,1), (2,1,2)$ and $(3,1,3)$, ie. when $j=1$, the sum in eq. \eqref{eq:chichargestructure} may be written as two terms
\begin{equation}
    \sum_{p=0}^{2(n+j)-1} d_{n,j,k}a^pb^{2(n+j)-p} f_{n,j,k}(v) = [f^{(v)}_{-}f^{(ab)}_{-+}+f^{(v)}_+f^{(ab)}_{+-}]_{n,j,k}
\end{equation}
where, again, $f^{(v)}$ and $f^{(ab)}$ are polynomials exclusively dependent on $v$ or $(a,b)$ respectively. Subscript $+,-$ signs indicate whether $v$, $a$ and $b$ appear as even or odd powers, ie. $f^{(ab)}_{+-}$ is even in $a$ but odd in $b$. The exact structure of polynomials $f$ depends on $n,j$ and $k$. Furthermore, $f^{(v)}$ and $f^{(ab)}$ share numerical coefficients of $v$ and $(a,b)$ terms. For instance, with $(n,j,k)=(2,1,2)$, the coefficients $\{1,5,10\}$ appear both in $f^{(v)}$ and $f^{(ab)}$. We expect this behavior to continue to higher orders in $n$ and $k$.\\
\\
It is similarly a straight forward matter to consider relativistic Coulomb scattering. The corresponding potential $U$ is given by the $G=a=0$ limit of eq. \eqref{eq:Ukerrnewman}. It only involves one coupling constant, $e$, in terms of which the scattering angle is expanded and presented in table \ref{tab:kerrnewmancharged}, A). Order by order comparisons match the small $Qe/J$ expansion of the well-known Coulomb scattering angle presented e.g. in ref. \cite{KOT2022},
\begin{equation}\label{eq:chiDarwin}
    \chi+\pi=\frac{J}{\sqrt{J^2-Q^2e^2}}\left(\pi-2\,\arctan\left(\frac{Qe}{v\sqrt{J^2-Q^2e^2}}\right)\right) \quad \quad \text{(Coulomb scattering)}.
\end{equation}
\section{Effective One-Body metric for Kerr-Schwarzschild black hole binaries}\label{sec:EOBdeformedKerr}
\subsection{The EOB formalism and construction of the EOB metric}
We now apply the method of ref. \cite{DHLV2022} to the full binary problem by means of Effective One-Body (EOB) theory. Dynamics of aligned, spinning binary black holes are mapped to an effective system consisting of a test-particle in an EOB metric $g_{\mu\nu}^{(eff)}$. This is achieved by directly relating equatorial scattering angles of the binary to those of the effective test-particle. We stress that our approach is not unique - different EOB metrics may be constructed by similar methods.\\
\\
The EOB formalism is first presented in the post-Minkowskian regime by reviewing ref. \cite{DV2021}. Consider equatorial scattering of binary Kerr-Schwarzschild black holes with masses $m_1$ and $m_2$, and spins $a_1= a$ and $a_2=0$. Center of mass coordinates may be used, in which $E$ denotes the total energy of the system, $v$ is the relative asymptotical velocity of the binary objects, $p_\infty$ is the asymptotical momentum of a single body, $L$ is the orbital angular momentum of the system, and $b$ denotes the impact parameter. These quantities are related by
\begin{equation}
v=|\mathbf v_1-\mathbf v_2|, \quad\quad p_\infty=\frac{(E^2-(m_1+m_2)^2)(E^2-(m_1-m_2)^2)}{4E^2}=\frac{m_1m_2}{E}\gamma v, \quad\quad L=bp_\infty,    
\end{equation}
where $\mathbf v_1$ and $\mathbf v_2$ are velocities of the individual black holes. For convenience we have defined the reduced mass $\mu$, total mass $M$, and asymptotical Lorentz contraction factor $\gamma$ as
\begin{equation}\label{eq:constdefsEOB}
    \mu\equiv\frac{m_1m_2}{m_1+m_2}, \quad M\equiv m_1+m_2, \quad \nu\equiv \frac{\mu}{M}, \quad \gamma=\frac{1}{\sqrt{1-v^2}}.
\end{equation}

This binary is now described by an effective system consisting of a test-particle of mass $\mu$ scattering on a metric $g_{\mu\nu}^{(eff)}$. Two-body quantities are related to effective ones by an EOB map, which we now present. Label with subscript "eff" quantities of the effective system. From kinematic considerations of the effective test-particle, one may establish the following EOB map
\begin{equation}\label{eq:eobmapgeneral}
    v_{eff}=v, \quad E=M\sqrt{1+2\nu\left(\frac{E_{eff}}{\mu}-1\right)}, \quad p_{eff}=\mu \gamma v=\frac{E}{M}p_\infty, \quad b_{eff}=b \Rightarrow L_{eff}=b_{eff}p_{eff}=L\frac{E}{M},
\end{equation}
where $E_{eff}=\mu\gamma$ is the test-particle energy, and $v_{eff},p_{eff},L_{eff}$ and $b_{eff}$ are the asymptotical test-particle velocity, asymptotical momentum, angular momentum and impact parameter respectively. Furthermore, the effective formalism should have some notion of spin, call it $a_{eff}$. EOB maps between spin have been discussed in detail in e.g. refs. \cite{Damour2001,BD2017,BD2018,Vines2018_scaspi}. We shall here use a simple map, namely
\begin{equation}\label{eq:eobmapspin}
    a_{eff}=a.
\end{equation}
Last, a map between scattering angles is required. Denote by $\chi$ and $\chi_{eff}$ the full two-body and effective test-particle scattering angles. The most natural mapping between these, as discussed in ref. \cite{Damour2016_grasca}, is simply
\begin{equation}\label{eq:eobmapchi}
    \chi_{eff}=\chi.
\end{equation}
Both angles are treated perturbatively in $G$, $\chi=\sum_{n=0}^\infty\chi_n G^n$. The calculation of $\chi_{eff}$ depends on $g_{\mu\nu}^{(eff)}$. With the full two-body system amplitude calculations give the individual components $\chi_n$. As stated before, ref. \cite{GOV2019} provides the scattering angle of a Kerr-Schwarzschild binary with spin parameter $a$ and respective masses $m_1$ and $m_2$,
\begin{equation}\label{eq:GOV_spins:a0_2PM}
    \begin{split}
    \chi^{(a,0)}=&
    \frac{2G E \left(-2 a v+b v^2+b\right)}{v^2 \left(b^2-a^2\right)}\\
    &\begin{split}
        +\frac{\pi E G^2}{32 b^6 v^4}\big[&5 a^4 \left(m_1 \left(35 v^4+180 v^2+24\right)+24 \left(4 m_2 v^2+m_2\right)\right)-96 a^3 b v \left(m_1 \left(5 v^2+4\right)+m_2 \left(2 v^2+3\right)\right)\\
        &+6 a^2 b^2 \left(m_1 \left(15 v^4+72 v^2+8\right)+4 m_2 \left(2 v^4+11 v^2+2\right)\right)-16 a b^3 v \left(3 v^2+2\right) (4 m_1+3 m_2)\\
        &+24 b^4 v^2 \left(v^2+4\right) (m_1+m_2)\big]+\mathcal O(a^5)\\
    \end{split}\\
    &+\mathcal O(G^3).
    \end{split}
\end{equation}
The spin configuration is indicated by $(a,0)$. As mentioned, the connection between eq. \eqref{eq:GOV_spins:a0_2PM} and Kerr black hole scattering has only been confirmed at $\mathcal O(G)$ for all orders in $a$, and at $\mathcal O(G^2)$ only up to $\mathcal O(a)$ \cite{GOV2019}. We restrict to these orders in the text, and provide a conjectural Kerr binary EOB metric matching the full $\mathcal O(a^4)$ result in table \ref{tab:kappaeobkerrgen2PM}.

The EOB metric $g_{\mu\nu}^{(eff)}$ is constructed in such a way that eq. \eqref{eq:eobmapchi} is satisfied. An ansatz is provided, the parameters of which are constrained by eq. \eqref{eq:eobmapchi}. As scattering is planar by construction, a rotationally symmetric, asymptotically flat ansatz is natural. We further demand that the metric reduces to Kerr in the test-particle limit $m_2\rightarrow0$.

One may thus naturally search for EOB metrics among generalizations of Kerr. It is instructive to review in short the results obtained by ref. \cite{DV2021}. Here binary Schwarzschild black holes were considered. The EOB metric had a generalized Schwarzschild form
\begin{equation}\label{eq:gdamgaardisotropicgen}
    ds^2=-\left(\frac{1-\alpha(r)}{1+\alpha(r)}\right)^2dt^2+\left(1+\alpha(r)\right)^4(dr^2+r^2d\Omega^2), \quad\quad \alpha(r)=\sum_{n=1}^\infty \alpha_n \frac{G^n}{r^n},
\end{equation}
which simply replaces $GM/(2r)\rightarrow \alpha(r)$ in the Schwarzschild spacetime written in isotropic coordinates. Adopting an isotropic calculation renders the scattering angle integrand, $d\phi/dr$ in a form comparable directly to amplitude calculations of refs. \cite{BCD2020,DV2021}. This allows specification of $\alpha_n$ directly from comparing scattering angle integrands, as opposed to merely the scattering angles. One finds, comparing up to 2PM,
\begin{equation}\label{eq:hdamgaard}
    \alpha_1=\frac{1}{2}E, \quad \alpha_2=-\frac{3(5\gamma^2-1)}{8(2\gamma^2-1)}\left(1-\frac{M}{E}\right)E^2.
\end{equation}
The Schwarzschild metric in isotropic coordinates, meaning $\alpha\rightarrow GM/(2r)$, is recovered in the test-particle limit.

Now turn to Kerr-Schwarzschild binaries. We provide an EOB metric ansatz written in non-isotropic coordinates, meaning $g_{rr}\ne r^2g_{\phi\phi}$. Our method differs from that of ref. \cite{DV2021} by explicitly computing the effective scattering angle, instead of comparing integrands. Comparing our result with eq. \eqref{eq:GOV_spins:a0_2PM} determines the EOB metric parameters. We choose an ansatz constructed from equatorial Kerr in Boyer-Lindquist coordinates by replacing $GM/r\rightarrow \kappa(r)$,
\begin{equation}\label{eq:geobkerrgen}
    g_{\mu\nu}^{(eff)}=
    \left(
\begin{array}{ccc}
 -\left(1-\kappa(r)\right) & 0 & -a\kappa(r) \\
 0 & \frac{r^2}{a^2+r^2 \left(1-\kappa(r)\right)} & 0 \\
 -a \kappa(r) & 0 & a^2 \left(1+\kappa(r)\right)+r^2 \\
\end{array}
\right), \quad \kappa(r)=\sum_{n=1}^\infty \kappa_n \frac{G^n}{r^n}
\end{equation}
parametrized with coordinates $\{t,r,\phi\}$. This choice of EOB metric is entirely arbitrary. A different ansatz could be equally viable, producing a different final result. The metric above resums orders in $a$, in a structure similar to Kerr(-Newman) metrics. Note however that $\kappa$ may also depend on $a$. Below we present $\kappa$ to $\mathcal O(a^4)$ as eq. \eqref{eq:GOV_spins:a0_2PM} naturally restricts hereto.

Equation \eqref{eq:chiscalar} readily yields the test-particle scattering angle. However, one should be careful about the parameter-dependence of $\kappa$, which may possibly depend on all test-particle quantities, namely energy $E_{eff}$, impact parameter $b_{eff}$, and angular momentum $L_{eff}$, including also asymptotical velocity $v$ and asymptotical momentum $p_\infty$. As an example, eq. \eqref{eq:gdamgaardisotropicgen} is dependent on $E$. However impact parameter, or equivalently, angular momentum dependence influences the application of $\frac{d}{db}$ in eq. \eqref{eq:chiscalar}. The result of calculating the post-Minkowskian scattering angle by eq. \eqref{eq:chiscalar} (see footnote \footnote{Note a subtlety with applying eq. \eqref{eq:chiscalar}. If $\kappa$ depends on $b$, $h(r)$ is no longer given by the usual scalar structure of eq. \eqref{eq:hrscalar}. However, as explained in ref. \cite{DHLV2022}, the scattering angle integral takes the form $d\phi/dr = dp_r/dL$ regardless. $L$ is angular momentum and $p_r$ is canonical radial momentum. Consequently eq. \eqref{eq:chiscalar} still applies}), with arbitrary b-dependent $\kappa$, is therefore an expression with first-order derivatives of $\kappa_n$ in $b$. In our case, it is sufficient to assume $\kappa_1$ is independent of $b$, which will yield consistent solutions. Other $\kappa_n\rightarrow \kappa_n(b)$ will remain unspecified functions of $b$. Imposing the EOB map of eqs. \eqref{eq:eobmapgeneral}-\eqref{eq:eobmapchi}, the effective test-particle scattering angle of the EOB metric from eq. \eqref{eq:geobkerrgen} becomes
\begin{equation}\label{eq:chieobkerrgen2pm}
    \begin{split}
    \chi_{eff}=&\frac{G \kappa _1 \left(-2 a v+b v^2+b\right)}{v^2 \left(b^2-a^2\right)}\\
    &
    \begin{split}
        +\frac{\pi  G^2}{128 b^6 v^4} &\Big(4 v^2 \big[\kappa _2(b) \left(5 a^4 \left(5 v^2+6\right)-48 a^3 b v+6 a^2 b^2 \left(3 v^2+4\right)-32 a b^3 v+8 b^4 \left(v^2+2\right)\right)\\
        &\hspace{4ex}-b \left(a^4 \left(5 v^2+6\right)-12 a^3 b v+2 a^2 b^2 \left(3 v^2+4\right)-16 a b^3 v+8 b^4 \left(v^2+2\right)\right) \frac{d\kappa_2(b)}{db}\big]\\
        &+\kappa _1^2 \big[5 a^4 \left(35 v^4+180 v^2+24\right)-96 a^3 b v \left(5 v^2+4\right)+6 a^2 b^2 \left(15 v^4+72 v^2+8\right)\\
        &\hspace{4ex}-64 a b^3 v \left(3 v^2+2\right)+24 b^4 v^2 \left(v^2+4\right)\big]\Big)+\mathcal O(a^5)
    \end{split}\\
    &+\mathcal O(G^3),
    \end{split}
\end{equation}
presented here up to $\mathcal O(G^2)$ and truncated to $\mathcal O(a^4)$ to facilitate direct comparison with eq. \eqref{eq:GOV_spins:a0_2PM}. Of course, the angle could be evaluated to any order in $G$ and $a$. At the current precision, equating eq. \eqref{eq:GOV_spins:a0_2PM} with eq. \eqref{eq:chieobkerrgen2pm}, $\kappa_n$ may be determined up to $\mathcal O(a^4)$. This produces a first order differential equation. The solutions will therefore naturally involve a $b$-independent integration constant $\mathcal C$. For brevity we will only present the $\mathcal O(a)$ result in the text, leaving the complete $\mathcal O(a^4)$ result to table \ref{tab:kappaeobkerrgen2PM}. One finds
\begin{equation}\label{eq:kappasolgenkerr}
    \begin{split}
    \kappa_1=&2E,\\
     \kappa_2=&\left(\frac{3 \left(v^2+4\right) E  (m_1+m_2-E )}{v^2+2}+\frac{b\,\mathcal C}{v^2+2}\right)\\
     &+a \left(\frac{E  \left[-2 (m_1-E) \left(3 v^4+4 v^2+8\right)-3 m_2 \left(v^4+4\right) \right]}{b v \left(v^2+2\right)^2}+\frac{2\mathcal C v}{(v^2+2)^2}\right)+\mathcal O(a^2).
    \end{split}
\end{equation}
The integration constant $\mathcal C$ may be set to $0$ by requiring $g_{\mu\nu}^{(eff)}\rightarrow \text{Kerr}$ in the test-particle limit of $m_2$. Contrary, the limit $m_1\rightarrow 0$ describes a spinning test-particle in Schwarzschild. Here the EOB metric does not reduce to Schwarzschild, as it also encodes spin of the probe.

We stress the simplicity of the EOB construction presented here. An EOB metric is readily found from an ansatz and EOB map, by directly matching scattering angles of the full two-body system with those of the effective test-particle. Non-metric, post-geodesic Finsler-type contributions of e.g. refs. \cite{Damour2020_claqua,BD2017} are not needed. Similar observations were made without spin in ref. \cite{DV2021}. We leave to future work the extension to higher Post Minkowskian orders, by the inclusion of $\kappa_{n>2}$ terms. We emphasize that the EOB metric found above is by no means unique. Other solutions, based on a different ansatz, may exist.

\begin{table}[H]
    \begin{tabular}{| c | p{0.80\paperwidth} |}
    \hline
    $n$ & $[\kappa_2$ at $\mathcal O(a^n)$] / $\frac{a^nE}{b^nv^{n}(v^2+2)^{n+1}}$\\\hline\hline
    $0$ & $3 \left(v^2+4\right) (m_1+m_2-E )$\\\hline
    $1$ & $-2 (m_1-E) \left(3 v^4+4 v^2+8\right)-3 m_2 \left(v^4+4\right)$\\\hline
    $2$ & $\frac{1}{4} (m_1-E) \left(6 v^8+18 v^6+148 v^4+96 v^2+32\right)+\frac{1}{4} m_2 \left(-v^8-14 v^6+72 v^4+16 v^2+32\right)$\\\hline
    $3$ & $-\frac{v^2}{4} (m_1-E) \left(12 v^8+144 v^6+176 v^4+352 v^2+192\right)+\frac{v^2}{4} m_2 \left(v^8-34 v^6+60 v^4-184 v^2-128\right)$\\\hline
    $4$ & $\frac{v^2}{16} (m_1-E) \left(22 v^{12}+230 v^{10}+576 v^8+3240 v^6+3456 v^4+1984 v^2+512\right)\newline
    +\frac{v^2}{16} m_2 \left(-27 v^{12}-156 v^{10}-484 v^8+1264 v^6+864 v^4+832 v^2+512\right)$\\\hline
    \end{tabular}
  \caption{Parameter $\kappa_2$ from the EOB metric of eq. \eqref{eq:geobkerrgen} computed up to $\mathcal O(a^4)$ by matching the scattering angle of the amplitude calculations from \cite{GOV2019}, provided in eq. \eqref{eq:GOV_spins:a0_2PM} above. Equation \eqref{eq:GOV_spins:a0_2PM} only matches black hole scattering at $\mathcal O(G^2)$ for $\mathcal O(a)$, with higher orders in $a$ matched only by conjecture. Each row, labelled by $n$ contains the $\mathcal O(a^n)$ contribution to $\kappa_2$. Expressions are given in terms of full two-body quantities. These solutions follow from equating eq. \eqref{eq:chieobkerrgen2pm} with eq. \eqref{eq:GOV_spins:a0_2PM} and inserting the 1PM result $\kappa_1=2E$ from eq. \eqref{eq:kappa2PMa1}. The arbitrary b-independent integration constant $\mathcal C$ associated with these solutions is set to $\mathcal C=0$.}
  \label{tab:kappaeobkerrgen2PM}
  \end{table}
\clearpage
\subsection{Comparison with earlier approaches without spin}
Our EOB metric is first compared to previous approaches without spin. Letting $a=0$ in the previous section, below our treatment is shown equivalent with that of refs. \cite{DV2021,Damour2020_claqua}. Particularly, the above results are related to ref. \cite{Damour2020_claqua} by a gauge transformation of the post-geodesic $\mathcal Q$ term and a coordinate-shift of the scattering angle integral to incorporate differing EOB maps. Ref. \cite{Damour2020_claqua} presents both Schwarzschild-like and isotropic EOB metrics. We will compare exclusively with the Schwarzschild type.\\
\\
Ref. \cite{Damour2020_claqua} uses an EOB map differing from eq. \eqref{eq:eobmapgeneral} and \eqref{eq:eobmapchi} purely by equating angular momenta and not impact parameters
\begin{equation}
    \mathllap{L=L_{eff}\quad}\Rightarrow\mathrlap{\quad b=b_{eff}\frac{E}{M}.\qquad\qquad\quad
        \text{(EOB map of ref. \cite{Damour2020_claqua})}}
\end{equation}
We have expressly indicated above that these are relations used in ref. \cite{Damour2020_claqua} and continue to denote $L$, $L_{eff}$ and $b$ as defined by eq. \eqref{eq:eobmapgeneral} in everything following below. This EOB map can readily be converted to eq. \eqref{eq:eobmapgeneral} as both maps employ $\chi=\chi_{eff}$. Equation \eqref{eq:chiscalar} therefore implies
\begin{equation}\label{eq:chiconnect}
    \frac{d}{dL}\int_{R_m}^\infty \! dR\, p_R = \frac{d}{dL_{eff}}\int_{r_m}^\infty\! dr\, p_r = \frac{M}{E}\frac{d}{dL}\int_{r_m}^\infty\! dr\, p_r,
\end{equation}
where the integral in $R$ is that of ref. \cite{Damour2020_claqua}, and the integral in $r$ that of the current paper. $p_R$ and $p_r$ are the corresponding canonical radial momenta.
This equality allows the natural identification, applicable when both formalisms use the same ansatz (Schwarzschild-like) metric,
\begin{equation}\label{eq:identcoords_EOBmaps}
    R=\frac{M}{E}r, \quad\quad p_R=p_r,
\end{equation}
amounting to a coordinate-shift of the scattering angle integral. This is exactly the identification made by ref. \cite{DV2021}, there interpreted as a canonical transformation between effective test-particle momentum of ref. \cite{Damour2020_claqua} and center of mass momentum of the full two-body system. Particularly, eq. (54) and (51) in ref. \cite{DV2021} is exactly eq. \eqref{eq:identcoords_EOBmaps} above.

We show below that the coordinate transformation of eq. \eqref{eq:identcoords_EOBmaps} yields a transformed momentum $p_R$ related to that found in ref. \cite{Damour2020_claqua} by a gauge-transformation in $\mathcal Q$. The formalisms are thus equivalent. To see this, consider the specific forms of $p_r$ and $p_R$. One finds
\begin{subequations}\label{eq:preobmine}
    \begin{equation}
    p_r=p_{eff}^2-\frac{L_{eff}^2}{r^2}-U(r) = p_{eff}^2-\frac{L^2}{R^2}-U(r),
    \end{equation}
with
    \begin{equation}\label{eq:myU}
        \begin{split}
    U(r)&=\frac{2 G M}{R^3} \left(L^2+\frac{\mu ^2 R^2 \left(v^2+1\right)}{v^2-1}\right)\\
    &+\frac{G^2 M^2 \left(L^2 \left(v^2-1\right) \left(3 M \left(v^2+4\right)+\left(v^2-4\right) E \right)+\mu ^2 R^2 \left(3 M \left(v^4+5 v^2+4\right)+\left(v^4+v^2+4\right) E \right)\right)}{R^4 \left(v^2-1\right) \left(v^2+2\right) E }\\
    &+\mathcal O(G^3),
        \end{split}
    \end{equation}
\end{subequations}
by inserting eq. \eqref{eq:geobkerrgen} in eq. \eqref{eq:hamscalar}, subsequently rewriting in terms of $R=\frac{M}{E}r$ and $L_{eff}=\frac{E}{M}L$.

Canonical momentum $p_R$ is calculated very similarly in ref. \cite{Damour2020_claqua}, from a modified Hamilton-Jacobi equation,
\begin{equation}\label{eq:eobmetricdamour}
    g_{D}^{\mu\nu}p_\mu p_\nu=-\mu^2-\mathcal Q,
\end{equation}
with post-geodesic correction $\mathcal Q$. Subscript D indicates quantities from ref. \cite{Damour2020_claqua}. It was shown in refs. \cite{Damour2017_higgra,Damour2016_grasca} that, i) the EOB metric $g^D_{\mu\nu}$ can be chosen to be a Schwarzschild metric with mass $M=m_1+m_2$, ii) $\mathcal Q$ starts at $G^2$, and iii) the scattering angle is invariant under certain gauge-like transformations of $\mathcal Q$. Furthermore, it was shown a suitable gauge could be chosen, such that $\mathcal Q$ only depends on $R$ and quantities relating to energy. We therefore write 
\begin{equation}
    \mathcal Q=\sum_{n=2}^\infty \mathcal Q_n(R) G^n
\end{equation}
without loss of generality. Truncating at $\mathcal O(G^2)$, one finds
\begin{subequations}\label{eq:preobDamour}
    \begin{equation}
        p_R=p_{eff}^2-\frac{L^2}{R^2}-U(R),
    \end{equation}
    \begin{equation}\label{eq:UD}
        U(R)=\frac{2 G M \left(L^2+\frac{\mu ^2 R^2 \left(v^2+1\right)}{v^2-1}\right)}{R^3}+G^2\mathcal Q_2+\frac{G^2 M^2 \left(4 L^2+R^2 \left(\frac{4 \mu ^2 \left(v^2+2\right)}{v^2-1}\right)\right)}{R^4}+\mathcal O(G^3),
    \end{equation}
\end{subequations}
where $U(R)$ denotes the corresponding potential, suitably identified as $g_{\mu\nu}^D$ is already in normal form. Note in particular that, in the potential, $\mathcal Q_2$ term appears isolated from other metric-dependent terms. This becomes important in a moment.

$\mathcal Q$ may be found from our EOB formalism by imposing eq. \eqref{eq:identcoords_EOBmaps} and inserting eqs. \eqref{eq:preobmine} and \eqref{eq:preobDamour}. One finds
\begin{equation}
    U(r)=U(R)
\end{equation}
to all orders in $G$. This equality is trivial up to 1PM as both EOB metrics are Schwarzschild-like. $\mathcal Q$ therefore starts at $\mathcal O(G^2)$ as expected. At 2PM the equality determines the lowest order coefficient of $\mathcal Q$
\begin{equation}\label{eq:q2my}
    \mathcal Q=\mathcal Q^M_2G^2+\mathcal O(G^3)=\frac{3 \left(v^2+4\right) (M-E ) \left(L^2 \left(v^2-1\right)+\mu ^2 R^2 \left(v^2+1\right)\right)}{R^2 \left(v^2-1\right) \left(v^2+2\right) E }\frac{G^2M^2}{R^2}+\mathcal O(G^3),
\end{equation}
corresponding to translating the EOB metric of eq. \eqref{eq:kappa2PMa1} to a Finsler-type post-geodesic form directly comparable with ref. \cite{Damour2017_higgra,Damour2016_grasca,Damour2020_claqua}. The obtained value of $\mathcal Q_2$ is dependent on angular momentum, and thus definitely not in the gauge used in ref. \cite{Damour2020_claqua}. Results for $\mathcal Q$ from refs. \cite{Damour2017_higgra,Damour2016_grasca,Damour2020_claqua} instead yield,
\begin{equation}\label{eq:q2Damour}
    \mathcal Q^D_2=\frac{3 \mu ^2 \left(v^2+4\right) (M-E )}{2 \left(v^2-1\right) E },
\end{equation}
which is different from eq. \eqref{eq:q2my} by
\begin{equation}\label{eq:deltaQ}
    \Delta \mathcal Q \equiv \mathcal Q_2^D-\mathcal Q^M_2 = -\frac{3 M^2 \left(v^2+4\right) (M-E ) \left(2 L^2 \left(v^2-1\right)+\mu ^2 R^2 v^2\right)}{2 R^4 \left(v^2-1\right) \left(v^2+2\right) E }.
\end{equation}
However, $\mathcal Q_2^M$ may be related to $\mathcal Q_2^D$ by a gauge-transformation. After all, the scattering angle calculated with each is the same. Gauge-transformations may be introduced by considering, as in ref. \cite{Damour2017_higgra}, the scattering angle integral eq. \eqref{eq:chiscalar}. Plug in eq. \eqref{eq:preobDamour} with unspecified $\mathcal Q_2$. Denoting by $\chi_{\mathcal Q}$ contributions to the scattering angle that come from $\mathcal Q$, one finds up to $\mathcal O(G^2)$
\begin{equation}\label{eq:chiq}
    \chi_{\mathcal Q}=\frac{1}{p_{eff}}\frac{d}{dL}\int_0^\infty\! du\,\mathcal Q_2(R)G^2+\mathcal O(G^3), \quad\quad R^2=u^2+L^2/p_{eff}^2=u^2+b^2\frac{M^2}{E^2},
\end{equation}
rewriting eq. \eqref{eq:chiscalar} in terms of $R$, imposing in effect the EOB map of ref. \cite{Damour2016_grasca}. Consider how the full scattering angle integral changes with $\mathcal Q^M_2$ compared to $\mathcal Q_2^D$, keeping the Schwarzschild metric $g_{\mu\nu}^D$. By construction, the scattering angles calculated in either case are equal. One therefore concludes
\begin{equation}\label{eq:gaugeQ}
    \frac{d}{dL}\int_0^\infty\! du\,\mathcal Q^D_2(R)=\frac{d}{dL}\int_0^\infty\! du\,\mathcal Q^M_2(R), \quad\quad R^2=u^2+b^2\frac{M^2}{E^2}.
\end{equation}
Subtracting the LHS from the RHS yields
\begin{equation}
    \frac{d}{dL}\int_0^\infty\! du\,\Delta \mathcal Q=0,
\end{equation}
as confirmed by an explicit calculation. One may thus interpret $\mathcal Q^M_2$ as being related to $\mathcal Q_2^D$ by a gauge-like transformation which keeps the scattering angle invariant,
\begin{equation}\label{eq:Q2gaugetransformation}
    \mathcal Q_2^D=\mathcal Q^M_2+\frac{d}{du} G(R), \quad R^2=u^2+b^2\frac{M^2}{E^2},
\end{equation}
with contributions from $G(R)$ vanishing in the integration limits of eq. \eqref{eq:chiq}. A very similar result was found in ref. \cite{Damour2017_higgra}, based on analogous considerations. Incidentally, an identical relation holds when imposing the EOB map of eq. \eqref{eq:eobmapgeneral}, with the replacements $b\rightarrow b\frac{E}{M}$, $R\rightarrow r$ and $u$ continuing to denote the integration parameter in eq. \eqref{eq:chiscalar}. For reference, $G(R)$ inferred from eq. \eqref{eq:deltaQ} is
\begin{equation}
    G(R)= \frac{3 G^2 \mu ^2 M^2 u v^2 \left(v^2+4\right) (M-E )}{2 \left(v^2-1\right) \left(v^2+2\right) E  \left(b^2\frac{M^2}{E^2}+u^2\right)}, \quad\quad u^2=R^2-b^2\frac{M^2}{E^2}.
\end{equation}
The gauge-transformation presented above is explicitly restricted to $\mathcal O(G)$. Similar arguments may be made at higher orders in $G$ from a more complicated gauge-relation derived from multiple terms of eq. \eqref{eq:chiscalar}.

Concluding, we may interpret our EOB metric without spin as a gauge-specific embedding of a Schwarzschild metric with post-geodesic $\mathcal Q$ contribution.\\
\\
We next compare our results with ref. \cite{DV2021}, which is related to ref. \cite{Damour2020_claqua} simply by the coordinate-shift of eq. \eqref{eq:identcoords_EOBmaps}. By means of the previous analysis, our result is therefore related to ref. \cite{DV2021} purely by the gauge-transformation in $\mathcal Q$. Apparent from isotropic and Schwarzschild EOB constructions of ref. \cite{Damour2020_claqua}, this transformation cannot be interpreted as a simple coordinate-shift between Schwarzschild and isotropic metrics, eqs. \eqref{eq:geobkerrgen} and \eqref{eq:gdamgaardisotropicgen}. Such behavior is to be expected since the EOB metric is by no means unique - multiple distinct metrics, ie. not related by coordinate transformations, may encode full binary dynamics. To see this, try bringing the isotropic EOB metric of ref. \cite{DV2021} to a Schwarzschild form which satisfies $g_{\phi\phi}=\rho^2$ and $g_{tt}=-1/g_{\rho\rho}$. Denoting the transformed coordinates $\{t,\rho,\phi\}$, and the transformed metric $\tilde g_{\mu\nu}$, one does not recover eq. \eqref{eq:geobkerrgen} with $a=0$, nor even Schwarzschild structure. Instead
\begin{subequations}
\begin{equation}\label{eq:gdamgaardschwarzschildform}
    \tilde g_{\mu\nu}=
    \left(
\begin{array}{ccc}
 -\left(1-\mathcal A\right) & 0 & 0 \\
 0 & \frac{1}{1-\widetilde{\mathcal A}} & 0 \\
0 & 0 & \rho^2 \\
\end{array}
\right),
\end{equation}
with 
\begin{equation}
    \mathcal A=\frac{4\alpha_1G}{\rho}+\frac{4\alpha_2G^2}{\rho^2}+\mathcal O(G^3),\quad\quad \widetilde{\mathcal A}=\frac{4\alpha_1G}{\rho}+\frac{8\alpha_2G^2}{\rho^2}+\mathcal O(G^3).
\end{equation}
$\mathcal A$ and $\widetilde{\mathcal A}$ are unequal starting at $\mathcal O(G^2)$, breaking the Schwarzschild-like characteristics. $\alpha_1$ and $\alpha_2$ are given in eq. \eqref{eq:hdamgaard}. We have used 
\begin{equation}
    r=\rho-2 \alpha_1 G-\left(\alpha_1^2+2 \alpha_2\right) G^2/\rho+\mathcal O(G^3)
\end{equation}
\end{subequations}
and neglected $\mathcal O(G^3)$ terms in the metric. This is warranted as $\alpha(r)$ is only specified up to $\mathcal O(G^2)$ anyway. A resummation in $G$ is presented merely to highlight deviations from the Schwarzschild form.
As a consequence, the gauge-transformation in $\mathcal Q$ of eq. \eqref{eq:Q2gaugetransformation}, does not correspond to a coordinate-transformation of the EOB metric.

\subsection{Comparison with earlier approaches with spin}
We now turn to consistency checks with earlier approaches including spin. Observations are similar to those without spin. At 1PM, our EOB metric is compared with that of Justin Vines ref. \cite{Vines2018_scaspi}, section III,b. In ref. \cite{Vines2018_scaspi} the EOB map of ref. \cite{Damour2020_claqua} is used, introducing a Kerr-like EOB metric with mass $M=m_1+m_2$ and spin $\tilde a=\frac{M}{E}a$. By an analysis identical to that without spin, one may propose the connection of eq. \eqref{eq:chiconnect} yielding eq. \eqref{eq:identcoords_EOBmaps}
\begin{equation}\label{eq:connectioneffformalisms}
    R=\frac{M}{E}r,\quad\quad p_R(R,\tilde a,L)=p_r(r,a,L_{eff}).
\end{equation}
This identification is indeed correct, as an explicit calculations of $p_r$ and $p_R$ shows. We remind the reader that $L$ and $L_{eff}$ are defined in eq. \eqref{eq:eobmapgeneral}. The simplicity of this coordinate connection is purely due to the Kerr-like structure of both metrics. Our result is thus consistent with ref. \cite{Vines2018_scaspi}.

\section{Spinning Binary EOB metrics and Newman-Janis Algorithm}\label{sec:EOBNJA}
Finally, we consider the Newman-Janis Algorithm (NJA), a remarkable procedure for introducing spin to a non-spinning, so-called seed metric. Although presented first as an ad-hoc observation by Newman and Janis in 1965 \cite{NJ1965}, its uniqueness has subsequently been investigated \cite{DS2000}. Beyond useful only for its original application in obtaining the Kerr metric from a Schwarzschild spacetime, it successfully produces also the Kerr-Newman metric from a Reissner Nordström seed.

As such it is interesting to explore the application of the NJA to EOB metrics. Namely, does one recover an EOB metric for aligned spinning Kerr black holes by applying the NJA to the EOB metric of two Schwarzschild black holes? We consider only equatorial, aligned spin scattering throughout. Adopting the EOB formalism above with a scalar effective test-particle of mass
\begin{equation}
    \mu=\frac{m_1m_2}{M},
\end{equation}
and EOB maps given by eqs. \eqref{eq:eobmapgeneral} and \eqref{eq:eobmapchi}, the requirement for such a metric is that it reproduces the aligned spinning binary scattering angles, eq. \eqref{eq:GOV2PMalignedspin}, order by order in $G$. We will concern ourselves with 1PM and 2PM scattering angles below.

We choose to construct our NJA metric from the non-spinning binary Schwarzschild EOB metric of eqs. \eqref{eq:gdamgaardisotropicgen} and \eqref{eq:hdamgaard}. We employ the NJA in the form discussed by refs. \cite{NJ1965,DS2000}, and restrict ourselves purely to the equatorial plane ($\theta=\pi/2$). Only the final result is presented in the main text, and specifics of the procedure are included in appendix A. A few remarks are worth noting. First, notice that eq. \eqref{eq:gdamgaardisotropicgen} is symmetric in binary masses $m_1$ and $m_2$. The NJA-transformed metric will naturally preserve this symmetry, and we therefore expect the metric to describe some equal-in-spin binary, for which the scattering angle of eq. \eqref{eq:GOV2PMalignedspin} is symmetric in masses. Furthermore, note we are only interested in scattering angles up to $\mathcal O(G^2)$, and therefore we take as the seed metric the $\mathcal O(G^2)$ accurate Schwarzschild-form of eq. \eqref{eq:gdamgaardisotropicgen} presented in eq. \eqref{eq:gdamgaardschwarzschildform}. In the application of the NJA, the transformed tetrads defining the transformed metric are similarly truncated to $\mathcal O(G^2)$ (see step 4 in appendix A).\\
\\
The NJA transformed metric of eq. \eqref{eq:gdamgaardschwarzschildform}, following the original procedure of refs. \cite{NJ1965,DS2000}, becomes
\begin{equation}\label{eq:gNJA}
    g_{\mu\nu}^\text{(NJA)}=
    \left(
\begin{array}{cccc}
 -\frac{r^2 \left(r (r-4 \alpha_1 G)-8 \alpha_2 G^2\right)}{\left(r^2-2 \alpha_2 G^2\right)^2} & 0 & -\frac{a r^2 (4 \alpha_1 G r+6 \alpha_2 G^2)}{\left(r^2-2 \alpha_2 G^2\right)^2} \\
 0 & \frac{r^2}{a^2+r (r-4 \alpha_1 G)-8 \alpha_2 G^2} & 0 \\
 -\frac{a r^2 (4 \alpha_1G r+6 \alpha_2 G^2)}{\left(r^2-2 \alpha_2 G^2\right)^2} & 0 & r^2 \left(\frac{a^2 \left(r (4 \alpha_1 G+r)+4 \alpha_2 G^2\right)}{\left(r^2-2 \alpha_2 G^2\right)^2}+1\right) \\
\end{array}
\right),
  \end{equation}
  which has Boyer-Lidquist structure reminiscent of a Kerr-Newman metric. Specifically, eq. \eqref{eq:gNJA} recovers the equatorial Kerr-Newman metric by setting $\alpha_1=\frac{M}{2}-\frac{Q^2}{4r}$ and $\alpha_2=0$, for which eq. \eqref{eq:gdamgaardschwarzschildform} becomes the Reissner-Nordström metric.

We now discuss EOB interpretation of such a metric, based on scattering data. The equatorial scattering angle of a scalar (effective) test-particle in $g_{\mu\nu}^{\text{(NJA)}}$ is readily calculated by eq. \eqref{eq:chiscalar},
  \begin{equation}\label{eq:chiEOBNJA}
    \begin{split}
    \chi^\text{NJA}_\text{EOB}=&\frac{2 G E  \left(-2 a v+b v^2+b\right)}{v^2 \left(b^2-a^2\right)}\\
    &
    +\frac{3 \pi  G^2 M \left(v^2+4\right) E }{4 b^2 v^2}-\frac{a \left(\pi  G^2 E  \left(9 M \left(v^2+4\right) v^2+\left(15 v^4+4 v^2+16\right) E \right)\right)}{4 \left(b^3 v^3 \left(v^2+1\right)\right)}+\mathcal O(a^2)\\
    &+\mathcal O(G^3),
    \end{split}
  \end{equation}
  after coordinate transformation $r^2\rightarrow\rho^2=r^2+a^2$ retrieving the normal form. $M=m_1+m_2$ denotes the total mass of the binary. To obtain EOB interpretation, this should reproduce a full binary scattering angle. We compare explicitly with the aligned Kerr binary from ref. \cite{GOV2019}, adopting the EOB map of eqs. \eqref{eq:eobmapgeneral} and \eqref{eq:eobmapchi}. The 2PM angle above has been truncated at $\mathcal O(a)$ to make such comparisons. Further analysis could equally be carried out at higher orders in $a$, however eq. \eqref{eq:GOV2PMalignedspin} is not yet confirmed to represent scattering of black holes at these orders. Consider all configurations $(a_1,a_2)$ of eq. \eqref{eq:GOV2PMalignedspin} conceivably described by the NJA. The 1PM scattering angle matches exactly that of eq. \eqref{eq:GOV2PMalignedspin} with configuration $a_1+a_2=a$. Spins $a_1$ and $a_2$ are further restricted by going to 2PM. Equation \eqref{eq:chiEOBNJA} is symmetric in masses $m_1$ and $m_2$. This symmetry is only recovered in equation \eqref{eq:GOV2PMalignedspin} if $a_1=a_2=a/2$. The corresponding full binary scattering angle from eq. \eqref{eq:GOV2PMalignedspin} is
  \begin{equation}\label{eq:GOV_spin:ahah_2PM}
    \begin{split}
    \chi^{(a/2,a/2)}=&\frac{2 G E  \left(-2 a v+b v^2+b\right)}{v^2 \left(b^2-a^2\right)}\\
    &
    +\frac{3 \pi  G^2 M \left(v^2+4\right) E }{4 b^2 v^2}-\frac{7 a \left(\pi  G^2 M \left(3 v^2+2\right) E \right)}{4 b^3 v^3}+\mathcal O(a^2)\\
    &+\mathcal O(G^3),
    \end{split}
  \end{equation}
which is the only binary Kerr scattering angle conceivably matched by eq. \eqref{eq:chiEOBNJA}. However comparing eq. \eqref{eq:chiEOBNJA} and eq. \eqref{eq:GOV_spin:ahah_2PM}, they do not match up. In the present setting, based on the non-spinning EOB metric of eq. \eqref{eq:gdamgaardisotropicgen} and eq. \eqref{eq:hdamgaard}, the NJA thus fails to produce an EOB metric for aligned spinning binaries. 
  
Although the present NJA interpretation fails, we have only considered scalar test-particles. One could equally consider scattering of a spinning test-particle on the metric of eq. \eqref{eq:gNJA}. Whether this combination has EOB interpretation, and produces correct two-body scattering angles, is not pursued here.
\section{Conclusion}\label{sec:conclusion}
Using the method of ref. \cite{DHLV2022}, eq. \eqref{eq:chigen}, scattering angles of systems involving both electromagnetic and gravitational interactions have been calculated. As a specific case-study, post-Minkowskian scattering angles in the Kerr-Newman metric have been computed for charged test-particles (table \ref{tab:kerrnewmanscalar} and table \ref{tab:kerrnewmancharged} B). Purely electromagnetic situations are also treatable - the well known Coulomb scattering angle is readily obtained in resummed form in the weak field limit (table \ref{tab:kerrnewmancharged} A). Charged test-particle scattering angles in Kerr-Newman show definite structure. At order $\mathcal O(G^n)$ and $\mathcal O(e^j)$,
\begin{equation}
    \chi=\sum_{n=0}^\infty\sum_{j=0}^\infty\sum_{k=0}^n \chi_{n,j}^{(k)}, \quad\quad \chi_n^{(k)}\sim \pi^{(n+j+k+1) \text{mod }2} \frac{G^{n} e^j Q^{2k+j} M^{n -k }}{m^j \gamma^{j}v^{2 (j+n) } \left(b^2-a^2\right)^{(3 (n+j) +k -1)/2}} f_{n,j,k}(v),
\end{equation}
where $f_{n,j,k}(v)$ is a polynomial in $v$. For $j=0$ and odd (even) $n$, $f_{n,j,k}(v)$ contains integer (fractional) orders of $a$ and $b$. When $j\ne 0$ only integer orders of $a$ and $b$ are present, with the structure
\begin{equation}
    f_{n,j,k}(v)\sim \sum_{p=0}^{2(n+j)} a^pb^{2(n+j)-p} \tilde f_{n,j,k}(v),
\end{equation}
where $\tilde f_{n,j,k}$ is some polynomial in $v$, independent of $a$ and $b$.\\
\\
The flexibility of the scattering angle formula of ref. \cite{DHLV2022} is aptly suited for EOB formalisms with planar orbits. An EOB metric describing full binary motion may be constructed by explicitly matching scattering angles and adopting the EOB map of ref. \cite{DV2021} (eq. \eqref{eq:eobmapgeneral} of the present paper). An ansatz,
\begin{equation}
    \begin{split}
    g_{\mu\nu}^{(eff)}=
    \left(
\begin{array}{ccc}
 -\left(1-\kappa(r)\right) & 0 & -a\kappa(r) \\
 0 & \frac{r^2}{a^2+r^2 \left(1-\kappa(r)\right)} & 0 \\
 -a \kappa(r) & 0 & a^2 \left(1+\kappa(r)\right)+r^2 \\
\end{array}
\right),\quad \kappa(r)=&\sum_{n=1}^\infty \kappa_n \frac{G^n}{r^n},
\end{split}
\end{equation}
for the EOB metric may be constructed from the Kerr metric in Boyer-Lindquist coordinates by replacing $2GM/r\rightarrow \kappa(r)$. The result is a resummed metric in spin parameter $a$ and Newtons gravitational constant $G$. $\kappa_n$ may be determined by matching the resulting test-particle scattering angle with that of a Kerr-Schwarzschild binary \cite{GOV2019}, order by order in $G$ and $a$. Our current treatment extends to 2PM and $\mathcal O(a)$, yielding
\begin{equation}\label{eq:kappa2PMa1}
    \begin{split}
    \kappa_1=& 2E,\\
     \kappa_2=&\left(\frac{3 \left(v^2+4\right) E  (m_1+m_2-E )}{v^2+2}\right)\\
     &+a \left(-\frac{E  \left(6 m_1 v^4+8 m_1 v^2+16 m_1+3 m_2 v^4+12 m_2-6 v^4 E -8 v^2 E -16 E \right)}{b v \left(v^2+2\right)^2}\right)+\mathcal O(a^2)
    \end{split}
\end{equation}
by requiring that the metric reduces to Kerr in the $m_2\rightarrow0$ limit. By identical comparisons with higher orders of $a$ in ref. \cite{GOV2019}, $\kappa_2$ is presented up to $\mathcal O(a^4)$ in table \ref{tab:kappaeobkerrgen2PM}. We see no obstruction in continuing the treatment to further orders in $G$ and $a$. In the non-spinning limit, our EOB formalism is related to previous approaches of refs. \cite{Damour2017_higgra,Damour2020_claqua,DV2021} through a gauge-transformation of the post-geodesic Finsler-type $\mathcal Q$ term. Including spin, at 1PM, our formalism is equivalent to that of ref. \cite{Vines2018_scaspi} after a coordinate-reparametrisation identical to the connection of ref. \cite{DV2021} with ref. \cite{Damour2020_claqua}. Furthermore, for spinning binaries, a subtlety with $b$ dependence of the metric was noted. Corresponding to angular momentum dependence, the EOB metric may readily depend on it, however requires extra care when evaluating test-particle scattering angles. In particular with the current ansatz metric, the scattering angle becomes a function of linear derivatives of $\kappa_n$ with respect to $b$.\\
\\
Supplementary to our EOB analysis, the Newman-Janis Algorithm was explored in its application to the non-spinning EOB metric of ref. \cite{DV2021}, see eq. \eqref{eq:gdamgaardisotropicgen}. The complexification technique and coordinate-transformations involved are those originally introduced by Newman and Janis \cite{NJ1965}, and the EOB map of eqs. \eqref{eq:eobmapgeneral} and \eqref{eq:eobmapchi} is assumed. Applying the NJA to eq. \eqref{eq:gdamgaardisotropicgen}, the result is
\begin{equation}
    g_{\mu\nu}^\text{(NJA)}=
    \left(
\begin{array}{cccc}
 -\frac{r^2 \left(r (r-4 \alpha_1 G)-8 \alpha_2 G^2\right)}{\left(r^2-2 \alpha_2 G^2\right)^2} & 0 & -\frac{a r^2 (4 \alpha_1 G r+6 \alpha_2 G^2)}{\left(r^2-2 \alpha_2 G^2\right)^2} \\
 0 & \frac{r^2}{a^2+r (r-4 \alpha_1 G)-8 \alpha_2 G^2} & 0 \\
 -\frac{a r^2 (4 \alpha_1G r+6 \alpha_2 G^2)}{\left(r^2-2 \alpha_2 G^2\right)^2} & 0 & r^2 \left(\frac{a^2 \left(r (4 \alpha_1 G+r)+4 \alpha_2 G^2\right)}{\left(r^2-2 \alpha_2 G^2\right)^2}+1\right) \\
\end{array}
\right),
  \end{equation}
derived with $\mathcal O(G^2)$-accurate manipulations. Scalar test-particle scattering angles in this metric are computed, and compared to different combinations of aligned spin $(a_1,a_2)$ of the full binary result \cite{GOV2019}. No spin map is a priori assumed. At 1PM we find that angles indeed match, due to natural Kerr-like structure in both cases. At 2PM, however, no spin configurations are possible for which the angles are equal. It is therefore, within the confines of the current construction, not possible to interpret the NJA transformed metric above as an EOB metric of aligned binary Kerr black holes. This analysis is based on a scalar effective test-particle. Alternatively, the same analysis including spin on the test-particle might conceivably have EOB interpretation. This is left to future work.\\
\\

\noindent{\sc Acknowledgements}: I thank Poul Henrik Damgaard and Andres Luna for helpful discussions. This work was supported in part by the Independent Research Fund Denmark, Grant No. 0135-00089B.
\newpage
\def\arraystretch{1.5}
\begin{table}[H]
    \begin{tabular}{| c | p{0.80\paperwidth} |}
    \hline
    $(n,k)$ & $\chi^{(k)}_n/\frac{G^n Q^{2k} M^{n-k}}{v^{2n}(b^2-a^2)^{(3n+k-1)/2}}$\\\hline\hline
    (1,0) & $2 \left(-2 a v+b v^2+b\right)$\\[3pt]
   (1,1) & $\pi/(2a^2) \left(2 a^3 v-a^2 \left(-v^2 \sqrt{b^2-a^2}+2 b v^2+b\right)+b^2 v^2 \left(b-\sqrt{b^2-a^2}\right)\right)$\\\hline
    (2,0) &  $
    \pi/(2 a^2)\big[-4 a^5 v-4 a^3 b^2 v \left(3 v^2+2\right)+2 a^2 b^2 v^2 \left(b \left(2 v^2+3\right)-v^2 \sqrt{b^2-a^2}\right)\newline+b^4 v^4 \left(\sqrt{b^2-a^2}-b\right)+a^4 \left(v^4 \sqrt{b^2-a^2}+3 b \left(4 v^2+1\right)\right)\big]$
    \\[3pt]
    (2,1) & $
    (8 a^3 + 24 a b^2) (1 + v^2)v + (-6 a^2 b - 2 b^3) (1 + 6 v^2 + 
      v^4)$
    \\[3pt]
    (2,2) & $
    -(3\pi)/(16 a^4)\big[4 a^7 \left(2 v^3+v\right)+4 a^5 b^2 v \left(3 v^2+4\right)+2 b^6 v^4 \left(b-\sqrt{b^2-a^2}\right)+a^2 b^4 v^4 \left(6 \sqrt{b^2-a^2}-7 b\right)\newline
    -a^6 \left(b \left(8 \left(v^2+3\right) v^2+3\right)-2 v^4 \sqrt{b^2-a^2}\right)+2 a^4 b^2 \left(b \left(4 v^4-3 v^2-1\right)-3 v^4 \sqrt{b^2-a^2}\right)\big]$\\\hline
    (3,0) & $
    \frac{1}{3} \big[4 a^5 v \left(3 v^4-10 v^2-9\right)+6 a^4 b \left(-v^6+15 v^4+45 v^2+5\right)-8 a^3 b^2 v \left(15 v^4+70 v^2+27\right)\newline
    +4 a^2 b^3 \left(11 v^6+135 v^4+105 v^2+5\right)-36 a b^4 v \left(5 \left(v^2+2\right) v^2+1\right)+2 b^5 \left(5 v^2 \left(v^4+9 v^2+3\right)-1\right)\big]$
    \\[3pt]
    (3,1) & $
    (3\pi)/(8 a^4)\big[2 a^9 v \left(20 v^2+9\right)+12 a^7 b^2 v \left(10 v^4+35 v^2+12\right)+6 a^5 b^4 v \left(15 v^4+40 v^2+8\right)+2 b^8 v^6 \left(b-\sqrt{b^2-a^2}\right)\newline
    +a^2 b^6 v^6 \left(8 \sqrt{b^2-a^2}-9 b\right)-a^8 \left(2 v^6 \sqrt{b^2-a^2}+15 b \left(8 v^4+12 v^2+1\right)\right)\newline
    -a^6 b^2 \left(5 b \left(8 v^6+72 v^4+63 v^2+4\right)-8 v^6 \sqrt{b^2-a^2}\right)
    +3 a^4 b^4 v^2 \left(b \left(4 v^4-15 v^2-10\right)-4 v^4 \sqrt{b^2-a^2}\right)\big]$
    \\[3pt]
    (3,2) & $
    (-4 a^5 - 40 a^3 b^2 - 20 a b^4) (3 + 10 v^2 + 
      3 v^4)v + (10 a^4 b + 20 a^2 b^3 + 2 b^5) (1 + 15 v^2 + 15 v^4 + 
      v^6)$
    \\[3pt]
    (3,3) & $
    (5\pi)/(128 a^6)\big[2 a^{11} v \left(24 v^4+40 v^2+9\right)+4 a^9 b^2 v \left(60 v^4+205 v^2+54\right)+18 a^7 b^4 v \left(5 v^4+20 v^2+8\right)\newline+8 b^{10} v^6 \left(b-\sqrt{b^2-a^2}\right)+4 a^2 b^8 v^6 \left(10 \sqrt{b^2-a^2}-11 b\right)-a^{10} \left(3 b \left(16 v^6+120 v^4+90 v^2+5\right)-8 v^6 \sqrt{b^2-a^2}\right)\newline+5 a^8 b^2 \left(b \left(8 v^6-108 v^4-123 v^2-8\right)-8 v^6 \sqrt{b^2-a^2}\right)-a^6 b^4 \left(b \left(118 v^6+45 v^4+60 v^2+8\right)-80 v^6 \sqrt{b^2-a^2}\right)\newline+a^4 b^6 v^6 \left(99 b-80 \sqrt{b^2-a^2}\right)\big]
    $ \\\hline
    (4,0) & $
    (3\pi)/(16 a^4)\big[-8 a^{11} v \left(14 v^2+5\right)-8 a^9 b^2 v \left(140 v^4+273 v^2+60\right)-8 a^7 b^4 v \left(70 v^6+455 v^4+392 v^2+40\right)\newline-56 a^5 b^6 v^3 \left(5 \left(v^2+4\right) v^2+8\right)+2 b^{10} v^8 \left(\sqrt{b^2-a^2}-b\right)+a^2 b^8 v^8 \left(11 b-10 \sqrt{b^2-a^2}\right)\newline+a^{10} \left(35 b \left(16 \left(v^4+v^2\right)+1\right)-2 v^8 \sqrt{b^2-a^2}\right)+10 a^8 b^2 \left(v^8 \sqrt{b^2-a^2}+7 b \left(16 v^6+56 v^4+26 v^2+1\right)\right)\newline+2 a^6 b^4 v^2 \left(7 b \left(8 v^6+120 v^4+195 v^2+40\right)-10 v^6 \sqrt{b^2-a^2}\right)+4 a^4 b^6 v^4 \left(5 v^4 \sqrt{b^2-a^2}+b \left(-4 v^4+35 v^2+35\right)\right)\big]$
    \\[3pt]
    (4,1) & $
    \frac{4}{3} \big[8 a^7 v \left(-v^6+7 v^4+21 v^2+5\right)+5 a^6 b \left(v^8-28 v^6-210 v^4-140 v^2-7\right)+24 a^5 b^2 v \left(7 \left(v^4+13 v^2+19\right) v^2+25\right)\newline-5 a^4 b^3 \left(13 v^8+420 v^6+1190 v^4+532 v^2+21\right)+40 a^3 b^4 v \left(7 \left(3 v^4+19 v^2+17\right) v^2+15\right)\newline-3 a^2 b^5 \left(31 v^8+700 v^6+1330 v^4+364 v^2+7\right)+40 a b^6 v \left(7 \left(v^4+5 v^2+3\right) v^2+1\right)\newline+b^7 \left(1-7 v^2 \left(v^6+20 v^4+30 v^2+4\right)\right)\big]$
    \\[3pt]
    (4,2) & $
    (15\pi)/(128 a^6)\big[-8 a^{13} v \left(56 v^4+84 v^2+15\right)-8 a^{11} b^2 v \left(7 \left(24 v^4+200 v^2+237\right) v^2+270\right)\newline-8 a^9 b^4 v \left(7 \left(60 v^4+395 v^2+372\right) v^2+360\right)-24 a^7 b^6 v \left(7 \left(5 \left(v^2+6\right) v^2+24\right) v^2+16\right)\newline+8 b^{12} v^8 \left(\sqrt{b^2-a^2}-b\right)+4 a^2 b^{10} v^8 \left(13 b-12 \sqrt{b^2-a^2}\right)+a^{12} \left(8 v^8 \sqrt{b^2-a^2}+21 b \left(8 \left(8 v^4+30 v^2+15\right) v^2+5\right)\right)\newline+4 a^{10} b^2 \left(7 b \left(16 v^8+360 v^6+930 v^4+395 v^2+15\right)-12 v^8 \sqrt{b^2-a^2}\right)\newline+2 a^8 b^4 \left(60 v^8 \sqrt{b^2-a^2}+7 b \left(8 v^8+540 v^6+1185 v^4+400 v^2+12\right)\right)\newline+4 a^6 b^6 v^2 \left(b \left(58 v^6+105 v^4+210 v^2+56\right)-40 v^6 \sqrt{b^2-a^2}\right)+a^4 b^8 v^8 \left(120 \sqrt{b^2-a^2}-143 b\right)\big]$
    \\[3pt]
    (4,3) & $
    (16 a^7 + 336 a^5 b^2 + 560 a^3 b^4 + 112 a b^6) (1 + 7 v^2 + 
      7 v^4 + v^6)v\newline+ (-14 a^6 b - 70 a^4 b^3 - 42 a^2 b^5 - 2 b^7) (1 + 
      28 v^2 + 70 v^4 + 28 v^6 + v^8)$
    \\[3pt]
    (4,4) & $
    (35\pi)/(2048 a^8)\big[-8 a^{15} v \left(2 v^2+5\right) \left(8 \left(v^4+v^2\right)+1\right)-24 a^{13} b^2 v \left(7 \left(8 v^4+52 v^2+45\right) v^2+40\right)\newline-120 a^{11} b^4 v \left(7 \left(2 v^4+15 v^2+16\right) v^2+16\right)-8 a^9 b^6 v \left(7 \left(5 \left(v^2+8\right) v^2+48\right) v^2+64\right)+16 b^{14} v^8 \left(\sqrt{b^2-a^2}-b\right)\newline+8 a^2 b^{12} v^8 \left(15 b-14 \sqrt{b^2-a^2}\right)+a^{14} \left(b \left(32 \left(4 v^6+56 v^4+105 v^2+35\right) v^2+35\right)-16 v^8 \sqrt{b^2-a^2}\right)\newline+14 a^{12} b^2 \left(8 v^8 \sqrt{b^2-a^2}+15 b \left(32 v^6+80 v^4+30 v^2+1\right)\right)\newline+42 a^{10} b^4 \left(b \left(16 v^8+80 v^6+225 v^4+104 v^2+4\right)-8 v^8 \sqrt{b^2-a^2}\right)\newline+4 a^8 b^6 \left(140 v^8 \sqrt{b^2-a^2}+b \left(-200 v^8+35 v^6+105 v^4+56 v^2+4\right)\right)+5 a^6 b^8 v^8 \left(143 b-112 \sqrt{b^2-a^2}\right)\newline+6 a^4 b^{10} v^8 \left(56 \sqrt{b^2-a^2}-65 b\right)\big]
    $ \\\hline
    \end{tabular}
  \caption{Scattering angle of a scalar test-particle in the Kerr-Newman metric, orbiting in the equatorial plane. $\chi^{(k)}_n$ is the $\mathcal O(G^n)$ contribution to the full scattering angle, of which the $Q^{2k}$-proportional part is taken. $k$ ranges from $0$ to $n$ owing to the $GQ^2$ charge-dependence of the metric, eq. \eqref{eq:kerrnewman}.}
  \label{tab:kerrnewmanscalar}
  \end{table}
  \def\arraystretch{1.5}
  \begin{table}[H]
    \begin{tabular}{| c | p{0.80\paperwidth} |}
    \multicolumn{2}{c}{\bf A) Coulomb scattering}\\\hline
    \multicolumn{2}{|l|}{$\chi=-\frac{2 Q \sqrt{1-v^2} e }{b m v^2}
    -\frac{\pi Q^2 \left(v^2-1\right) e ^2}{2 \left(b^2 m^2 v^2\right)}
    -\frac{2 Q^3 \left(1-v^2\right)^{3/2} \left(3 v^2-1\right)  e^3}{3 b^3 m^3 v^6}+\cdots=\frac{J}{\sqrt{J^2-Q^2e^2}}\left(\pi-2\,arctan\left(\frac{Qe}{v\sqrt{J^2-Q^2e^2}}\right)\right)-\pi$}\\\hline\multicolumn{2}{c}{}\\[-3mm]
    \multicolumn{2}{c}{\bf B) Kerr-Newman equatorial scattering of charged test-particle}\\\hline
    $(n,j,k)$ & $\chi^{(k)}_{n,j}/\frac{G^{n} e^j Q^{2k+j} M^{n -k }}{m^j v^{2 (j+n) } \left(b^2-a^2\right)^{(3 (n+j) +k -1)/2}}$\\\hline\hline
    (0,1,0) & $2 \sqrt{1-v^2} (a v-b)$\\[3pt]
    (0,2,0) & $\frac{1}{2} \pi  \left(v^2-1\right) (a-b v) \left(2 a^2 v-3 a b+b^2 v\right)$\\[3pt]
    (0,3,0) & $\frac{2}{3} \left(1-v^2\right)^{3/2} \big[a^5 v \left(v^2+9\right)-3 a^4 b \left(9 v^2+5\right)+2 a^3 b^2 v \left(7 v^2+27\right)-2 a^2 b^3 \left(21 v^2+5\right)+9 a b^4 v \left(v^2+1\right)+b^5 \left(1-3 v^2\right)\big]$\\\hline
    (1,1,0) & $3 \pi  \sqrt{1-v^2} (a v-b) (a-b v)^2$\\[3pt]
    (1,1,1) & $2\left( v^2+3\right) \left(-a^3 v-3 a b^2 v\right)+2\left(2 v^2+1\right) \left(3 a^2 b+b^3\right)$\\\hline
    (1,2,0) & $2 \left(v^2-1\right) \big[4 a^5 v \left(v^2+3\right)-3 a^4 b \left(v^4+18 v^2+5\right)+8 a^3 b^2 v \left(7 v^2+9\right)-2 a^2 b^3 \left(9 v^4+42 v^2+5\right)\newline+12 a b^4 v \left(3 v^2+1\right)+b^5 \left(1-3 v^2 \left(v^2+2\right)\right)\big]$\\[3pt]
    (1,2,1) & $-\frac{3}{8} \pi  \left(v^2-1\right) (a-b v) \left[4 a^4 v \left(2 v^2+3\right)-15 a^3 \left(4 b v^2+b\right)+3 a^2 b^2 v \left(8 v^2+27\right)-5 a b^3 \left(9 v^2+4\right)+3 b^4 v \left(v^2+4\right)\right]$\\\hline
    (1,3,0) & $\frac{15}{4} \pi  \left(1-v^2\right)^{3/2} (a-b v)^2 \big[a^5 v \left(4 v^2+5\right)-a^4 b \left(30 v^2+7\right)+a^3 b^2 v \left(13 v^2+46\right)\newline-a^2 b^3 \left(31 v^2+14\right)+4 a b^4 v \left(v^2+3\right)-2 b^5 v^2\big]$\\[3pt]
    (1,3,1) & $-\frac{4}{3} \left(1-v^2\right)^{3/2} \big[a^7 v \left(v^4+30 v^2+25\right)-5 a^6 b \left(15 v^4+50 v^2+7\right)+3 a^5 b^2 v \left(13 v^4+190 v^2+125\right)\newline-5 a^4 b^3 \left(85 v^4+190 v^2+21\right)+5 a^3 b^4 v \left(19 v^4+170 v^2+75\right)-3 a^2 b^5 \left(95 v^4+130 v^2+7\right)+25 a b^6 v \left(v^4+6 v^2+1\right)\newline+b^7 \left(1-5 v^2 \left(3 v^2+2\right)\right)\big]$\\\hline\hline
    (2,1,0) & $-2 \sqrt{1-v^2} \big[a^5 v \left(v^4-10 v^2-15\right)+15 a^4 b \left(v^4+6 v^2+1\right)-10 a^3 b^2 v \left(v^4+14 v^2+9\right)+10 a^2 b^3 \left(9 v^4+14 v^2+1\right)\newline-15 a b^4 v \left(v^4+6 v^2+1\right)+b^5 \left(15 v^4+10 v^2-1\right)\big]$\\[3pt]
    (2,1,1) & $-\frac{15}{4} \pi  \sqrt{1-v^2} (a v-b) (a-b v)^2 \left(a^2 \left(4 v^2+3\right)-14 a b v+b^2 \left(3 v^2+4\right)\right)$\\[3pt]
    (2,1,2) & $2 v\left(v^4+10 v^2+5\right) \sqrt{1-v^2} \left(a^5+10 a^3 b^2+5 a b^4\right)+2 \left(5 v^4+10 v^2+1\right) \sqrt{1-v^2} \left(-5 a^4 b-10 a^2 b^3-b^5\right)$\\\hline
    (2,2,0) & $\frac{45}{8} \pi  \left(v^2-1\right) (a-b v)^3 \left[a^4 \left(8 v^3+6 v\right)-7 a^3 \left(6 b v^2+b\right)+3 a^2 b^2 v \left(4 v^2+17\right)-7 a b^3 \left(3 v^2+2\right)+b^4 v \left(v^2+6\right)\right]$\\[3pt]
    (2,2,1) & $4 (v-1) (v+1) \big[-6 a^7 v \left(v^4+10 v^2+5\right)+5 a^6 b \left(v^6+45 v^4+75 v^2+7\right)-6 a^5 b^2 v \left(39 v^4+190 v^2+75\right)\newline +15 a^4 b^3 \left(5 v^6+85 v^4+95 v^2+7\right)-10 a^3 b^4 v \left(57 v^4+170 v^2+45\right)+3 a^2 b^5 \left(5 \left(5 v^4+57 v^2+39\right) v^2+7\right)\newline-30 a b^6 v \left(5 \left(v^2+2\right) v^2+1\right)+b^7 \left(5 v^2 \left(v^4+9 v^2+3\right)-1\right)\big]$\\
    (2,2,2) & $\frac{45}{128} \pi  \left(v^2-1\right) (a-b v) \big[2 a^6 v \left(8 \left(v^2+5\right) v^2+15\right)-35 a^5 b \left(8 v^4+12 v^2+1\right)+5 a^4 b^2 v \left(24 v^4+232 v^2+101\right)\newline-70 a^3 b^3 \left(v^2+2\right) \left(10 v^2+1\right)+10 a^2 b^4 v \left(9 v^4+101 v^2+58\right)-7 a b^5 \left(25 v^4+60 v^2+8\right)+5 b^6 v \left(v^4+12 v^2+8\right)\big]$\\\hline\hline
    (3,1,0) & $\frac{105}{4} \pi  \sqrt{1-v^2} (a v-b) (a-b v)^4 \left[a^2 \left(2 v^2+1\right)-6 a b v+b^2 \left(v^2+2\right)\right]$\\[3pt]
    (3,1,1) & $4 \sqrt{1-v^2} \big[a^7 v \left(v^6-21 v^4-105 v^2-35\right)+35 a^6 b \left(v^2+1\right) \left(v^4+14 v^2+1\right)-21 a^5 b^2 v \left(v^6+39 v^4+95 v^2+25\right)\newline+35 a^4 b^3 \left(15 v^6+85 v^4+57 v^2+3\right)-35 a^3 b^4 v \left(3 v^6+57 v^4+85 v^2+15\right)+21 a^2 b^5 \left(25 v^6+95 v^4+39 v^2+1\right)\newline-35 a b^6 v \left(v^2+1\right) \left(v^4+14 v^2+1\right)+b^7 \left(7 v^2 \left(5 \left(v^2+3\right) v^2+3\right)-1\right)\big]$\\[3pt]
    (3,1,2) & $\frac{315}{64} \pi  \sqrt{1-v^2} (a v-b) (a-b v)^2 \big[a^4 \left(8 v^4+20 v^2+5\right)-12 a^3 b v \left(6 v^2+5\right)+2 a^2 b^2 \left(10 v^4+79 v^2+10\right)\newline-12 a b^3 v \left(5 v^2+6\right)+b^4 \left(5 \left(v^2+4\right) v^2+8\right)\big]$\\[3pt]
    (3,1,3) & $2 v\left(v^6+21 v^4+35 v^2+7\right) \sqrt{1-v^2} \left(-a^7-21 a^5 b^2-35 a^3 b^4-7 a b^6\right)\newline
    +2 \left(7 \left(v^4+5 v^2+3\right) v^2+1\right) \sqrt{1-v^2} \left(7 a^6 b+35 a^4 b^3+21 a^2 b^5+b^7\right)$\\\hline
    \end{tabular}
  \caption{Weak-field scattering angle of a charged test-particle; A) off another stationary charge $Q$ (Coulomb scattering), and B) in the equatorial plane of the Kerr-Newman metric. $\chi^{(k)}_{n,j}$ is the scattering angle at $\mathcal O(G^n)$ and $\mathcal O(e^j)$. Results are decomposed into $Q^{2k}$-proportional pieces with $0\leq k\leq n$. Purely gravitational contributions to the scattering angle, $\chi_{n,0}^{(k)}$, may be found in table \ref{tab:kerrnewmanscalar} and are not displayed here.}
  \label{tab:kerrnewmancharged}
  \end{table}
\section*{Appendix}
\subsection{The Newman-Janis Algorithm and its application to equation \eqref{eq:gdamgaardschwarzschildform}} \label{app:newmanjanis}
For completeness, we first present the Newman-Janis algorithm with a review of ref. \cite{DS2000}, before discussing its specific application to eq. \eqref{eq:gdamgaardschwarzschildform}. The Newman-Janis Algorithm is a five-step procedure, proceeding as follows;
\begin{enumerate}
    \item Consider a given spherically symmetric seed metric to which spin should be endowed. Write it in advanced null coordinates, generically
    \begin{equation}\label{eq:nullcoorgenespacetime}
        ds^2=-e^{2\Phi(r)}du^2-2e^{\Phi(r)+\lambda(r)}dudr+r^2d\Omega^2,
    \end{equation}
    specifying functions $\Phi(r)$ and $\lambda(r)$ on a case-to-case basis.
    \item Invert the metric and express it in terms of a null tetrad of vectors $\{l^\mu,n^\mu,m^\mu\}$,
    \begin{equation}
      g^{\mu\nu}=-l^\mu n^\nu-n^\mu l^\nu+m^\mu \bar m^\nu + \bar m^\mu m^\nu,
    \end{equation}
    where $m^\mu$ can be complex valued, and $\bar m^\mu$ is its complex conjugate, such that the metric is real-valued. Furthermore $l,m,n$ obey 
    \begin{equation}
        l_\mu l^\mu = m_\mu m^\mu = n_\mu n^\mu = 0, \quad l_\mu n^\mu = -m_\mu \bar m^\mu = 1, \quad l_\mu m^\mu = n_\mu m^\mu = 0
        .
    \end{equation}
    One may abbreviate notation by introducing $Z_a^\mu=(l^\mu,n^\mu,m^\mu,\bar m^\mu)$.
    The metric of eq. \eqref{eq:nullcoorgenespacetime} has
  \begin{subequations}
    \begin{align}
    l^\mu &= \delta^\mu_1,\\
    n^\mu &= e^{-\lambda(r)-\Phi(r)}\delta_0^\mu-\frac{1}{2}e^{-2\lambda(r)}\delta_1^\mu,
    \\
     m^\mu&=\frac{1}{\sqrt2r}(\delta_2^\mu+\frac{i}{\sin\theta}\delta_3^\mu).
  \end{align}
  \end{subequations}
  \item Complexify coordinates by letting $x^\rho\rightarrow x'^\rho=x^\rho +iy^\rho \in \mathbb C$ and $Z^\mu_a\rightarrow Z'^\mu_a$. The transformation only requires $ g'_{\mu\nu} \in \mathbb R$ and $Z'=Z$ when $x'=\bar{x'}$, a bar denoting complex conjugation. Multiple complexification transformations obey this requirement, and as such the Newman-Janis Algorithm is ambiguous. The transformation adopted here is that originally introduced by Newman and Janis,
  \begin{equation}\label{eq:1/rtransformations}
    \frac{1}{r}\rightarrow \frac{1}{2}(\frac{1}{r'}+\frac{1}{\bar{r}'}), \quad\quad \frac{1}{r^2}\rightarrow \frac{1}{r'\bar{r}'}.
  \end{equation}
  Note in particular that $1/r^2$ terms transform differently from $1/r$ terms. Justification of this fact was given in ref. \cite{DS2000}. However, as will be demonstrated, this ambiguity vanishes in the equatorial plane.
  \item Perform a complex coordinate transformation $x'^\mu= x^\mu+i y^\mu(x)$ whereby $Z'^\mu_a$ transforms as
  \begin{equation}\label{eq:tetradtransform}
    Z^\mu_a=\frac{\partial x^\mu}{\partial x'^\nu} Z'^\nu_a.
\end{equation}
The new coordinates $x^\mu\in \{u,r,\theta,\phi\}$ are real-valued. To avoid clutter, names of coordinates have been reused from step 1. The particular transformation used by Newman and Janis is
\begin{equation}
  x^\mu = x'^\mu - ia\cos\theta(\delta_0^\mu-\delta_1^\mu),
\end{equation}
which introduces the spin parameter $a$. This transformation will be adopted here as well. In the equatorial plane $x^\mu=x'^{\mu}$, and step 3 becomes unambiguous with $1/r$ and $1/r^2$ transforming in the same way.
\item Finally, it is assumed that one can convert the metric to a Kerr-like structure by a coordinate transformation in $u$ and $\phi$ of the form $u=t+F(r)$, $\phi=\psi+G(r)$. The purpose is to remove all off-diagonal terms except $g_{t\phi}$.
\end{enumerate}
\hrule\vspace{1em}
We now apply the NJA to the seed metric of eqs. \eqref{eq:gdamgaardisotropicgen} and \eqref{eq:hdamgaard}. $\alpha_1$ and $\alpha_2$ are kept symbolic, their values only reinstated when analysis is complete.

Null coordinates may readily be identified from eq. \eqref{eq:gdamgaardschwarzschildform}, yielding the line element (step 1 of the NJA)
\begin{subequations}
\begin{equation}\label{eq:NJAEOBnullform}
    ds^2=-(1-\mathcal A)du^2-2\sqrt{\frac{1-\mathcal A}{1-\widetilde{\mathcal A}}}\,\,dud\rho+\rho^2d\Omega^2,
\end{equation}
where
\begin{equation}
    \mathcal A=\frac{4\alpha_1G}{\rho}+\frac{4\alpha_2G^2}{\rho^2}+\mathcal O(G^3)\quad\text{and}\quad \widetilde{\mathcal A}=\frac{4\alpha_1G}{\rho}+\frac{8\alpha_2G^2}{\rho^2}+\mathcal O(G^3).
\end{equation}
\end{subequations}

As mentioned in ref. \cite{DS2000}, the same analysis with a Reissner-Nordström metric yields instead 
\begin{equation}
    \mathcal A=\widetilde{\mathcal A}=2GM/\rho-GQ^2/\rho^2, \quad\quad \text{(Reissner Nordström metric)}
\end{equation}
simplifying the $dud\rho$ component. This difference follows from the observation of eq. \eqref{eq:gdamgaardschwarzschildform}; eq. \eqref{eq:gdamgaardisotropicgen} cannot be brought to exact Schwarzschild form, meaning $g_{\rho\rho}\ne-1/g_{tt}$. Were this the case, the NJA-transformed metric would simply have Kerr-Newman structure.

The line-element of eq. \eqref{eq:NJAEOBnullform} is written in terms of null-tetrads (step 2 of the NJA), which may easily by identified by comparing eq. \eqref{eq:NJAEOBnullform} with eq. \eqref{eq:nullcoorgenespacetime}
\begin{subequations}
    \begin{equation}
      e^{\Phi(\rho)}=(1-\mathcal A)^{1/2}, \quad e^{\lambda(\rho)}=(1-\widetilde{\mathcal A})^{-1/2},
    \end{equation}
    \begin{equation}
      l^\mu=\delta_1^\mu,\quad\quad
      n^\mu=\sqrt{\frac{1-\widetilde{\mathcal A}}{1-\mathcal A}}\delta_0^\mu-\frac{1}{2}(1-\widetilde{\mathcal A})\delta_1^\mu,\quad\quad
      m^\mu=\frac{1}{\sqrt2\rho}(\delta_2^\mu+\frac{i}{\sin\theta}\delta_3^\mu),
    \end{equation}
\end{subequations}
and coordinates are complexified (step 3 of the NJA) by replacing 
\begin{equation}
    1/\rho\rightarrow \frac{1}{2}(1/r'+1/\bar{r}') \text{ and } 1/\rho^2\rightarrow 1/(r'\bar{r}').
\end{equation}

A coordinate transformation $x^\mu=x'^\mu-ia\cos\theta(\delta_0^\mu-\delta_1^\mu)$ is now performed (step 4 of the NJA). Reuse notation by denoting $x^\mu=\{u,r,\theta,\phi\}$. The tetrads transform according to eq. \eqref{eq:tetradtransform}. In particular, expressions are reduced by expanding $\tilde Z^\mu_a$ to $\mathcal O(G^2)$.
The transformed metric in the equatorial ($\theta=\pi/2$) plane reads
\begin{align}\nonumber \label{eq:spin_metric}
    \tilde g_{\mu\nu}^\text{(NJA)}&=- \tilde l_\mu \tilde n_\nu-\tilde n_\mu \tilde l_\nu+\tilde m_\mu \bar{\tilde m}_\nu + \bar{\tilde m}_\mu \tilde m_\nu\\
    &=\left(
        \begin{array}{cccc}
         -(1-\mathcal A) & -\frac{\sqrt{1-\mathcal{A}}}{\sqrt{1-\mathcal{\widetilde A}}} & -a \left(\frac{\sqrt{1-\mathcal{A}}}{\sqrt{1-\mathcal{\widetilde A}}}+\mathcal A-1\right) \\
         -\frac{\sqrt{1-\mathcal{A}}}{\sqrt{1-\mathcal{\widetilde A}}} & 0 & \frac{a \sqrt{1-\mathcal{A}}}{\sqrt{1-\mathcal{\widetilde A}}} \\
         -a \left(\frac{\sqrt{1-\mathcal{A}}}{\sqrt{1-\mathcal{\widetilde A}}}+\mathcal A-1\right) & \frac{a \sqrt{1-\mathcal{A}}}{\sqrt{1-\mathcal{\widetilde A}}} & r^2-a^2 \left(-\frac{2 \sqrt{1-\mathcal A}}{\sqrt{1-\mathcal{\widetilde A}}}-\mathcal A+1\right) \\
        \end{array}
        \right)\\\nonumber
  &=
  \left(
  \begin{array}{cccc}
   -\frac{r^2 \left(r (r-4 \alpha_1 G)-8 \alpha_2 G^2\right)}{\left(r^2-2 \alpha_2 G^2\right)^2} & -\frac{r^2}{r^2-2 \alpha_2 G^2} & -\frac{a G r^2 (4 \alpha_1 r+6 \alpha_2 G)}{\left(r^2-2 \alpha_2 G^2\right)^2} \\
   -\frac{r^2}{r^2-2 \alpha_2 G^2} & 0 & -\frac{a r^2}{2 \alpha_2 G^2-r^2} \\
   -\frac{a G r^2 (4 \alpha_1 r+6 \alpha_2 G)}{\left(r^2-2 \alpha_2 G^2\right)^2} & -\frac{a r^2}{2 \alpha_2 G^2-r^2} & -r^2 \left(-\frac{a^2 \left(r (4 \alpha_1 G+r)+4 \alpha_2 G^2\right)}{\left(r^2-2 \alpha_2 G^2\right)^2}-1\right) \\
  \end{array}
  \right),
  \end{align}
  where the first expression uses all-order expressions of $\tilde Z^\mu_a$ and the second uses their expansion to 2PM.

  Last (step 5 of the NJA), $g_{tr}$ terms may be removed by the coordinate transformation $u\rightarrow t$ and $\phi\rightarrow \psi$
  \begin{equation}
  \begin{gathered}
    dt=du+D_{01}(r)dr,\quad\quad d\psi=d\phi+D_{31}(r)dr,\\
    D_{01}=\frac{a^2-2 \alpha_2 G^2+r^2}{a^2-4 \alpha_1 G r-8 \alpha_2 G^2+r^2}, \quad\quad D_{31}= \frac{a}{a^2-4 \alpha_1 G r-8 \alpha_2 G^2+r^2}
  \end{gathered}
\end{equation}
producing the final result
\begin{equation}
    g_{\mu\nu}^\text{(NJA)}=
    \left(
\begin{array}{cccc}
 -\frac{r^2 \left(r (r-4 \alpha_1 G)-8 \alpha_2 G^2\right)}{\left(r^2-2 \alpha_2 G^2\right)^2} & 0 & -\frac{a r^2 (4 \alpha_1 G r+6 \alpha_2 G^2)}{\left(r^2-2 \alpha_2 G^2\right)^2} \\
 0 & \frac{r^2}{a^2+r (r-4 \alpha_1 G)-8 \alpha_2 G^2} & 0 \\
 -\frac{a r^2 (4 \alpha_1G r+6 \alpha_2 G^2)}{\left(r^2-2 \alpha_2 G^2\right)^2} & 0 & r^2 \left(\frac{a^2 \left(r (4 \alpha_1 G+r)+4 \alpha_2 G^2\right)}{\left(r^2-2 \alpha_2 G^2\right)^2}+1\right) \\
\end{array}
\right),
  \end{equation}
which is the NJA metric from the seed of eq. \eqref{eq:gdamgaardschwarzschildform}, presented in eq. \eqref{eq:gNJA} of the main text.

\bibliography{Scattering_KerrNewman.bib}

\begin{thebibliography}{140}%
\makeatletter
\providecommand \@ifxundefined [1]{%
 \@ifx{#1\undefined}
}%
\providecommand \@ifnum [1]{%
 \ifnum #1\expandafter \@firstoftwo
 \else \expandafter \@secondoftwo
 \fi
}%
\providecommand \@ifx [1]{%
 \ifx #1\expandafter \@firstoftwo
 \else \expandafter \@secondoftwo
 \fi
}%
\providecommand \natexlab [1]{#1}%
\providecommand \enquote  [1]{``#1''}%
\providecommand \bibnamefont  [1]{#1}%
\providecommand \bibfnamefont [1]{#1}%
\providecommand \citenamefont [1]{#1}%
\providecommand \href@noop [0]{\@secondoftwo}%
\providecommand \href [0]{\begingroup \@sanitize@url \@href}%
\providecommand \@href[1]{\@@startlink{#1}\@@href}%
\providecommand \@@href[1]{\endgroup#1\@@endlink}%
\providecommand \@sanitize@url [0]{\catcode `\\12\catcode `\$12\catcode
  `\&12\catcode `\#12\catcode `\^12\catcode `\_12\catcode `\%12\relax}%
\providecommand \@@startlink[1]{}%
\providecommand \@@endlink[0]{}%
\providecommand \url  [0]{\begingroup\@sanitize@url \@url }%
\providecommand \@url [1]{\endgroup\@href {#1}{\urlprefix }}%
\providecommand \urlprefix  [0]{URL }%
\providecommand \Eprint [0]{\href }%
\providecommand \doibase [0]{http://dx.doi.org/}%
\providecommand \selectlanguage [0]{\@gobble}%
\providecommand \bibinfo  [0]{\@secondoftwo}%
\providecommand \bibfield  [0]{\@secondoftwo}%
\providecommand \translation [1]{[#1]}%
\providecommand \BibitemOpen [0]{}%
\providecommand \bibitemStop [0]{}%
\providecommand \bibitemNoStop [0]{.\EOS\space}%
\providecommand \EOS [0]{\spacefactor3000\relax}%
\providecommand \BibitemShut  [1]{\csname bibitem#1\endcsname}%
\let\auto@bib@innerbib\@empty
\bibitem [{\citenamefont {Abbott}\ \emph {et~al.}(2016)\citenamefont {Abbott}
  \emph {et~al.}}]{Abbott.etal.2016_GW150914}%
  \BibitemOpen
  \bibfield  {author} {\bibinfo {author} {\bibfnamefont {B.~P.}\ \bibnamefont
  {Abbott}} \emph {et~al.} (\bibinfo {collaboration} {LIGO Scientific,
  Virgo}),\ }\href {\doibase 10.1103/PhysRevLett.116.221101} {\bibfield
  {journal} {\bibinfo  {journal} {Phys. Rev. Lett.}\ }\textbf {\bibinfo
  {volume} {116}},\ \bibinfo {pages} {221101} (\bibinfo {year} {2016})},\
  \bibinfo {note} {[Erratum: Phys.Rev.Lett. 121, 129902 (2018)]},\ \Eprint
  {http://arxiv.org/abs/1602.03841} {arXiv:1602.03841 [gr-qc]} \BibitemShut
  {NoStop}%
\bibitem [{\citenamefont {L\"ammerzahl}\ and\ \citenamefont
  {Hackmann}(2016)}]{LH2016}%
  \BibitemOpen
  \bibfield  {author} {\bibinfo {author} {\bibfnamefont {C.}~\bibnamefont
  {L\"ammerzahl}}\ and\ \bibinfo {author} {\bibfnamefont {E.}~\bibnamefont
  {Hackmann}},\ }\href {\doibase 10.1007/978-3-319-20046-0_5} {\bibfield
  {journal} {\bibinfo  {journal} {Springer Proc. Phys.}\ }\textbf {\bibinfo
  {volume} {170}},\ \bibinfo {pages} {43} (\bibinfo {year} {2016})},\ \Eprint
  {http://arxiv.org/abs/1506.01572} {arXiv:1506.01572 [gr-qc]} \BibitemShut
  {NoStop}%
\bibitem [{\citenamefont {{Hagihara}}(1930)}]{Hagihara1931_therel}%
  \BibitemOpen
  \bibfield  {author} {\bibinfo {author} {\bibfnamefont {Y.}~\bibnamefont
  {{Hagihara}}},\ }\href@noop {} {\bibfield  {journal} {\bibinfo  {journal}
  {Japanese Journal of Astronomy and Geophysics}\ }\textbf {\bibinfo {volume}
  {8}},\ \bibinfo {pages} {67} (\bibinfo {year} {1930})}\BibitemShut {NoStop}%
\bibitem [{\citenamefont {{Darwin}}(1959)}]{Darwin1959}%
  \BibitemOpen
  \bibfield  {author} {\bibinfo {author} {\bibfnamefont {C.}~\bibnamefont
  {{Darwin}}},\ }\href {\doibase 10.1098/rspa.1959.0015} {\bibfield  {journal}
  {\bibinfo  {journal} {Proceedings of the Royal Society of London Series A}\
  }\textbf {\bibinfo {volume} {249}},\ \bibinfo {pages} {180} (\bibinfo {year}
  {1959})}\BibitemShut {NoStop}%
\bibitem [{\citenamefont {{Darwin}}(1961)}]{Darwin1961}%
  \BibitemOpen
  \bibfield  {author} {\bibinfo {author} {\bibfnamefont {C.}~\bibnamefont
  {{Darwin}}},\ }\href {\doibase 10.1098/rspa.1961.0142} {\bibfield  {journal}
  {\bibinfo  {journal} {Proceedings of the Royal Society of London Series A}\
  }\textbf {\bibinfo {volume} {263}},\ \bibinfo {pages} {39} (\bibinfo {year}
  {1961})}\BibitemShut {NoStop}%
\bibitem [{\citenamefont {Gackstatter}(1983)}]{Gackstatter1983_orbray}%
  \BibitemOpen
  \bibfield  {author} {\bibinfo {author} {\bibfnamefont {F.}~\bibnamefont
  {Gackstatter}},\ }\href@noop {} {\bibfield  {journal} {\bibinfo  {journal}
  {Annalen der Physik (Leipzig)}\ }\textbf {\bibinfo {volume} {40}},\ \bibinfo
  {pages} {352} (\bibinfo {year} {1983})}\BibitemShut {NoStop}%
\bibitem [{\citenamefont {Grunau}\ and\ \citenamefont
  {Kagramanova}(2011)}]{GK2011}%
  \BibitemOpen
  \bibfield  {author} {\bibinfo {author} {\bibfnamefont {S.}~\bibnamefont
  {Grunau}}\ and\ \bibinfo {author} {\bibfnamefont {V.}~\bibnamefont
  {Kagramanova}},\ }\href {\doibase 10.1103/PhysRevD.83.044009} {\bibfield
  {journal} {\bibinfo  {journal} {Phys. Rev. D}\ }\textbf {\bibinfo {volume}
  {83}},\ \bibinfo {pages} {044009} (\bibinfo {year} {2011})},\ \Eprint
  {http://arxiv.org/abs/1011.5399} {arXiv:1011.5399 [gr-qc]} \BibitemShut
  {NoStop}%
\bibitem [{\citenamefont {Carter}(1968)}]{Carter1968}%
  \BibitemOpen
  \bibfield  {author} {\bibinfo {author} {\bibfnamefont {B.}~\bibnamefont
  {Carter}},\ }\href {\doibase 10.1103/PhysRev.174.1559} {\bibfield  {journal}
  {\bibinfo  {journal} {Phys. Rev.}\ }\textbf {\bibinfo {volume} {174}},\
  \bibinfo {pages} {1559} (\bibinfo {year} {1968})}\BibitemShut {NoStop}%
\bibitem [{\citenamefont {Mino}(2003)}]{Mino2003_perapp}%
  \BibitemOpen
  \bibfield  {author} {\bibinfo {author} {\bibfnamefont {Y.}~\bibnamefont
  {Mino}},\ }\href {\doibase 10.1103/PhysRevD.67.084027} {\bibfield  {journal}
  {\bibinfo  {journal} {Phys. Rev. D}\ }\textbf {\bibinfo {volume} {67}},\
  \bibinfo {pages} {084027} (\bibinfo {year} {2003})},\ \Eprint
  {http://arxiv.org/abs/gr-qc/0302075} {arXiv:gr-qc/0302075} \BibitemShut
  {NoStop}%
\bibitem [{\citenamefont {{\v{C}}ade{\v{z}}}\ \emph {et~al.}(1998)\citenamefont
  {{\v{C}}ade{\v{z}}}, \citenamefont {Fanton},\ and\ \citenamefont
  {Calvani}}]{CFC1998_linemi}%
  \BibitemOpen
  \bibfield  {author} {\bibinfo {author} {\bibfnamefont {A.}~\bibnamefont
  {{\v{C}}ade{\v{z}}}}, \bibinfo {author} {\bibfnamefont {C.}~\bibnamefont
  {Fanton}}, \ and\ \bibinfo {author} {\bibfnamefont {M.}~\bibnamefont
  {Calvani}},\ }\href@noop {} {\bibfield  {journal} {\bibinfo  {journal} {New
  Astronomy}\ }\textbf {\bibinfo {volume} {3}},\ \bibinfo {pages} {647}
  (\bibinfo {year} {1998})}\BibitemShut {NoStop}%
\bibitem [{\citenamefont {Fujita}\ and\ \citenamefont
  {Hikida}(2009)}]{FH2009_anasol}%
  \BibitemOpen
  \bibfield  {author} {\bibinfo {author} {\bibfnamefont {R.}~\bibnamefont
  {Fujita}}\ and\ \bibinfo {author} {\bibfnamefont {W.}~\bibnamefont
  {Hikida}},\ }\href {\doibase 10.1088/0264-9381/26/13/135002} {\bibfield
  {journal} {\bibinfo  {journal} {Class. Quant. Grav.}\ }\textbf {\bibinfo
  {volume} {26}},\ \bibinfo {pages} {135002} (\bibinfo {year} {2009})},\
  \Eprint {http://arxiv.org/abs/0906.1420} {arXiv:0906.1420 [gr-qc]}
  \BibitemShut {NoStop}%
\bibitem [{\citenamefont {Dyson}\ and\ \citenamefont {van~de
  Meent}(2023)}]{DM2023}%
  \BibitemOpen
  \bibfield  {author} {\bibinfo {author} {\bibfnamefont {C.}~\bibnamefont
  {Dyson}}\ and\ \bibinfo {author} {\bibfnamefont {M.}~\bibnamefont {van~de
  Meent}},\ }\href@noop {} {\  (\bibinfo {year} {2023})},\ \Eprint
  {http://arxiv.org/abs/2302.03704} {arXiv:2302.03704 [gr-qc]} \BibitemShut
  {NoStop}%
\bibitem [{\citenamefont {Sharp}(1979)}]{Sharp1979_geobla}%
  \BibitemOpen
  \bibfield  {author} {\bibinfo {author} {\bibfnamefont {N.~A.}\ \bibnamefont
  {Sharp}},\ }\href {\doibase 10.1007/BF00756902} {\bibfield  {journal}
  {\bibinfo  {journal} {Gen. Rel. Grav.}\ }\textbf {\bibinfo {volume} {10}},\
  \bibinfo {pages} {659} (\bibinfo {year} {1979})}\BibitemShut {NoStop}%
\bibitem [{\citenamefont {Kraniotis}(2005)}]{Kraniotis2005_fradra}%
  \BibitemOpen
  \bibfield  {author} {\bibinfo {author} {\bibfnamefont {G.~V.}\ \bibnamefont
  {Kraniotis}},\ }\href {\doibase 10.1088/0264-9381/22/21/001} {\bibfield
  {journal} {\bibinfo  {journal} {Class. Quant. Grav.}\ }\textbf {\bibinfo
  {volume} {22}},\ \bibinfo {pages} {4391} (\bibinfo {year} {2005})},\ \Eprint
  {http://arxiv.org/abs/gr-qc/0507056} {arXiv:gr-qc/0507056} \BibitemShut
  {NoStop}%
\bibitem [{\citenamefont {Hackmann}\ and\ \citenamefont {Xu}(2013)}]{HX2013}%
  \BibitemOpen
  \bibfield  {author} {\bibinfo {author} {\bibfnamefont {E.}~\bibnamefont
  {Hackmann}}\ and\ \bibinfo {author} {\bibfnamefont {H.}~\bibnamefont {Xu}},\
  }\href {\doibase 10.1103/PhysRevD.87.124030} {\bibfield  {journal} {\bibinfo
  {journal} {Phys. Rev. D}\ }\textbf {\bibinfo {volume} {87}},\ \bibinfo
  {pages} {124030} (\bibinfo {year} {2013})},\ \Eprint
  {http://arxiv.org/abs/1304.2142} {arXiv:1304.2142 [gr-qc]} \BibitemShut
  {NoStop}%
\bibitem [{\citenamefont {Hackmann}(2010)}]{Hackman2010_thesis}%
  \BibitemOpen
  \bibfield  {author} {\bibinfo {author} {\bibfnamefont {E.}~\bibnamefont
  {Hackmann}},\ }\href@noop {} {\enquote {\bibinfo {title} {Geodesic equations
  in black hole space-times with cosmological constant},}\ } (\bibinfo {year}
  {2010})\BibitemShut {NoStop}%
\bibitem [{\citenamefont {Kol}\ \emph {et~al.}(2022)\citenamefont {Kol},
  \citenamefont {O'connell},\ and\ \citenamefont {Telem}}]{KOT2022}%
  \BibitemOpen
  \bibfield  {author} {\bibinfo {author} {\bibfnamefont {U.}~\bibnamefont
  {Kol}}, \bibinfo {author} {\bibfnamefont {D.}~\bibnamefont {O'connell}}, \
  and\ \bibinfo {author} {\bibfnamefont {O.}~\bibnamefont {Telem}},\ }\href
  {\doibase 10.1007/JHEP03(2022)141} {\bibfield  {journal} {\bibinfo  {journal}
  {JHEP}\ }\textbf {\bibinfo {volume} {03}},\ \bibinfo {pages} {141} (\bibinfo
  {year} {2022})},\ \Eprint {http://arxiv.org/abs/2109.12092} {arXiv:2109.12092
  [hep-th]} \BibitemShut {NoStop}%
\bibitem [{\citenamefont {Damour}(2016)}]{Damour2016_grasca}%
  \BibitemOpen
  \bibfield  {author} {\bibinfo {author} {\bibfnamefont {T.}~\bibnamefont
  {Damour}},\ }\href {\doibase 10.1103/PhysRevD.94.104015} {\bibfield
  {journal} {\bibinfo  {journal} {Phys. Rev. D}\ }\textbf {\bibinfo {volume}
  {94}},\ \bibinfo {pages} {104015} (\bibinfo {year} {2016})},\ \Eprint
  {http://arxiv.org/abs/1609.00354} {arXiv:1609.00354 [gr-qc]} \BibitemShut
  {NoStop}%
\bibitem [{\citenamefont {Damour}(2020{\natexlab{a}})}]{Damour2020_claqua}%
  \BibitemOpen
  \bibfield  {author} {\bibinfo {author} {\bibfnamefont {T.}~\bibnamefont
  {Damour}},\ }\href {\doibase 10.1103/PhysRevD.102.024060} {\bibfield
  {journal} {\bibinfo  {journal} {Phys. Rev. D}\ }\textbf {\bibinfo {volume}
  {102}},\ \bibinfo {pages} {024060} (\bibinfo {year} {2020}{\natexlab{a}})},\
  \Eprint {http://arxiv.org/abs/1912.02139} {arXiv:1912.02139 [gr-qc]}
  \BibitemShut {NoStop}%
\bibitem [{\citenamefont {Damgaard}\ \emph {et~al.}(2022)\citenamefont
  {Damgaard}, \citenamefont {Hoogeveen}, \citenamefont {Luna},\ and\
  \citenamefont {Vines}}]{DHLV2022}%
  \BibitemOpen
  \bibfield  {author} {\bibinfo {author} {\bibfnamefont {P.~H.}\ \bibnamefont
  {Damgaard}}, \bibinfo {author} {\bibfnamefont {J.}~\bibnamefont {Hoogeveen}},
  \bibinfo {author} {\bibfnamefont {A.}~\bibnamefont {Luna}}, \ and\ \bibinfo
  {author} {\bibfnamefont {J.}~\bibnamefont {Vines}},\ }\href {\doibase
  10.1103/PhysRevD.106.124030} {\bibfield  {journal} {\bibinfo  {journal}
  {Phys. Rev. D}\ }\textbf {\bibinfo {volume} {106}},\ \bibinfo {pages}
  {124030} (\bibinfo {year} {2022})},\ \Eprint
  {http://arxiv.org/abs/2208.11028} {arXiv:2208.11028 [hep-th]} \BibitemShut
  {NoStop}%
\bibitem [{\citenamefont {Wallace}(1973)}]{Wallace1973}%
  \BibitemOpen
  \bibfield  {author} {\bibinfo {author} {\bibfnamefont {S.~J.}\ \bibnamefont
  {Wallace}},\ }\href {\doibase 10.1016/0003-4916(73)90008-0} {\bibfield
  {journal} {\bibinfo  {journal} {Annals Phys.}\ }\textbf {\bibinfo {volume}
  {78}},\ \bibinfo {pages} {190} (\bibinfo {year} {1973})}\BibitemShut
  {NoStop}%
\bibitem [{\citenamefont {Bjerrum-Bohr}\ \emph {et~al.}(2020)\citenamefont
  {Bjerrum-Bohr}, \citenamefont {Cristofoli},\ and\ \citenamefont
  {Damgaard}}]{BCD2020}%
  \BibitemOpen
  \bibfield  {author} {\bibinfo {author} {\bibfnamefont {N.~E.~J.}\
  \bibnamefont {Bjerrum-Bohr}}, \bibinfo {author} {\bibfnamefont
  {A.}~\bibnamefont {Cristofoli}}, \ and\ \bibinfo {author} {\bibfnamefont
  {P.~H.}\ \bibnamefont {Damgaard}},\ }\href {\doibase 10.1007/JHEP08(2020)038}
  {\bibfield  {journal} {\bibinfo  {journal} {JHEP}\ }\textbf {\bibinfo
  {volume} {08}},\ \bibinfo {pages} {038} (\bibinfo {year} {2020})},\ \Eprint
  {http://arxiv.org/abs/1910.09366} {arXiv:1910.09366 [hep-th]} \BibitemShut
  {NoStop}%
\bibitem [{\citenamefont {Bl\"umlein}\ \emph {et~al.}(2020)\citenamefont
  {Bl\"umlein}, \citenamefont {Maier}, \citenamefont {Marquard},\ and\
  \citenamefont {Sch\"afer}}]{BMMS2020}%
  \BibitemOpen
  \bibfield  {author} {\bibinfo {author} {\bibfnamefont {J.}~\bibnamefont
  {Bl\"umlein}}, \bibinfo {author} {\bibfnamefont {A.}~\bibnamefont {Maier}},
  \bibinfo {author} {\bibfnamefont {P.}~\bibnamefont {Marquard}}, \ and\
  \bibinfo {author} {\bibfnamefont {G.}~\bibnamefont {Sch\"afer}},\ }\href
  {\doibase 10.1016/j.physletb.2020.135496} {\bibfield  {journal} {\bibinfo
  {journal} {Phys. Lett. B}\ }\textbf {\bibinfo {volume} {807}},\ \bibinfo
  {pages} {135496} (\bibinfo {year} {2020})},\ \Eprint
  {http://arxiv.org/abs/2003.07145} {arXiv:2003.07145 [gr-qc]} \BibitemShut
  {NoStop}%
\bibitem [{\citenamefont {Bini}\ \emph
  {et~al.}(2020{\natexlab{a}})\citenamefont {Bini}, \citenamefont {Damour},\
  and\ \citenamefont {Geralico}}]{BDG2020}%
  \BibitemOpen
  \bibfield  {author} {\bibinfo {author} {\bibfnamefont {D.}~\bibnamefont
  {Bini}}, \bibinfo {author} {\bibfnamefont {T.}~\bibnamefont {Damour}}, \ and\
  \bibinfo {author} {\bibfnamefont {A.}~\bibnamefont {Geralico}},\ }\href
  {\doibase 10.1103/PhysRevD.102.084047} {\bibfield  {journal} {\bibinfo
  {journal} {Phys. Rev. D}\ }\textbf {\bibinfo {volume} {102}},\ \bibinfo
  {pages} {084047} (\bibinfo {year} {2020}{\natexlab{a}})},\ \Eprint
  {http://arxiv.org/abs/2007.11239} {arXiv:2007.11239 [gr-qc]} \BibitemShut
  {NoStop}%
\bibitem [{\citenamefont {Bini}\ \emph
  {et~al.}(2020{\natexlab{b}})\citenamefont {Bini}, \citenamefont {Damour},\
  and\ \citenamefont {Geralico}}]{BGD2020_secondpaper}%
  \BibitemOpen
  \bibfield  {author} {\bibinfo {author} {\bibfnamefont {D.}~\bibnamefont
  {Bini}}, \bibinfo {author} {\bibfnamefont {T.}~\bibnamefont {Damour}}, \ and\
  \bibinfo {author} {\bibfnamefont {A.}~\bibnamefont {Geralico}},\ }\href
  {\doibase 10.1103/PhysRevD.102.024061} {\bibfield  {journal} {\bibinfo
  {journal} {Phys. Rev. D}\ }\textbf {\bibinfo {volume} {102}},\ \bibinfo
  {pages} {024061} (\bibinfo {year} {2020}{\natexlab{b}})},\ \Eprint
  {http://arxiv.org/abs/2004.05407} {arXiv:2004.05407 [gr-qc]} \BibitemShut
  {NoStop}%
\bibitem [{\citenamefont {Bl{\"u}mlein}\ \emph {et~al.}(2021)\citenamefont
  {Bl{\"u}mlein}, \citenamefont {Maier}, \citenamefont {Marquard},\ and\
  \citenamefont {Sch{\"a}fer}}]{BMMS2021}%
  \BibitemOpen
  \bibfield  {author} {\bibinfo {author} {\bibfnamefont {J.}~\bibnamefont
  {Bl{\"u}mlein}}, \bibinfo {author} {\bibfnamefont {A.}~\bibnamefont {Maier}},
  \bibinfo {author} {\bibfnamefont {P.}~\bibnamefont {Marquard}}, \ and\
  \bibinfo {author} {\bibfnamefont {G.}~\bibnamefont {Sch{\"a}fer}},\ }\href
  {\doibase 10.1016/j.nuclphysb.2021.115352} {\bibfield  {journal} {\bibinfo
  {journal} {Nuclear Physics B}\ }\textbf {\bibinfo {volume} {965}},\ \bibinfo
  {pages} {115352} (\bibinfo {year} {2021})}\BibitemShut {NoStop}%
\bibitem [{\citenamefont {Foffa}\ \emph {et~al.}(2021)\citenamefont {Foffa},
  \citenamefont {Sturani},\ and\ \citenamefont {Torres~Bobadilla}}]{FRT2021}%
  \BibitemOpen
  \bibfield  {author} {\bibinfo {author} {\bibfnamefont {S.}~\bibnamefont
  {Foffa}}, \bibinfo {author} {\bibfnamefont {R.}~\bibnamefont {Sturani}}, \
  and\ \bibinfo {author} {\bibfnamefont {W.~J.}\ \bibnamefont
  {Torres~Bobadilla}},\ }\href {\doibase 10.1007/JHEP02(2021)165} {\bibfield
  {journal} {\bibinfo  {journal} {JHEP}\ }\textbf {\bibinfo {volume} {02}},\
  \bibinfo {pages} {165} (\bibinfo {year} {2021})},\ \Eprint
  {http://arxiv.org/abs/2010.13730} {arXiv:2010.13730 [gr-qc]} \BibitemShut
  {NoStop}%
\bibitem [{\citenamefont {Bl\"umlein}\ \emph {et~al.}(2021)\citenamefont
  {Bl\"umlein}, \citenamefont {Maier}, \citenamefont {Marquard},\ and\
  \citenamefont {Sch\"afer}}]{BMM2021}%
  \BibitemOpen
  \bibfield  {author} {\bibinfo {author} {\bibfnamefont {J.}~\bibnamefont
  {Bl\"umlein}}, \bibinfo {author} {\bibfnamefont {A.}~\bibnamefont {Maier}},
  \bibinfo {author} {\bibfnamefont {P.}~\bibnamefont {Marquard}}, \ and\
  \bibinfo {author} {\bibfnamefont {G.}~\bibnamefont {Sch\"afer}},\ }\href
  {\doibase 10.1016/j.physletb.2021.136260} {\bibfield  {journal} {\bibinfo
  {journal} {Phys. Lett. B}\ }\textbf {\bibinfo {volume} {816}},\ \bibinfo
  {pages} {136260} (\bibinfo {year} {2021})},\ \Eprint
  {http://arxiv.org/abs/2101.08630} {arXiv:2101.08630 [gr-qc]} \BibitemShut
  {NoStop}%
\bibitem [{\citenamefont {K\"alin}\ \emph
  {et~al.}(2020{\natexlab{a}})\citenamefont {K\"alin}, \citenamefont {Liu},\
  and\ \citenamefont {Porto}}]{KLP2020}%
  \BibitemOpen
  \bibfield  {author} {\bibinfo {author} {\bibfnamefont {G.}~\bibnamefont
  {K\"alin}}, \bibinfo {author} {\bibfnamefont {Z.}~\bibnamefont {Liu}}, \ and\
  \bibinfo {author} {\bibfnamefont {R.~A.}\ \bibnamefont {Porto}},\ }\href
  {\doibase 10.1103/PhysRevLett.125.261103} {\bibfield  {journal} {\bibinfo
  {journal} {Phys. Rev. Lett.}\ }\textbf {\bibinfo {volume} {125}},\ \bibinfo
  {pages} {261103} (\bibinfo {year} {2020}{\natexlab{a}})},\ \Eprint
  {http://arxiv.org/abs/2007.04977} {arXiv:2007.04977 [hep-th]} \BibitemShut
  {NoStop}%
\bibitem [{\citenamefont {K\"alin}\ \emph
  {et~al.}(2020{\natexlab{b}})\citenamefont {K\"alin}, \citenamefont {Liu},\
  and\ \citenamefont {Porto}}]{KLP2020_contid}%
  \BibitemOpen
  \bibfield  {author} {\bibinfo {author} {\bibfnamefont {G.}~\bibnamefont
  {K\"alin}}, \bibinfo {author} {\bibfnamefont {Z.}~\bibnamefont {Liu}}, \ and\
  \bibinfo {author} {\bibfnamefont {R.~A.}\ \bibnamefont {Porto}},\ }\href
  {\doibase 10.1103/PhysRevD.102.124025} {\bibfield  {journal} {\bibinfo
  {journal} {Phys. Rev. D}\ }\textbf {\bibinfo {volume} {102}},\ \bibinfo
  {pages} {124025} (\bibinfo {year} {2020}{\natexlab{b}})},\ \Eprint
  {http://arxiv.org/abs/2008.06047} {arXiv:2008.06047 [hep-th]} \BibitemShut
  {NoStop}%
\bibitem [{\citenamefont {Dlapa}\ \emph
  {et~al.}(2022{\natexlab{a}})\citenamefont {Dlapa}, \citenamefont {K\"alin},
  \citenamefont {Liu},\ and\ \citenamefont {Porto}}]{PKLP2022}%
  \BibitemOpen
  \bibfield  {author} {\bibinfo {author} {\bibfnamefont {C.}~\bibnamefont
  {Dlapa}}, \bibinfo {author} {\bibfnamefont {G.}~\bibnamefont {K\"alin}},
  \bibinfo {author} {\bibfnamefont {Z.}~\bibnamefont {Liu}}, \ and\ \bibinfo
  {author} {\bibfnamefont {R.~A.}\ \bibnamefont {Porto}},\ }\href {\doibase
  10.1103/PhysRevLett.128.161104} {\bibfield  {journal} {\bibinfo  {journal}
  {Phys. Rev. Lett.}\ }\textbf {\bibinfo {volume} {128}},\ \bibinfo {pages}
  {161104} (\bibinfo {year} {2022}{\natexlab{a}})},\ \Eprint
  {http://arxiv.org/abs/2112.11296} {arXiv:2112.11296 [hep-th]} \BibitemShut
  {NoStop}%
\bibitem [{\citenamefont {Levi}\ \emph
  {et~al.}(2021{\natexlab{a}})\citenamefont {Levi}, \citenamefont {Mcleod},\
  and\ \citenamefont {Von~Hippel}}]{LMH2021}%
  \BibitemOpen
  \bibfield  {author} {\bibinfo {author} {\bibfnamefont {M.}~\bibnamefont
  {Levi}}, \bibinfo {author} {\bibfnamefont {A.~J.}\ \bibnamefont {Mcleod}}, \
  and\ \bibinfo {author} {\bibfnamefont {M.}~\bibnamefont {Von~Hippel}},\
  }\href {\doibase 10.1007/JHEP07(2021)115} {\bibfield  {journal} {\bibinfo
  {journal} {JHEP}\ }\textbf {\bibinfo {volume} {07}},\ \bibinfo {pages} {115}
  (\bibinfo {year} {2021}{\natexlab{a}})},\ \Eprint
  {http://arxiv.org/abs/2003.02827} {arXiv:2003.02827 [hep-th]} \BibitemShut
  {NoStop}%
\bibitem [{\citenamefont {Kim}\ \emph {et~al.}(2022{\natexlab{a}})\citenamefont
  {Kim}, \citenamefont {Levi},\ and\ \citenamefont {Yin}}]{KLY2022}%
  \BibitemOpen
  \bibfield  {author} {\bibinfo {author} {\bibfnamefont {J.-W.}\ \bibnamefont
  {Kim}}, \bibinfo {author} {\bibfnamefont {M.}~\bibnamefont {Levi}}, \ and\
  \bibinfo {author} {\bibfnamefont {Z.}~\bibnamefont {Yin}},\ }\href@noop {} {\
   (\bibinfo {year} {2022}{\natexlab{a}})},\ \Eprint
  {http://arxiv.org/abs/2208.14949} {arXiv:2208.14949 [hep-th]} \BibitemShut
  {NoStop}%
\bibitem [{\citenamefont {Mandal}\ \emph {et~al.}(2022)\citenamefont {Mandal},
  \citenamefont {Mastrolia}, \citenamefont {Patil},\ and\ \citenamefont
  {Steinhoff}}]{MMPS2022}%
  \BibitemOpen
  \bibfield  {author} {\bibinfo {author} {\bibfnamefont {M.~K.}\ \bibnamefont
  {Mandal}}, \bibinfo {author} {\bibfnamefont {P.}~\bibnamefont {Mastrolia}},
  \bibinfo {author} {\bibfnamefont {R.}~\bibnamefont {Patil}}, \ and\ \bibinfo
  {author} {\bibfnamefont {J.}~\bibnamefont {Steinhoff}},\ }\href@noop {} {\
  (\bibinfo {year} {2022})},\ \Eprint {http://arxiv.org/abs/2210.09176}
  {arXiv:2210.09176 [hep-th]} \BibitemShut {NoStop}%
\bibitem [{\citenamefont {Levi}(2012)}]{Levi2012_bindyn}%
  \BibitemOpen
  \bibfield  {author} {\bibinfo {author} {\bibfnamefont {M.}~\bibnamefont
  {Levi}},\ }\href {\doibase 10.1103/PhysRevD.85.064043} {\bibfield  {journal}
  {\bibinfo  {journal} {Phys. Rev. D}\ }\textbf {\bibinfo {volume} {85}},\
  \bibinfo {pages} {064043} (\bibinfo {year} {2012})},\ \Eprint
  {http://arxiv.org/abs/1107.4322} {arXiv:1107.4322 [gr-qc]} \BibitemShut
  {NoStop}%
\bibitem [{\citenamefont {Levi}\ and\ \citenamefont
  {Steinhoff}(2014)}]{LS2014}%
  \BibitemOpen
  \bibfield  {author} {\bibinfo {author} {\bibfnamefont {M.}~\bibnamefont
  {Levi}}\ and\ \bibinfo {author} {\bibfnamefont {J.}~\bibnamefont
  {Steinhoff}},\ }\href {\doibase 10.1088/1475-7516/2014/12/003} {\bibfield
  {journal} {\bibinfo  {journal} {JCAP}\ }\textbf {\bibinfo {volume} {12}},\
  \bibinfo {pages} {003} (\bibinfo {year} {2014})},\ \Eprint
  {http://arxiv.org/abs/1408.5762} {arXiv:1408.5762 [gr-qc]} \BibitemShut
  {NoStop}%
\bibitem [{\citenamefont {Levi}\ and\ \citenamefont
  {Steinhoff}(2016)}]{LS2016}%
  \BibitemOpen
  \bibfield  {author} {\bibinfo {author} {\bibfnamefont {M.}~\bibnamefont
  {Levi}}\ and\ \bibinfo {author} {\bibfnamefont {J.}~\bibnamefont
  {Steinhoff}},\ }\href {\doibase 10.1088/1475-7516/2016/01/008} {\bibfield
  {journal} {\bibinfo  {journal} {JCAP}\ }\textbf {\bibinfo {volume} {01}},\
  \bibinfo {pages} {008} (\bibinfo {year} {2016})},\ \Eprint
  {http://arxiv.org/abs/1506.05794} {arXiv:1506.05794 [gr-qc]} \BibitemShut
  {NoStop}%
\bibitem [{\citenamefont {Levi}\ and\ \citenamefont
  {Steinhoff}(2021)}]{LS2021}%
  \BibitemOpen
  \bibfield  {author} {\bibinfo {author} {\bibfnamefont {M.}~\bibnamefont
  {Levi}}\ and\ \bibinfo {author} {\bibfnamefont {J.}~\bibnamefont
  {Steinhoff}},\ }\href {\doibase 10.1088/1475-7516/2021/09/029} {\bibfield
  {journal} {\bibinfo  {journal} {JCAP}\ }\textbf {\bibinfo {volume} {09}},\
  \bibinfo {pages} {029} (\bibinfo {year} {2021})},\ \Eprint
  {http://arxiv.org/abs/1607.04252} {arXiv:1607.04252 [gr-qc]} \BibitemShut
  {NoStop}%
\bibitem [{\citenamefont {Kim}\ \emph {et~al.}(2022{\natexlab{b}})\citenamefont
  {Kim}, \citenamefont {Levi},\ and\ \citenamefont {Yin}}]{KLY2022_quaspi}%
  \BibitemOpen
  \bibfield  {author} {\bibinfo {author} {\bibfnamefont {J.-W.}\ \bibnamefont
  {Kim}}, \bibinfo {author} {\bibfnamefont {M.}~\bibnamefont {Levi}}, \ and\
  \bibinfo {author} {\bibfnamefont {Z.}~\bibnamefont {Yin}},\ }\href {\doibase
  10.1016/j.physletb.2022.137410} {\bibfield  {journal} {\bibinfo  {journal}
  {Phys. Lett. B}\ }\textbf {\bibinfo {volume} {834}},\ \bibinfo {pages}
  {137410} (\bibinfo {year} {2022}{\natexlab{b}})},\ \Eprint
  {http://arxiv.org/abs/2112.01509} {arXiv:2112.01509 [hep-th]} \BibitemShut
  {NoStop}%
\bibitem [{\citenamefont {Levi}\ \emph
  {et~al.}(2021{\natexlab{b}})\citenamefont {Levi}, \citenamefont {Mcleod},\
  and\ \citenamefont {Von~Hippel}}]{LMH2021_graqua}%
  \BibitemOpen
  \bibfield  {author} {\bibinfo {author} {\bibfnamefont {M.}~\bibnamefont
  {Levi}}, \bibinfo {author} {\bibfnamefont {A.~J.}\ \bibnamefont {Mcleod}}, \
  and\ \bibinfo {author} {\bibfnamefont {M.}~\bibnamefont {Von~Hippel}},\
  }\href {\doibase 10.1007/JHEP07(2021)116} {\bibfield  {journal} {\bibinfo
  {journal} {JHEP}\ }\textbf {\bibinfo {volume} {07}},\ \bibinfo {pages} {116}
  (\bibinfo {year} {2021}{\natexlab{b}})},\ \Eprint
  {http://arxiv.org/abs/2003.07890} {arXiv:2003.07890 [hep-th]} \BibitemShut
  {NoStop}%
\bibitem [{\citenamefont {Cho}\ \emph {et~al.}(2022{\natexlab{a}})\citenamefont
  {Cho}, \citenamefont {Porto},\ and\ \citenamefont {Yang}}]{CPY2022}%
  \BibitemOpen
  \bibfield  {author} {\bibinfo {author} {\bibfnamefont {G.}~\bibnamefont
  {Cho}}, \bibinfo {author} {\bibfnamefont {R.~A.}\ \bibnamefont {Porto}}, \
  and\ \bibinfo {author} {\bibfnamefont {Z.}~\bibnamefont {Yang}},\ }\href
  {\doibase 10.1103/PhysRevD.106.L101501} {\bibfield  {journal} {\bibinfo
  {journal} {Phys. Rev. D}\ }\textbf {\bibinfo {volume} {106}},\ \bibinfo
  {pages} {L101501} (\bibinfo {year} {2022}{\natexlab{a}})},\ \Eprint
  {http://arxiv.org/abs/2201.05138} {arXiv:2201.05138 [gr-qc]} \BibitemShut
  {NoStop}%
\bibitem [{\citenamefont {Kim}\ \emph {et~al.}(2022{\natexlab{c}})\citenamefont
  {Kim}, \citenamefont {Levi},\ and\ \citenamefont
  {Yin}}]{KLY2022_secondpaper}%
  \BibitemOpen
  \bibfield  {author} {\bibinfo {author} {\bibfnamefont {J.-W.}\ \bibnamefont
  {Kim}}, \bibinfo {author} {\bibfnamefont {M.}~\bibnamefont {Levi}}, \ and\
  \bibinfo {author} {\bibfnamefont {Z.}~\bibnamefont {Yin}},\ }\href@noop {} {\
   (\bibinfo {year} {2022}{\natexlab{c}})},\ \Eprint
  {http://arxiv.org/abs/2209.09235} {arXiv:2209.09235 [hep-th]} \BibitemShut
  {NoStop}%
\bibitem [{\citenamefont {Levi}\ and\ \citenamefont {Yin}(2022)}]{LY2022}%
  \BibitemOpen
  \bibfield  {author} {\bibinfo {author} {\bibfnamefont {M.}~\bibnamefont
  {Levi}}\ and\ \bibinfo {author} {\bibfnamefont {Z.}~\bibnamefont {Yin}},\
  }\href@noop {} {\  (\bibinfo {year} {2022})},\ \Eprint
  {http://arxiv.org/abs/2211.14018} {arXiv:2211.14018 [hep-th]} \BibitemShut
  {NoStop}%
\bibitem [{\citenamefont {Vines}(2018)}]{Vines2018_scaspi}%
  \BibitemOpen
  \bibfield  {author} {\bibinfo {author} {\bibfnamefont {J.}~\bibnamefont
  {Vines}},\ }\href {\doibase 10.1088/1361-6382/aaa3a8} {\bibfield  {journal}
  {\bibinfo  {journal} {Class. Quant. Grav.}\ }\textbf {\bibinfo {volume}
  {35}},\ \bibinfo {pages} {084002} (\bibinfo {year} {2018})},\ \Eprint
  {http://arxiv.org/abs/1709.06016} {arXiv:1709.06016 [gr-qc]} \BibitemShut
  {NoStop}%
\bibitem [{\citenamefont {Bini}\ and\ \citenamefont {Damour}(2017)}]{BD2017}%
  \BibitemOpen
  \bibfield  {author} {\bibinfo {author} {\bibfnamefont {D.}~\bibnamefont
  {Bini}}\ and\ \bibinfo {author} {\bibfnamefont {T.}~\bibnamefont {Damour}},\
  }\href {\doibase 10.1103/PhysRevD.96.104038} {\bibfield  {journal} {\bibinfo
  {journal} {Phys. Rev. D}\ }\textbf {\bibinfo {volume} {96}},\ \bibinfo
  {pages} {104038} (\bibinfo {year} {2017})},\ \Eprint
  {http://arxiv.org/abs/1709.00590} {arXiv:1709.00590 [gr-qc]} \BibitemShut
  {NoStop}%
\bibitem [{\citenamefont {Bini}\ and\ \citenamefont {Damour}(2018)}]{BD2018}%
  \BibitemOpen
  \bibfield  {author} {\bibinfo {author} {\bibfnamefont {D.}~\bibnamefont
  {Bini}}\ and\ \bibinfo {author} {\bibfnamefont {T.}~\bibnamefont {Damour}},\
  }\href {\doibase 10.1103/PhysRevD.98.044036} {\bibfield  {journal} {\bibinfo
  {journal} {Phys. Rev. D}\ }\textbf {\bibinfo {volume} {98}},\ \bibinfo
  {pages} {044036} (\bibinfo {year} {2018})},\ \Eprint
  {http://arxiv.org/abs/1805.10809} {arXiv:1805.10809 [gr-qc]} \BibitemShut
  {NoStop}%
\bibitem [{\citenamefont {Febres~Cordero}\ \emph {et~al.}(2023)\citenamefont
  {Febres~Cordero}, \citenamefont {Kraus}, \citenamefont {Lin}, \citenamefont
  {Ruf},\ and\ \citenamefont {Zeng}}]{FKLRZ2022}%
  \BibitemOpen
  \bibfield  {author} {\bibinfo {author} {\bibfnamefont {F.}~\bibnamefont
  {Febres~Cordero}}, \bibinfo {author} {\bibfnamefont {M.}~\bibnamefont
  {Kraus}}, \bibinfo {author} {\bibfnamefont {G.}~\bibnamefont {Lin}}, \bibinfo
  {author} {\bibfnamefont {M.~S.}\ \bibnamefont {Ruf}}, \ and\ \bibinfo
  {author} {\bibfnamefont {M.}~\bibnamefont {Zeng}},\ }\href {\doibase
  10.1103/PhysRevLett.130.021601} {\bibfield  {journal} {\bibinfo  {journal}
  {Phys. Rev. Lett.}\ }\textbf {\bibinfo {volume} {130}},\ \bibinfo {pages}
  {021601} (\bibinfo {year} {2023})},\ \Eprint
  {http://arxiv.org/abs/2205.07357} {arXiv:2205.07357 [hep-th]} \BibitemShut
  {NoStop}%
\bibitem [{\citenamefont {Jakobsen}\ and\ \citenamefont
  {Mogull}(2022{\natexlab{a}})}]{JM2022_linres}%
  \BibitemOpen
  \bibfield  {author} {\bibinfo {author} {\bibfnamefont {G.~U.}\ \bibnamefont
  {Jakobsen}}\ and\ \bibinfo {author} {\bibfnamefont {G.}~\bibnamefont
  {Mogull}},\ }\href@noop {} {\  (\bibinfo {year} {2022}{\natexlab{a}})},\
  \Eprint {http://arxiv.org/abs/2210.06451} {arXiv:2210.06451 [hep-th]}
  \BibitemShut {NoStop}%
\bibitem [{\citenamefont {Buonanno}\ \emph {et~al.}(2022)\citenamefont
  {Buonanno}, \citenamefont {Khalil}, \citenamefont {O'Connell}, \citenamefont
  {Roiban}, \citenamefont {Solon},\ and\ \citenamefont {Zeng}}]{BKORS2022}%
  \BibitemOpen
  \bibfield  {author} {\bibinfo {author} {\bibfnamefont {A.}~\bibnamefont
  {Buonanno}}, \bibinfo {author} {\bibfnamefont {M.}~\bibnamefont {Khalil}},
  \bibinfo {author} {\bibfnamefont {D.}~\bibnamefont {O'Connell}}, \bibinfo
  {author} {\bibfnamefont {R.}~\bibnamefont {Roiban}}, \bibinfo {author}
  {\bibfnamefont {M.~P.}\ \bibnamefont {Solon}}, \ and\ \bibinfo {author}
  {\bibfnamefont {M.}~\bibnamefont {Zeng}},\ }in\ \href@noop {} {\emph
  {\bibinfo {booktitle} {{2022 Snowmass Summer Study}}}}\ (\bibinfo {year}
  {2022})\ \Eprint {http://arxiv.org/abs/2204.05194} {arXiv:2204.05194
  [hep-th]} \BibitemShut {NoStop}%
\bibitem [{\citenamefont {Bjerrum-Bohr}\ \emph
  {et~al.}(2022{\natexlab{a}})\citenamefont {Bjerrum-Bohr}, \citenamefont
  {Damgaard}, \citenamefont {Plante},\ and\ \citenamefont
  {Vanhove}}]{BDPV2022}%
  \BibitemOpen
  \bibfield  {author} {\bibinfo {author} {\bibfnamefont {N.~E.~J.}\
  \bibnamefont {Bjerrum-Bohr}}, \bibinfo {author} {\bibfnamefont {P.~H.}\
  \bibnamefont {Damgaard}}, \bibinfo {author} {\bibfnamefont {L.}~\bibnamefont
  {Plante}}, \ and\ \bibinfo {author} {\bibfnamefont {P.}~\bibnamefont
  {Vanhove}},\ }\href {\doibase 10.1088/1751-8121/ac7a78} {\bibfield  {journal}
  {\bibinfo  {journal} {J. Phys. A}\ }\textbf {\bibinfo {volume} {55}},\
  \bibinfo {pages} {443014} (\bibinfo {year} {2022}{\natexlab{a}})},\ \Eprint
  {http://arxiv.org/abs/2203.13024} {arXiv:2203.13024 [hep-th]} \BibitemShut
  {NoStop}%
\bibitem [{\citenamefont {Kosower}\ \emph {et~al.}(2022)\citenamefont
  {Kosower}, \citenamefont {Monteiro},\ and\ \citenamefont
  {O'Connell}}]{KMO2022}%
  \BibitemOpen
  \bibfield  {author} {\bibinfo {author} {\bibfnamefont {D.~A.}\ \bibnamefont
  {Kosower}}, \bibinfo {author} {\bibfnamefont {R.}~\bibnamefont {Monteiro}}, \
  and\ \bibinfo {author} {\bibfnamefont {D.}~\bibnamefont {O'Connell}},\ }\href
  {\doibase 10.1088/1751-8121/ac8846} {\bibfield  {journal} {\bibinfo
  {journal} {J. Phys. A}\ }\textbf {\bibinfo {volume} {55}},\ \bibinfo {pages}
  {443015} (\bibinfo {year} {2022})},\ \Eprint
  {http://arxiv.org/abs/2203.13025} {arXiv:2203.13025 [hep-th]} \BibitemShut
  {NoStop}%
\bibitem [{\citenamefont {Vines}\ \emph {et~al.}(2019)\citenamefont {Vines},
  \citenamefont {Steinhoff},\ and\ \citenamefont {Buonanno}}]{VSB2019}%
  \BibitemOpen
  \bibfield  {author} {\bibinfo {author} {\bibfnamefont {J.}~\bibnamefont
  {Vines}}, \bibinfo {author} {\bibfnamefont {J.}~\bibnamefont {Steinhoff}}, \
  and\ \bibinfo {author} {\bibfnamefont {A.}~\bibnamefont {Buonanno}},\ }\href
  {\doibase 10.1103/PhysRevD.99.064054} {\bibfield  {journal} {\bibinfo
  {journal} {Phys. Rev. D}\ }\textbf {\bibinfo {volume} {99}},\ \bibinfo
  {pages} {064054} (\bibinfo {year} {2019})},\ \Eprint
  {http://arxiv.org/abs/1812.00956} {arXiv:1812.00956 [gr-qc]} \BibitemShut
  {NoStop}%
\bibitem [{\citenamefont {Damour}(2018)}]{Damour2017_higgra}%
  \BibitemOpen
  \bibfield  {author} {\bibinfo {author} {\bibfnamefont {T.}~\bibnamefont
  {Damour}},\ }\href {\doibase 10.1103/PhysRevD.97.044038} {\bibfield
  {journal} {\bibinfo  {journal} {Phys. Rev. D}\ }\textbf {\bibinfo {volume}
  {97}},\ \bibinfo {pages} {044038} (\bibinfo {year} {2018})},\ \Eprint
  {http://arxiv.org/abs/1710.10599} {arXiv:1710.10599 [gr-qc]} \BibitemShut
  {NoStop}%
\bibitem [{\citenamefont {Westpfahl}(1985)}]{Westpfahl1985}%
  \BibitemOpen
  \bibfield  {author} {\bibinfo {author} {\bibfnamefont {K.}~\bibnamefont
  {Westpfahl}},\ }\href {\doibase 10.1002/prop.2190330802} {\bibfield
  {journal} {\bibinfo  {journal} {Fortsch. Phys.}\ }\textbf {\bibinfo {volume}
  {33}},\ \bibinfo {pages} {417} (\bibinfo {year} {1985})}\BibitemShut
  {NoStop}%
\bibitem [{\citenamefont {Mogull}\ \emph {et~al.}(2021)\citenamefont {Mogull},
  \citenamefont {Plefka},\ and\ \citenamefont {Steinhoff}}]{MPS2021}%
  \BibitemOpen
  \bibfield  {author} {\bibinfo {author} {\bibfnamefont {G.}~\bibnamefont
  {Mogull}}, \bibinfo {author} {\bibfnamefont {J.}~\bibnamefont {Plefka}}, \
  and\ \bibinfo {author} {\bibfnamefont {J.}~\bibnamefont {Steinhoff}},\ }\href
  {\doibase 10.1007/JHEP02(2021)048} {\bibfield  {journal} {\bibinfo  {journal}
  {JHEP}\ }\textbf {\bibinfo {volume} {02}},\ \bibinfo {pages} {048} (\bibinfo
  {year} {2021})},\ \Eprint {http://arxiv.org/abs/2010.02865} {arXiv:2010.02865
  [hep-th]} \BibitemShut {NoStop}%
\bibitem [{\citenamefont {Jakobsen}\ \emph {et~al.}(2021)\citenamefont
  {Jakobsen}, \citenamefont {Mogull}, \citenamefont {Plefka},\ and\
  \citenamefont {Steinhoff}}]{JMPS2021}%
  \BibitemOpen
  \bibfield  {author} {\bibinfo {author} {\bibfnamefont {G.~U.}\ \bibnamefont
  {Jakobsen}}, \bibinfo {author} {\bibfnamefont {G.}~\bibnamefont {Mogull}},
  \bibinfo {author} {\bibfnamefont {J.}~\bibnamefont {Plefka}}, \ and\ \bibinfo
  {author} {\bibfnamefont {J.}~\bibnamefont {Steinhoff}},\ }\href {\doibase
  10.1103/PhysRevLett.126.201103} {\bibfield  {journal} {\bibinfo  {journal}
  {Phys. Rev. Lett.}\ }\textbf {\bibinfo {volume} {126}},\ \bibinfo {pages}
  {201103} (\bibinfo {year} {2021})},\ \Eprint
  {http://arxiv.org/abs/2101.12688} {arXiv:2101.12688 [gr-qc]} \BibitemShut
  {NoStop}%
\bibitem [{\citenamefont {Jakobsen}\ \emph
  {et~al.}(2022{\natexlab{a}})\citenamefont {Jakobsen}, \citenamefont {Mogull},
  \citenamefont {Plefka},\ and\ \citenamefont {Steinhoff}}]{JMPS2022}%
  \BibitemOpen
  \bibfield  {author} {\bibinfo {author} {\bibfnamefont {G.~U.}\ \bibnamefont
  {Jakobsen}}, \bibinfo {author} {\bibfnamefont {G.}~\bibnamefont {Mogull}},
  \bibinfo {author} {\bibfnamefont {J.}~\bibnamefont {Plefka}}, \ and\ \bibinfo
  {author} {\bibfnamefont {J.}~\bibnamefont {Steinhoff}},\ }\href {\doibase
  10.1103/PhysRevLett.128.011101} {\bibfield  {journal} {\bibinfo  {journal}
  {Phys. Rev. Lett.}\ }\textbf {\bibinfo {volume} {128}},\ \bibinfo {pages}
  {011101} (\bibinfo {year} {2022}{\natexlab{a}})},\ \Eprint
  {http://arxiv.org/abs/2106.10256} {arXiv:2106.10256 [hep-th]} \BibitemShut
  {NoStop}%
\bibitem [{\citenamefont {Jakobsen}\ \emph
  {et~al.}(2022{\natexlab{b}})\citenamefont {Jakobsen}, \citenamefont {Mogull},
  \citenamefont {Plefka},\ and\ \citenamefont {Steinhoff}}]{JMPS2022_SUSY}%
  \BibitemOpen
  \bibfield  {author} {\bibinfo {author} {\bibfnamefont {G.~U.}\ \bibnamefont
  {Jakobsen}}, \bibinfo {author} {\bibfnamefont {G.}~\bibnamefont {Mogull}},
  \bibinfo {author} {\bibfnamefont {J.}~\bibnamefont {Plefka}}, \ and\ \bibinfo
  {author} {\bibfnamefont {J.}~\bibnamefont {Steinhoff}},\ }\href {\doibase
  10.1007/JHEP01(2022)027} {\bibfield  {journal} {\bibinfo  {journal} {JHEP}\
  }\textbf {\bibinfo {volume} {01}},\ \bibinfo {pages} {027} (\bibinfo {year}
  {2022}{\natexlab{b}})},\ \Eprint {http://arxiv.org/abs/2109.04465}
  {arXiv:2109.04465 [hep-th]} \BibitemShut {NoStop}%
\bibitem [{\citenamefont {Jakobsen}\ and\ \citenamefont
  {Mogull}(2022{\natexlab{b}})}]{JM2022}%
  \BibitemOpen
  \bibfield  {author} {\bibinfo {author} {\bibfnamefont {G.~U.}\ \bibnamefont
  {Jakobsen}}\ and\ \bibinfo {author} {\bibfnamefont {G.}~\bibnamefont
  {Mogull}},\ }\href {\doibase 10.1103/PhysRevLett.128.141102} {\bibfield
  {journal} {\bibinfo  {journal} {Phys. Rev. Lett.}\ }\textbf {\bibinfo
  {volume} {128}},\ \bibinfo {pages} {141102} (\bibinfo {year}
  {2022}{\natexlab{b}})},\ \Eprint {http://arxiv.org/abs/2201.07778}
  {arXiv:2201.07778 [hep-th]} \BibitemShut {NoStop}%
\bibitem [{\citenamefont {Jakobsen}\ \emph
  {et~al.}(2022{\natexlab{c}})\citenamefont {Jakobsen}, \citenamefont {Mogull},
  \citenamefont {Plefka},\ and\ \citenamefont {Sauer}}]{JMPS2022_allthi}%
  \BibitemOpen
  \bibfield  {author} {\bibinfo {author} {\bibfnamefont {G.~U.}\ \bibnamefont
  {Jakobsen}}, \bibinfo {author} {\bibfnamefont {G.}~\bibnamefont {Mogull}},
  \bibinfo {author} {\bibfnamefont {J.}~\bibnamefont {Plefka}}, \ and\ \bibinfo
  {author} {\bibfnamefont {B.}~\bibnamefont {Sauer}},\ }\href {\doibase
  10.1007/JHEP10(2022)128} {\bibfield  {journal} {\bibinfo  {journal} {JHEP}\
  }\textbf {\bibinfo {volume} {10}},\ \bibinfo {pages} {128} (\bibinfo {year}
  {2022}{\natexlab{c}})},\ \Eprint {http://arxiv.org/abs/2207.00569}
  {arXiv:2207.00569 [hep-th]} \BibitemShut {NoStop}%
\bibitem [{\citenamefont {Saketh}\ and\ \citenamefont {Vines}(2022)}]{SV2022}%
  \BibitemOpen
  \bibfield  {author} {\bibinfo {author} {\bibfnamefont {M.~V.~S.}\
  \bibnamefont {Saketh}}\ and\ \bibinfo {author} {\bibfnamefont
  {J.}~\bibnamefont {Vines}},\ }\href {\doibase 10.1103/PhysRevD.106.124026}
  {\bibfield  {journal} {\bibinfo  {journal} {Phys. Rev. D}\ }\textbf {\bibinfo
  {volume} {106}},\ \bibinfo {pages} {124026} (\bibinfo {year} {2022})},\
  \Eprint {http://arxiv.org/abs/2208.03170} {arXiv:2208.03170 [gr-qc]}
  \BibitemShut {NoStop}%
\bibitem [{\citenamefont {Bastianelli}\ \emph {et~al.}(2022)\citenamefont
  {Bastianelli}, \citenamefont {Comberiati},\ and\ \citenamefont {de~la
  Cruz}}]{BCC2022}%
  \BibitemOpen
  \bibfield  {author} {\bibinfo {author} {\bibfnamefont {F.}~\bibnamefont
  {Bastianelli}}, \bibinfo {author} {\bibfnamefont {F.}~\bibnamefont
  {Comberiati}}, \ and\ \bibinfo {author} {\bibfnamefont {L.}~\bibnamefont
  {de~la Cruz}},\ }\href {\doibase 10.1007/JHEP02(2022)209} {\bibfield
  {journal} {\bibinfo  {journal} {JHEP}\ }\textbf {\bibinfo {volume} {02}},\
  \bibinfo {pages} {209} (\bibinfo {year} {2022})},\ \Eprint
  {http://arxiv.org/abs/2112.05013} {arXiv:2112.05013 [hep-th]} \BibitemShut
  {NoStop}%
\bibitem [{\citenamefont {Foffa}(2014)}]{Foffa2014}%
  \BibitemOpen
  \bibfield  {author} {\bibinfo {author} {\bibfnamefont {S.}~\bibnamefont
  {Foffa}},\ }\href {\doibase 10.1103/PhysRevD.89.024019} {\bibfield  {journal}
  {\bibinfo  {journal} {Phys. Rev. D}\ }\textbf {\bibinfo {volume} {89}},\
  \bibinfo {pages} {024019} (\bibinfo {year} {2014})},\ \Eprint
  {http://arxiv.org/abs/1309.3956} {arXiv:1309.3956 [gr-qc]} \BibitemShut
  {NoStop}%
\bibitem [{\citenamefont {K\"alin}\ and\ \citenamefont
  {Porto}(2020)}]{KP2020_posmin}%
  \BibitemOpen
  \bibfield  {author} {\bibinfo {author} {\bibfnamefont {G.}~\bibnamefont
  {K\"alin}}\ and\ \bibinfo {author} {\bibfnamefont {R.~A.}\ \bibnamefont
  {Porto}},\ }\href {\doibase 10.1007/JHEP11(2020)106} {\bibfield  {journal}
  {\bibinfo  {journal} {JHEP}\ }\textbf {\bibinfo {volume} {11}},\ \bibinfo
  {pages} {106} (\bibinfo {year} {2020})},\ \Eprint
  {http://arxiv.org/abs/2006.01184} {arXiv:2006.01184 [hep-th]} \BibitemShut
  {NoStop}%
\bibitem [{\citenamefont {K{\"a}lin}\ and\ \citenamefont
  {Porto}(2020)}]{KP2020}%
  \BibitemOpen
  \bibfield  {author} {\bibinfo {author} {\bibfnamefont {G.}~\bibnamefont
  {K{\"a}lin}}\ and\ \bibinfo {author} {\bibfnamefont {R.~A.}\ \bibnamefont
  {Porto}},\ }\href {\doibase 10.1007/JHEP01(2020)072} {\bibfield  {journal}
  {\bibinfo  {journal} {Journal of High Energy Physics}\ }\textbf {\bibinfo
  {volume} {2020}},\ \bibinfo {pages} {72} (\bibinfo {year}
  {2020})}\BibitemShut {NoStop}%
\bibitem [{\citenamefont {K\"alin}\ and\ \citenamefont
  {Porto}(2020)}]{KP2020_II}%
  \BibitemOpen
  \bibfield  {author} {\bibinfo {author} {\bibfnamefont {G.}~\bibnamefont
  {K\"alin}}\ and\ \bibinfo {author} {\bibfnamefont {R.~A.}\ \bibnamefont
  {Porto}},\ }\href {\doibase 10.1007/JHEP02(2020)120} {\bibfield  {journal}
  {\bibinfo  {journal} {JHEP}\ }\textbf {\bibinfo {volume} {02}},\ \bibinfo
  {pages} {120} (\bibinfo {year} {2020})},\ \Eprint
  {http://arxiv.org/abs/1911.09130} {arXiv:1911.09130 [hep-th]} \BibitemShut
  {NoStop}%
\bibitem [{\citenamefont {Liu}\ \emph {et~al.}(2021)\citenamefont {Liu},
  \citenamefont {Porto},\ and\ \citenamefont {Yang}}]{LPY2021}%
  \BibitemOpen
  \bibfield  {author} {\bibinfo {author} {\bibfnamefont {Z.}~\bibnamefont
  {Liu}}, \bibinfo {author} {\bibfnamefont {R.~A.}\ \bibnamefont {Porto}}, \
  and\ \bibinfo {author} {\bibfnamefont {Z.}~\bibnamefont {Yang}},\ }\href
  {\doibase 10.1007/JHEP06(2021)012} {\bibfield  {journal} {\bibinfo  {journal}
  {JHEP}\ }\textbf {\bibinfo {volume} {06}},\ \bibinfo {pages} {012} (\bibinfo
  {year} {2021})},\ \Eprint {http://arxiv.org/abs/2102.10059} {arXiv:2102.10059
  [hep-th]} \BibitemShut {NoStop}%
\bibitem [{\citenamefont {Cho}\ \emph {et~al.}(2022{\natexlab{b}})\citenamefont
  {Cho}, \citenamefont {K\"alin},\ and\ \citenamefont {Porto}}]{CKP2022}%
  \BibitemOpen
  \bibfield  {author} {\bibinfo {author} {\bibfnamefont {G.}~\bibnamefont
  {Cho}}, \bibinfo {author} {\bibfnamefont {G.}~\bibnamefont {K\"alin}}, \ and\
  \bibinfo {author} {\bibfnamefont {R.~A.}\ \bibnamefont {Porto}},\ }\href
  {\doibase 10.1007/JHEP04(2022)154} {\bibfield  {journal} {\bibinfo  {journal}
  {JHEP}\ }\textbf {\bibinfo {volume} {04}},\ \bibinfo {pages} {154} (\bibinfo
  {year} {2022}{\natexlab{b}})},\ \bibinfo {note} {[Erratum: JHEP 07, 002
  (2022)]},\ \Eprint {http://arxiv.org/abs/2112.03976} {arXiv:2112.03976
  [hep-th]} \BibitemShut {NoStop}%
\bibitem [{\citenamefont {Dlapa}\ \emph
  {et~al.}(2022{\natexlab{b}})\citenamefont {Dlapa}, \citenamefont {K\"alin},
  \citenamefont {Liu},\ and\ \citenamefont {Porto}}]{DKLP2022}%
  \BibitemOpen
  \bibfield  {author} {\bibinfo {author} {\bibfnamefont {C.}~\bibnamefont
  {Dlapa}}, \bibinfo {author} {\bibfnamefont {G.}~\bibnamefont {K\"alin}},
  \bibinfo {author} {\bibfnamefont {Z.}~\bibnamefont {Liu}}, \ and\ \bibinfo
  {author} {\bibfnamefont {R.~A.}\ \bibnamefont {Porto}},\ }\href {\doibase
  10.1016/j.physletb.2022.137203} {\bibfield  {journal} {\bibinfo  {journal}
  {Phys. Lett. B}\ }\textbf {\bibinfo {volume} {831}},\ \bibinfo {pages}
  {137203} (\bibinfo {year} {2022}{\natexlab{b}})},\ \Eprint
  {http://arxiv.org/abs/2106.08276} {arXiv:2106.08276 [hep-th]} \BibitemShut
  {NoStop}%
\bibitem [{\citenamefont {K\"alin}\ \emph {et~al.}(2022)\citenamefont
  {K\"alin}, \citenamefont {Neef},\ and\ \citenamefont {Porto}}]{KNP2022}%
  \BibitemOpen
  \bibfield  {author} {\bibinfo {author} {\bibfnamefont {G.}~\bibnamefont
  {K\"alin}}, \bibinfo {author} {\bibfnamefont {J.}~\bibnamefont {Neef}}, \
  and\ \bibinfo {author} {\bibfnamefont {R.~A.}\ \bibnamefont {Porto}},\
  }\href@noop {} {\  (\bibinfo {year} {2022})},\ \Eprint
  {http://arxiv.org/abs/2207.00580} {arXiv:2207.00580 [hep-th]} \BibitemShut
  {NoStop}%
\bibitem [{\citenamefont {Bern}\ \emph {et~al.}(2019)\citenamefont {Bern},
  \citenamefont {Cheung}, \citenamefont {Roiban}, \citenamefont {Shen},
  \citenamefont {Solon},\ and\ \citenamefont {Zeng}}]{ZCRC2019}%
  \BibitemOpen
  \bibfield  {author} {\bibinfo {author} {\bibfnamefont {Z.}~\bibnamefont
  {Bern}}, \bibinfo {author} {\bibfnamefont {C.}~\bibnamefont {Cheung}},
  \bibinfo {author} {\bibfnamefont {R.}~\bibnamefont {Roiban}}, \bibinfo
  {author} {\bibfnamefont {C.-H.}\ \bibnamefont {Shen}}, \bibinfo {author}
  {\bibfnamefont {M.~P.}\ \bibnamefont {Solon}}, \ and\ \bibinfo {author}
  {\bibfnamefont {M.}~\bibnamefont {Zeng}},\ }\href {\doibase
  10.1103/PhysRevLett.122.201603} {\bibfield  {journal} {\bibinfo  {journal}
  {Phys. Rev. Lett.}\ }\textbf {\bibinfo {volume} {122}},\ \bibinfo {pages}
  {201603} (\bibinfo {year} {2019})},\ \Eprint
  {http://arxiv.org/abs/1901.04424} {arXiv:1901.04424 [hep-th]} \BibitemShut
  {NoStop}%
\bibitem [{\citenamefont {Antonelli}\ \emph {et~al.}(2019)\citenamefont
  {Antonelli}, \citenamefont {Buonanno}, \citenamefont {Steinhoff},
  \citenamefont {van~de Meent},\ and\ \citenamefont {Vines}}]{ABSMV2019}%
  \BibitemOpen
  \bibfield  {author} {\bibinfo {author} {\bibfnamefont {A.}~\bibnamefont
  {Antonelli}}, \bibinfo {author} {\bibfnamefont {A.}~\bibnamefont {Buonanno}},
  \bibinfo {author} {\bibfnamefont {J.}~\bibnamefont {Steinhoff}}, \bibinfo
  {author} {\bibfnamefont {M.}~\bibnamefont {van~de Meent}}, \ and\ \bibinfo
  {author} {\bibfnamefont {J.}~\bibnamefont {Vines}},\ }\href {\doibase
  10.1103/PhysRevD.99.104004} {\bibfield  {journal} {\bibinfo  {journal} {Phys.
  Rev. D}\ }\textbf {\bibinfo {volume} {99}},\ \bibinfo {pages} {104004}
  (\bibinfo {year} {2019})},\ \Eprint {http://arxiv.org/abs/1901.07102}
  {arXiv:1901.07102 [gr-qc]} \BibitemShut {NoStop}%
\bibitem [{\citenamefont {Parra-Martinez}\ \emph {et~al.}(2020)\citenamefont
  {Parra-Martinez}, \citenamefont {Ruf},\ and\ \citenamefont {Zeng}}]{PRZ2020}%
  \BibitemOpen
  \bibfield  {author} {\bibinfo {author} {\bibfnamefont {J.}~\bibnamefont
  {Parra-Martinez}}, \bibinfo {author} {\bibfnamefont {M.~S.}\ \bibnamefont
  {Ruf}}, \ and\ \bibinfo {author} {\bibfnamefont {M.}~\bibnamefont {Zeng}},\
  }\href {\doibase 10.1007/JHEP11(2020)023} {\bibfield  {journal} {\bibinfo
  {journal} {JHEP}\ }\textbf {\bibinfo {volume} {11}},\ \bibinfo {pages} {023}
  (\bibinfo {year} {2020})},\ \Eprint {http://arxiv.org/abs/2005.04236}
  {arXiv:2005.04236 [hep-th]} \BibitemShut {NoStop}%
\bibitem [{\citenamefont {Di~Vecchia}\ \emph {et~al.}(2020)\citenamefont
  {Di~Vecchia}, \citenamefont {Heissenberg}, \citenamefont {Russo},\ and\
  \citenamefont {Veneziano}}]{VHRV2020}%
  \BibitemOpen
  \bibfield  {author} {\bibinfo {author} {\bibfnamefont {P.}~\bibnamefont
  {Di~Vecchia}}, \bibinfo {author} {\bibfnamefont {C.}~\bibnamefont
  {Heissenberg}}, \bibinfo {author} {\bibfnamefont {R.}~\bibnamefont {Russo}},
  \ and\ \bibinfo {author} {\bibfnamefont {G.}~\bibnamefont {Veneziano}},\
  }\href {\doibase 10.1016/j.physletb.2020.135924} {\bibfield  {journal}
  {\bibinfo  {journal} {Phys. Lett. B}\ }\textbf {\bibinfo {volume} {811}},\
  \bibinfo {pages} {135924} (\bibinfo {year} {2020})},\ \Eprint
  {http://arxiv.org/abs/2008.12743} {arXiv:2008.12743 [hep-th]} \BibitemShut
  {NoStop}%
\bibitem [{\citenamefont {Damour}(2020{\natexlab{b}})}]{Damour2020_radcon}%
  \BibitemOpen
  \bibfield  {author} {\bibinfo {author} {\bibfnamefont {T.}~\bibnamefont
  {Damour}},\ }\href {\doibase 10.1103/PhysRevD.102.124008} {\bibfield
  {journal} {\bibinfo  {journal} {Phys. Rev. D}\ }\textbf {\bibinfo {volume}
  {102}},\ \bibinfo {pages} {124008} (\bibinfo {year} {2020}{\natexlab{b}})},\
  \Eprint {http://arxiv.org/abs/2010.01641} {arXiv:2010.01641 [gr-qc]}
  \BibitemShut {NoStop}%
\bibitem [{\citenamefont {Mougiakakos}\ \emph {et~al.}(2021)\citenamefont
  {Mougiakakos}, \citenamefont {Riva},\ and\ \citenamefont
  {Vernizzi}}]{MRV2021}%
  \BibitemOpen
  \bibfield  {author} {\bibinfo {author} {\bibfnamefont {S.}~\bibnamefont
  {Mougiakakos}}, \bibinfo {author} {\bibfnamefont {M.~M.}\ \bibnamefont
  {Riva}}, \ and\ \bibinfo {author} {\bibfnamefont {F.}~\bibnamefont
  {Vernizzi}},\ }\href {\doibase 10.1103/PhysRevD.104.024041} {\bibfield
  {journal} {\bibinfo  {journal} {Phys. Rev. D}\ }\textbf {\bibinfo {volume}
  {104}},\ \bibinfo {pages} {024041} (\bibinfo {year} {2021})},\ \Eprint
  {http://arxiv.org/abs/2102.08339} {arXiv:2102.08339 [gr-qc]} \BibitemShut
  {NoStop}%
\bibitem [{\citenamefont {Herrmann}\ \emph {et~al.}(2021)\citenamefont
  {Herrmann}, \citenamefont {Parra-Martinez}, \citenamefont {Ruf},\ and\
  \citenamefont {Zeng}}]{HPRZ2021}%
  \BibitemOpen
  \bibfield  {author} {\bibinfo {author} {\bibfnamefont {E.}~\bibnamefont
  {Herrmann}}, \bibinfo {author} {\bibfnamefont {J.}~\bibnamefont
  {Parra-Martinez}}, \bibinfo {author} {\bibfnamefont {M.~S.}\ \bibnamefont
  {Ruf}}, \ and\ \bibinfo {author} {\bibfnamefont {M.}~\bibnamefont {Zeng}},\
  }\href {\doibase 10.1007/JHEP10(2021)148} {\bibfield  {journal} {\bibinfo
  {journal} {JHEP}\ }\textbf {\bibinfo {volume} {10}},\ \bibinfo {pages} {148}
  (\bibinfo {year} {2021})},\ \Eprint {http://arxiv.org/abs/2104.03957}
  {arXiv:2104.03957 [hep-th]} \BibitemShut {NoStop}%
\bibitem [{\citenamefont {Di~Vecchia}\ \emph
  {et~al.}(2021{\natexlab{a}})\citenamefont {Di~Vecchia}, \citenamefont
  {Heissenberg}, \citenamefont {Russo},\ and\ \citenamefont
  {Veneziano}}]{VHRV2021_radrea}%
  \BibitemOpen
  \bibfield  {author} {\bibinfo {author} {\bibfnamefont {P.}~\bibnamefont
  {Di~Vecchia}}, \bibinfo {author} {\bibfnamefont {C.}~\bibnamefont
  {Heissenberg}}, \bibinfo {author} {\bibfnamefont {R.}~\bibnamefont {Russo}},
  \ and\ \bibinfo {author} {\bibfnamefont {G.}~\bibnamefont {Veneziano}},\
  }\href {\doibase 10.1016/j.physletb.2021.136379} {\bibfield  {journal}
  {\bibinfo  {journal} {Phys. Lett. B}\ }\textbf {\bibinfo {volume} {818}},\
  \bibinfo {pages} {136379} (\bibinfo {year} {2021}{\natexlab{a}})},\ \Eprint
  {http://arxiv.org/abs/2101.05772} {arXiv:2101.05772 [hep-th]} \BibitemShut
  {NoStop}%
\bibitem [{\citenamefont {Di~Vecchia}\ \emph
  {et~al.}(2021{\natexlab{b}})\citenamefont {Di~Vecchia}, \citenamefont
  {Heissenberg}, \citenamefont {Russo},\ and\ \citenamefont
  {Veneziano}}]{VHRV2021}%
  \BibitemOpen
  \bibfield  {author} {\bibinfo {author} {\bibfnamefont {P.}~\bibnamefont
  {Di~Vecchia}}, \bibinfo {author} {\bibfnamefont {C.}~\bibnamefont
  {Heissenberg}}, \bibinfo {author} {\bibfnamefont {R.}~\bibnamefont {Russo}},
  \ and\ \bibinfo {author} {\bibfnamefont {G.}~\bibnamefont {Veneziano}},\
  }\href {\doibase 10.1007/JHEP07(2021)169} {\bibfield  {journal} {\bibinfo
  {journal} {JHEP}\ }\textbf {\bibinfo {volume} {07}},\ \bibinfo {pages} {169}
  (\bibinfo {year} {2021}{\natexlab{b}})},\ \Eprint
  {http://arxiv.org/abs/2104.03256} {arXiv:2104.03256 [hep-th]} \BibitemShut
  {NoStop}%
\bibitem [{\citenamefont {Bjerrum-Bohr}\ \emph
  {et~al.}(2021{\natexlab{a}})\citenamefont {Bjerrum-Bohr}, \citenamefont
  {Damgaard}, \citenamefont {Plant\'e},\ and\ \citenamefont
  {Vanhove}}]{BDPV2021}%
  \BibitemOpen
  \bibfield  {author} {\bibinfo {author} {\bibfnamefont {N.~E.~J.}\
  \bibnamefont {Bjerrum-Bohr}}, \bibinfo {author} {\bibfnamefont {P.~H.}\
  \bibnamefont {Damgaard}}, \bibinfo {author} {\bibfnamefont {L.}~\bibnamefont
  {Plant\'e}}, \ and\ \bibinfo {author} {\bibfnamefont {P.}~\bibnamefont
  {Vanhove}},\ }\href {\doibase 10.1103/PhysRevD.104.026009} {\bibfield
  {journal} {\bibinfo  {journal} {Phys. Rev. D}\ }\textbf {\bibinfo {volume}
  {104}},\ \bibinfo {pages} {026009} (\bibinfo {year} {2021}{\natexlab{a}})},\
  \Eprint {http://arxiv.org/abs/2104.04510} {arXiv:2104.04510 [hep-th]}
  \BibitemShut {NoStop}%
\bibitem [{\citenamefont {Bjerrum-Bohr}\ \emph
  {et~al.}(2021{\natexlab{b}})\citenamefont {Bjerrum-Bohr}, \citenamefont
  {Damgaard}, \citenamefont {Plant\'e},\ and\ \citenamefont
  {Vanhove}}]{DBPV2021_ampcla}%
  \BibitemOpen
  \bibfield  {author} {\bibinfo {author} {\bibfnamefont {N.~E.~J.}\
  \bibnamefont {Bjerrum-Bohr}}, \bibinfo {author} {\bibfnamefont {P.~H.}\
  \bibnamefont {Damgaard}}, \bibinfo {author} {\bibfnamefont {L.}~\bibnamefont
  {Plant\'e}}, \ and\ \bibinfo {author} {\bibfnamefont {P.}~\bibnamefont
  {Vanhove}},\ }\href {\doibase 10.1007/JHEP08(2021)172} {\bibfield  {journal}
  {\bibinfo  {journal} {JHEP}\ }\textbf {\bibinfo {volume} {08}},\ \bibinfo
  {pages} {172} (\bibinfo {year} {2021}{\natexlab{b}})},\ \Eprint
  {http://arxiv.org/abs/2105.05218} {arXiv:2105.05218 [hep-th]} \BibitemShut
  {NoStop}%
\bibitem [{\citenamefont {Damgaard}\ \emph {et~al.}(2021)\citenamefont
  {Damgaard}, \citenamefont {Plante},\ and\ \citenamefont {Vanhove}}]{DPV2021}%
  \BibitemOpen
  \bibfield  {author} {\bibinfo {author} {\bibfnamefont {P.~H.}\ \bibnamefont
  {Damgaard}}, \bibinfo {author} {\bibfnamefont {L.}~\bibnamefont {Plante}}, \
  and\ \bibinfo {author} {\bibfnamefont {P.}~\bibnamefont {Vanhove}},\ }\href
  {\doibase 10.1007/JHEP11(2021)213} {\bibfield  {journal} {\bibinfo  {journal}
  {JHEP}\ }\textbf {\bibinfo {volume} {11}},\ \bibinfo {pages} {213} (\bibinfo
  {year} {2021})},\ \Eprint {http://arxiv.org/abs/2107.12891} {arXiv:2107.12891
  [hep-th]} \BibitemShut {NoStop}%
\bibitem [{\citenamefont {Brandhuber}\ \emph {et~al.}(2021)\citenamefont
  {Brandhuber}, \citenamefont {Chen}, \citenamefont {Travaglini},\ and\
  \citenamefont {Wen}}]{BCTW2021}%
  \BibitemOpen
  \bibfield  {author} {\bibinfo {author} {\bibfnamefont {A.}~\bibnamefont
  {Brandhuber}}, \bibinfo {author} {\bibfnamefont {G.}~\bibnamefont {Chen}},
  \bibinfo {author} {\bibfnamefont {G.}~\bibnamefont {Travaglini}}, \ and\
  \bibinfo {author} {\bibfnamefont {C.}~\bibnamefont {Wen}},\ }\href {\doibase
  10.1007/JHEP10(2021)118} {\bibfield  {journal} {\bibinfo  {journal} {JHEP}\
  }\textbf {\bibinfo {volume} {10}},\ \bibinfo {pages} {118} (\bibinfo {year}
  {2021})},\ \Eprint {http://arxiv.org/abs/2108.04216} {arXiv:2108.04216
  [hep-th]} \BibitemShut {NoStop}%
\bibitem [{\citenamefont {Bern}\ \emph
  {et~al.}(2022{\natexlab{a}})\citenamefont {Bern}, \citenamefont
  {Parra-Martinez}, \citenamefont {Roiban}, \citenamefont {Ruf}, \citenamefont
  {Shen}, \citenamefont {Solon},\ and\ \citenamefont {Zeng}}]{BPRRSSZ2021}%
  \BibitemOpen
  \bibfield  {author} {\bibinfo {author} {\bibfnamefont {Z.}~\bibnamefont
  {Bern}}, \bibinfo {author} {\bibfnamefont {J.}~\bibnamefont
  {Parra-Martinez}}, \bibinfo {author} {\bibfnamefont {R.}~\bibnamefont
  {Roiban}}, \bibinfo {author} {\bibfnamefont {M.~S.}\ \bibnamefont {Ruf}},
  \bibinfo {author} {\bibfnamefont {C.-H.}\ \bibnamefont {Shen}}, \bibinfo
  {author} {\bibfnamefont {M.~P.}\ \bibnamefont {Solon}}, \ and\ \bibinfo
  {author} {\bibfnamefont {M.}~\bibnamefont {Zeng}},\ }\href {\doibase
  10.22323/1.416.0051} {\bibfield  {journal} {\bibinfo  {journal} {PoS}\
  }\textbf {\bibinfo {volume} {LL2022}},\ \bibinfo {pages} {051} (\bibinfo
  {year} {2022}{\natexlab{a}})}\BibitemShut {NoStop}%
\bibitem [{\citenamefont {Bern}\ \emph
  {et~al.}(2022{\natexlab{b}})\citenamefont {Bern}, \citenamefont
  {Parra-Martinez}, \citenamefont {Roiban}, \citenamefont {Ruf}, \citenamefont
  {Shen}, \citenamefont {Solon},\ and\ \citenamefont {Zeng}}]{BPRSS2022}%
  \BibitemOpen
  \bibfield  {author} {\bibinfo {author} {\bibfnamefont {Z.}~\bibnamefont
  {Bern}}, \bibinfo {author} {\bibfnamefont {J.}~\bibnamefont
  {Parra-Martinez}}, \bibinfo {author} {\bibfnamefont {R.}~\bibnamefont
  {Roiban}}, \bibinfo {author} {\bibfnamefont {M.~S.}\ \bibnamefont {Ruf}},
  \bibinfo {author} {\bibfnamefont {C.-H.}\ \bibnamefont {Shen}}, \bibinfo
  {author} {\bibfnamefont {M.~P.}\ \bibnamefont {Solon}}, \ and\ \bibinfo
  {author} {\bibfnamefont {M.}~\bibnamefont {Zeng}},\ }\href {\doibase
  10.1103/PhysRevLett.128.161103} {\bibfield  {journal} {\bibinfo  {journal}
  {Phys. Rev. Lett.}\ }\textbf {\bibinfo {volume} {128}},\ \bibinfo {pages}
  {161103} (\bibinfo {year} {2022}{\natexlab{b}})},\ \Eprint
  {http://arxiv.org/abs/2112.10750} {arXiv:2112.10750 [hep-th]} \BibitemShut
  {NoStop}%
\bibitem [{\citenamefont {Bjerrum-Bohr}\ \emph
  {et~al.}(2022{\natexlab{b}})\citenamefont {Bjerrum-Bohr}, \citenamefont
  {Plant\'e},\ and\ \citenamefont {Vanhove}}]{BPV2022}%
  \BibitemOpen
  \bibfield  {author} {\bibinfo {author} {\bibfnamefont {N.~E.~J.}\
  \bibnamefont {Bjerrum-Bohr}}, \bibinfo {author} {\bibfnamefont
  {L.}~\bibnamefont {Plant\'e}}, \ and\ \bibinfo {author} {\bibfnamefont
  {P.}~\bibnamefont {Vanhove}},\ }\href {\doibase 10.1007/JHEP03(2022)071}
  {\bibfield  {journal} {\bibinfo  {journal} {JHEP}\ }\textbf {\bibinfo
  {volume} {03}},\ \bibinfo {pages} {071} (\bibinfo {year}
  {2022}{\natexlab{b}})},\ \Eprint {http://arxiv.org/abs/2111.02976}
  {arXiv:2111.02976 [hep-th]} \BibitemShut {NoStop}%
\bibitem [{\citenamefont {Guevara}(2019)}]{Guevara2019_holcla}%
  \BibitemOpen
  \bibfield  {author} {\bibinfo {author} {\bibfnamefont {A.}~\bibnamefont
  {Guevara}},\ }\href {\doibase 10.1007/JHEP04(2019)033} {\bibfield  {journal}
  {\bibinfo  {journal} {JHEP}\ }\textbf {\bibinfo {volume} {04}},\ \bibinfo
  {pages} {033} (\bibinfo {year} {2019})},\ \Eprint
  {http://arxiv.org/abs/1706.02314} {arXiv:1706.02314 [hep-th]} \BibitemShut
  {NoStop}%
\bibitem [{\citenamefont {Damgaard}\ \emph {et~al.}(2019)\citenamefont
  {Damgaard}, \citenamefont {Haddad},\ and\ \citenamefont {Helset}}]{DHH2019}%
  \BibitemOpen
  \bibfield  {author} {\bibinfo {author} {\bibfnamefont {P.~H.}\ \bibnamefont
  {Damgaard}}, \bibinfo {author} {\bibfnamefont {K.}~\bibnamefont {Haddad}}, \
  and\ \bibinfo {author} {\bibfnamefont {A.}~\bibnamefont {Helset}},\ }\href
  {\doibase 10.1007/JHEP11(2019)070} {\bibfield  {journal} {\bibinfo  {journal}
  {JHEP}\ }\textbf {\bibinfo {volume} {11}},\ \bibinfo {pages} {070} (\bibinfo
  {year} {2019})},\ \Eprint {http://arxiv.org/abs/1908.10308} {arXiv:1908.10308
  [hep-ph]} \BibitemShut {NoStop}%
\bibitem [{\citenamefont {Guevara}\ \emph
  {et~al.}(2019{\natexlab{a}})\citenamefont {Guevara}, \citenamefont
  {Ochirov},\ and\ \citenamefont {Vines}}]{GOV2019}%
  \BibitemOpen
  \bibfield  {author} {\bibinfo {author} {\bibfnamefont {A.}~\bibnamefont
  {Guevara}}, \bibinfo {author} {\bibfnamefont {A.}~\bibnamefont {Ochirov}}, \
  and\ \bibinfo {author} {\bibfnamefont {J.}~\bibnamefont {Vines}},\ }\href
  {\doibase 10.1007/JHEP09(2019)056} {\bibfield  {journal} {\bibinfo  {journal}
  {JHEP}\ }\textbf {\bibinfo {volume} {09}},\ \bibinfo {pages} {056} (\bibinfo
  {year} {2019}{\natexlab{a}})},\ \Eprint {http://arxiv.org/abs/1812.06895}
  {arXiv:1812.06895 [hep-th]} \BibitemShut {NoStop}%
\bibitem [{\citenamefont {Guevara}\ \emph
  {et~al.}(2019{\natexlab{b}})\citenamefont {Guevara}, \citenamefont
  {Ochirov},\ and\ \citenamefont {Vines}}]{GOV2019_blasca}%
  \BibitemOpen
  \bibfield  {author} {\bibinfo {author} {\bibfnamefont {A.}~\bibnamefont
  {Guevara}}, \bibinfo {author} {\bibfnamefont {A.}~\bibnamefont {Ochirov}}, \
  and\ \bibinfo {author} {\bibfnamefont {J.}~\bibnamefont {Vines}},\ }\href
  {\doibase 10.1103/PhysRevD.100.104024} {\bibfield  {journal} {\bibinfo
  {journal} {Phys. Rev. D}\ }\textbf {\bibinfo {volume} {100}},\ \bibinfo
  {pages} {104024} (\bibinfo {year} {2019}{\natexlab{b}})},\ \Eprint
  {http://arxiv.org/abs/1906.10071} {arXiv:1906.10071 [hep-th]} \BibitemShut
  {NoStop}%
\bibitem [{\citenamefont {Bautista}\ and\ \citenamefont
  {Guevara}(2019)}]{BG2019}%
  \BibitemOpen
  \bibfield  {author} {\bibinfo {author} {\bibfnamefont {Y.~F.}\ \bibnamefont
  {Bautista}}\ and\ \bibinfo {author} {\bibfnamefont {A.}~\bibnamefont
  {Guevara}},\ }\href@noop {} {\  (\bibinfo {year} {2019})},\ \Eprint
  {http://arxiv.org/abs/1903.12419} {arXiv:1903.12419 [hep-th]} \BibitemShut
  {NoStop}%
\bibitem [{\citenamefont {Maybee}\ \emph {et~al.}(2019)\citenamefont {Maybee},
  \citenamefont {O'Connell},\ and\ \citenamefont {Vines}}]{MOV2019}%
  \BibitemOpen
  \bibfield  {author} {\bibinfo {author} {\bibfnamefont {B.}~\bibnamefont
  {Maybee}}, \bibinfo {author} {\bibfnamefont {D.}~\bibnamefont {O'Connell}}, \
  and\ \bibinfo {author} {\bibfnamefont {J.}~\bibnamefont {Vines}},\ }\href
  {\doibase 10.1007/JHEP12(2019)156} {\bibfield  {journal} {\bibinfo  {journal}
  {JHEP}\ }\textbf {\bibinfo {volume} {12}},\ \bibinfo {pages} {156} (\bibinfo
  {year} {2019})},\ \Eprint {http://arxiv.org/abs/1906.09260} {arXiv:1906.09260
  [hep-th]} \BibitemShut {NoStop}%
\bibitem [{\citenamefont {Chung}\ \emph {et~al.}(2019)\citenamefont {Chung},
  \citenamefont {Huang}, \citenamefont {Kim},\ and\ \citenamefont
  {Lee}}]{CHKL2019}%
  \BibitemOpen
  \bibfield  {author} {\bibinfo {author} {\bibfnamefont {M.-Z.}\ \bibnamefont
  {Chung}}, \bibinfo {author} {\bibfnamefont {Y.-T.}\ \bibnamefont {Huang}},
  \bibinfo {author} {\bibfnamefont {J.-W.}\ \bibnamefont {Kim}}, \ and\
  \bibinfo {author} {\bibfnamefont {S.}~\bibnamefont {Lee}},\ }\href {\doibase
  10.1007/JHEP04(2019)156} {\bibfield  {journal} {\bibinfo  {journal} {JHEP}\
  }\textbf {\bibinfo {volume} {04}},\ \bibinfo {pages} {156} (\bibinfo {year}
  {2019})},\ \Eprint {http://arxiv.org/abs/1812.08752} {arXiv:1812.08752
  [hep-th]} \BibitemShut {NoStop}%
\bibitem [{\citenamefont {Arkani-Hamed}\ \emph {et~al.}(2020)\citenamefont
  {Arkani-Hamed}, \citenamefont {Huang},\ and\ \citenamefont
  {O'Connell}}]{AHO2020}%
  \BibitemOpen
  \bibfield  {author} {\bibinfo {author} {\bibfnamefont {N.}~\bibnamefont
  {Arkani-Hamed}}, \bibinfo {author} {\bibfnamefont {Y.-t.}\ \bibnamefont
  {Huang}}, \ and\ \bibinfo {author} {\bibfnamefont {D.}~\bibnamefont
  {O'Connell}},\ }\href {\doibase 10.1007/JHEP01(2020)046} {\bibfield
  {journal} {\bibinfo  {journal} {JHEP}\ }\textbf {\bibinfo {volume} {01}},\
  \bibinfo {pages} {046} (\bibinfo {year} {2020})},\ \Eprint
  {http://arxiv.org/abs/1906.10100} {arXiv:1906.10100 [hep-th]} \BibitemShut
  {NoStop}%
\bibitem [{\citenamefont {Siemonsen}\ and\ \citenamefont
  {Vines}(2020)}]{SV2020}%
  \BibitemOpen
  \bibfield  {author} {\bibinfo {author} {\bibfnamefont {N.}~\bibnamefont
  {Siemonsen}}\ and\ \bibinfo {author} {\bibfnamefont {J.}~\bibnamefont
  {Vines}},\ }\href {\doibase 10.1103/PhysRevD.101.064066} {\bibfield
  {journal} {\bibinfo  {journal} {Phys. Rev. D}\ }\textbf {\bibinfo {volume}
  {101}},\ \bibinfo {pages} {064066} (\bibinfo {year} {2020})},\ \Eprint
  {http://arxiv.org/abs/1909.07361} {arXiv:1909.07361 [gr-qc]} \BibitemShut
  {NoStop}%
\bibitem [{\citenamefont {Chung}\ \emph
  {et~al.}(2020{\natexlab{a}})\citenamefont {Chung}, \citenamefont {Huang},
  \citenamefont {Kim},\ and\ \citenamefont {Lee}}]{CHKL2020}%
  \BibitemOpen
  \bibfield  {author} {\bibinfo {author} {\bibfnamefont {M.-Z.}\ \bibnamefont
  {Chung}}, \bibinfo {author} {\bibfnamefont {Y.-t.}\ \bibnamefont {Huang}},
  \bibinfo {author} {\bibfnamefont {J.-W.}\ \bibnamefont {Kim}}, \ and\
  \bibinfo {author} {\bibfnamefont {S.}~\bibnamefont {Lee}},\ }\href {\doibase
  10.1007/JHEP05(2020)105} {\bibfield  {journal} {\bibinfo  {journal} {JHEP}\
  }\textbf {\bibinfo {volume} {05}},\ \bibinfo {pages} {105} (\bibinfo {year}
  {2020}{\natexlab{a}})},\ \Eprint {http://arxiv.org/abs/2003.06600}
  {arXiv:2003.06600 [hep-th]} \BibitemShut {NoStop}%
\bibitem [{\citenamefont {Chung}\ \emph
  {et~al.}(2020{\natexlab{b}})\citenamefont {Chung}, \citenamefont {Huang},\
  and\ \citenamefont {Kim}}]{CHK2020}%
  \BibitemOpen
  \bibfield  {author} {\bibinfo {author} {\bibfnamefont {M.-Z.}\ \bibnamefont
  {Chung}}, \bibinfo {author} {\bibfnamefont {Y.-T.}\ \bibnamefont {Huang}}, \
  and\ \bibinfo {author} {\bibfnamefont {J.-W.}\ \bibnamefont {Kim}},\ }\href
  {\doibase 10.1007/JHEP09(2020)074} {\bibfield  {journal} {\bibinfo  {journal}
  {JHEP}\ }\textbf {\bibinfo {volume} {09}},\ \bibinfo {pages} {074} (\bibinfo
  {year} {2020}{\natexlab{b}})},\ \Eprint {http://arxiv.org/abs/1908.08463}
  {arXiv:1908.08463 [hep-th]} \BibitemShut {NoStop}%
\bibitem [{\citenamefont {Chung}\ \emph
  {et~al.}(2020{\natexlab{c}})\citenamefont {Chung}, \citenamefont {Huang},\
  and\ \citenamefont {Kim}}]{CHK2020_kernew}%
  \BibitemOpen
  \bibfield  {author} {\bibinfo {author} {\bibfnamefont {M.-Z.}\ \bibnamefont
  {Chung}}, \bibinfo {author} {\bibfnamefont {Y.-T.}\ \bibnamefont {Huang}}, \
  and\ \bibinfo {author} {\bibfnamefont {J.-W.}\ \bibnamefont {Kim}},\ }\href
  {\doibase 10.1007/JHEP12(2020)103} {\bibfield  {journal} {\bibinfo  {journal}
  {JHEP}\ }\textbf {\bibinfo {volume} {12}},\ \bibinfo {pages} {103} (\bibinfo
  {year} {2020}{\natexlab{c}})},\ \Eprint {http://arxiv.org/abs/1911.12775}
  {arXiv:1911.12775 [hep-th]} \BibitemShut {NoStop}%
\bibitem [{\citenamefont {Aoude}\ \emph {et~al.}(2020)\citenamefont {Aoude},
  \citenamefont {Haddad},\ and\ \citenamefont {Helset}}]{AHH2020}%
  \BibitemOpen
  \bibfield  {author} {\bibinfo {author} {\bibfnamefont {R.}~\bibnamefont
  {Aoude}}, \bibinfo {author} {\bibfnamefont {K.}~\bibnamefont {Haddad}}, \
  and\ \bibinfo {author} {\bibfnamefont {A.}~\bibnamefont {Helset}},\ }\href
  {\doibase 10.1007/JHEP05(2020)051} {\bibfield  {journal} {\bibinfo  {journal}
  {JHEP}\ }\textbf {\bibinfo {volume} {05}},\ \bibinfo {pages} {051} (\bibinfo
  {year} {2020})},\ \Eprint {http://arxiv.org/abs/2001.09164} {arXiv:2001.09164
  [hep-th]} \BibitemShut {NoStop}%
\bibitem [{\citenamefont {Guevara}\ \emph {et~al.}(2021)\citenamefont
  {Guevara}, \citenamefont {Maybee}, \citenamefont {Ochirov}, \citenamefont
  {O'connell},\ and\ \citenamefont {Vines}}]{GMOV2021}%
  \BibitemOpen
  \bibfield  {author} {\bibinfo {author} {\bibfnamefont {A.}~\bibnamefont
  {Guevara}}, \bibinfo {author} {\bibfnamefont {B.}~\bibnamefont {Maybee}},
  \bibinfo {author} {\bibfnamefont {A.}~\bibnamefont {Ochirov}}, \bibinfo
  {author} {\bibfnamefont {D.}~\bibnamefont {O'connell}}, \ and\ \bibinfo
  {author} {\bibfnamefont {J.}~\bibnamefont {Vines}},\ }\href {\doibase
  10.1007/JHEP03(2021)201} {\bibfield  {journal} {\bibinfo  {journal} {JHEP}\
  }\textbf {\bibinfo {volume} {03}},\ \bibinfo {pages} {201} (\bibinfo {year}
  {2021})},\ \Eprint {http://arxiv.org/abs/2012.11570} {arXiv:2012.11570
  [hep-th]} \BibitemShut {NoStop}%
\bibitem [{\citenamefont {Cristofoli}\ \emph {et~al.}(2021)\citenamefont
  {Cristofoli}, \citenamefont {Gonzo}, \citenamefont {Moynihan}, \citenamefont
  {O'Connell}, \citenamefont {Ross}, \citenamefont {Sergola},\ and\
  \citenamefont {White}}]{CGMORSW2021}%
  \BibitemOpen
  \bibfield  {author} {\bibinfo {author} {\bibfnamefont {A.}~\bibnamefont
  {Cristofoli}}, \bibinfo {author} {\bibfnamefont {R.}~\bibnamefont {Gonzo}},
  \bibinfo {author} {\bibfnamefont {N.}~\bibnamefont {Moynihan}}, \bibinfo
  {author} {\bibfnamefont {D.}~\bibnamefont {O'Connell}}, \bibinfo {author}
  {\bibfnamefont {A.}~\bibnamefont {Ross}}, \bibinfo {author} {\bibfnamefont
  {M.}~\bibnamefont {Sergola}}, \ and\ \bibinfo {author} {\bibfnamefont
  {C.~D.}\ \bibnamefont {White}},\ }\href@noop {} {\  (\bibinfo {year}
  {2021})},\ \Eprint {http://arxiv.org/abs/2112.07556} {arXiv:2112.07556
  [hep-th]} \BibitemShut {NoStop}%
\bibitem [{\citenamefont {Chen}\ \emph
  {et~al.}(2022{\natexlab{a}})\citenamefont {Chen}, \citenamefont {Chung},
  \citenamefont {Huang},\ and\ \citenamefont {Kim}}]{CCHK2022}%
  \BibitemOpen
  \bibfield  {author} {\bibinfo {author} {\bibfnamefont {W.-M.}\ \bibnamefont
  {Chen}}, \bibinfo {author} {\bibfnamefont {M.-Z.}\ \bibnamefont {Chung}},
  \bibinfo {author} {\bibfnamefont {Y.-t.}\ \bibnamefont {Huang}}, \ and\
  \bibinfo {author} {\bibfnamefont {J.-W.}\ \bibnamefont {Kim}},\ }\href
  {\doibase 10.1007/JHEP08(2022)148} {\bibfield  {journal} {\bibinfo  {journal}
  {JHEP}\ }\textbf {\bibinfo {volume} {08}},\ \bibinfo {pages} {148} (\bibinfo
  {year} {2022}{\natexlab{a}})},\ \Eprint {http://arxiv.org/abs/2111.13639}
  {arXiv:2111.13639 [hep-th]} \BibitemShut {NoStop}%
\bibitem [{\citenamefont {Chen}\ \emph
  {et~al.}(2022{\natexlab{b}})\citenamefont {Chen}, \citenamefont {Chung},
  \citenamefont {Huang},\ and\ \citenamefont {Kim}}]{CCHK2022_lenthi}%
  \BibitemOpen
  \bibfield  {author} {\bibinfo {author} {\bibfnamefont {W.-M.}\ \bibnamefont
  {Chen}}, \bibinfo {author} {\bibfnamefont {M.-Z.}\ \bibnamefont {Chung}},
  \bibinfo {author} {\bibfnamefont {Y.-t.}\ \bibnamefont {Huang}}, \ and\
  \bibinfo {author} {\bibfnamefont {J.-W.}\ \bibnamefont {Kim}},\ }\href
  {\doibase 10.1007/JHEP12(2022)058} {\bibfield  {journal} {\bibinfo  {journal}
  {JHEP}\ }\textbf {\bibinfo {volume} {12}},\ \bibinfo {pages} {058} (\bibinfo
  {year} {2022}{\natexlab{b}})},\ \Eprint {http://arxiv.org/abs/2205.07305}
  {arXiv:2205.07305 [hep-th]} \BibitemShut {NoStop}%
\bibitem [{\citenamefont {Kim}(2022)}]{Kim2022_quacor}%
  \BibitemOpen
  \bibfield  {author} {\bibinfo {author} {\bibfnamefont {J.-W.}\ \bibnamefont
  {Kim}},\ }\href {\doibase 10.1103/PhysRevD.106.L081901} {\bibfield  {journal}
  {\bibinfo  {journal} {Phys. Rev. D}\ }\textbf {\bibinfo {volume} {106}},\
  \bibinfo {pages} {L081901} (\bibinfo {year} {2022})},\ \Eprint
  {http://arxiv.org/abs/2207.04970} {arXiv:2207.04970 [hep-th]} \BibitemShut
  {NoStop}%
\bibitem [{\citenamefont {Haddad}\ and\ \citenamefont {Helset}(2020)}]{HH2020}%
  \BibitemOpen
  \bibfield  {author} {\bibinfo {author} {\bibfnamefont {K.}~\bibnamefont
  {Haddad}}\ and\ \bibinfo {author} {\bibfnamefont {A.}~\bibnamefont
  {Helset}},\ }\href {\doibase 10.1103/PhysRevLett.125.181603} {\bibfield
  {journal} {\bibinfo  {journal} {Phys. Rev. Lett.}\ }\textbf {\bibinfo
  {volume} {125}},\ \bibinfo {pages} {181603} (\bibinfo {year} {2020})},\
  \Eprint {http://arxiv.org/abs/2005.13897} {arXiv:2005.13897 [hep-th]}
  \BibitemShut {NoStop}%
\bibitem [{\citenamefont {Haddad}(2022)}]{Haddad2022_explea}%
  \BibitemOpen
  \bibfield  {author} {\bibinfo {author} {\bibfnamefont {K.}~\bibnamefont
  {Haddad}},\ }\href {\doibase 10.1103/PhysRevD.105.026004} {\bibfield
  {journal} {\bibinfo  {journal} {Phys. Rev. D}\ }\textbf {\bibinfo {volume}
  {105}},\ \bibinfo {pages} {026004} (\bibinfo {year} {2022})},\ \Eprint
  {http://arxiv.org/abs/2109.04427} {arXiv:2109.04427 [hep-th]} \BibitemShut
  {NoStop}%
\bibitem [{\citenamefont {Aoude}\ \emph
  {et~al.}(2022{\natexlab{a}})\citenamefont {Aoude}, \citenamefont {Haddad},\
  and\ \citenamefont {Helset}}]{AHH2022}%
  \BibitemOpen
  \bibfield  {author} {\bibinfo {author} {\bibfnamefont {R.}~\bibnamefont
  {Aoude}}, \bibinfo {author} {\bibfnamefont {K.}~\bibnamefont {Haddad}}, \
  and\ \bibinfo {author} {\bibfnamefont {A.}~\bibnamefont {Helset}},\ }\href
  {\doibase 10.1007/JHEP07(2022)072} {\bibfield  {journal} {\bibinfo  {journal}
  {JHEP}\ }\textbf {\bibinfo {volume} {07}},\ \bibinfo {pages} {072} (\bibinfo
  {year} {2022}{\natexlab{a}})},\ \Eprint {http://arxiv.org/abs/2203.06197}
  {arXiv:2203.06197 [hep-th]} \BibitemShut {NoStop}%
\bibitem [{\citenamefont {Aoude}\ \emph
  {et~al.}(2022{\natexlab{b}})\citenamefont {Aoude}, \citenamefont {Haddad},\
  and\ \citenamefont {Helset}}]{AHH2022_clagra}%
  \BibitemOpen
  \bibfield  {author} {\bibinfo {author} {\bibfnamefont {R.}~\bibnamefont
  {Aoude}}, \bibinfo {author} {\bibfnamefont {K.}~\bibnamefont {Haddad}}, \
  and\ \bibinfo {author} {\bibfnamefont {A.}~\bibnamefont {Helset}},\ }\href
  {\doibase 10.1103/PhysRevLett.129.141102} {\bibfield  {journal} {\bibinfo
  {journal} {Phys. Rev. Lett.}\ }\textbf {\bibinfo {volume} {129}},\ \bibinfo
  {pages} {141102} (\bibinfo {year} {2022}{\natexlab{b}})},\ \Eprint
  {http://arxiv.org/abs/2205.02809} {arXiv:2205.02809 [hep-th]} \BibitemShut
  {NoStop}%
\bibitem [{\citenamefont {Bern}\ \emph {et~al.}(2021)\citenamefont {Bern},
  \citenamefont {Luna}, \citenamefont {Roiban}, \citenamefont {Shen},\ and\
  \citenamefont {Zeng}}]{BLRSZ2021}%
  \BibitemOpen
  \bibfield  {author} {\bibinfo {author} {\bibfnamefont {Z.}~\bibnamefont
  {Bern}}, \bibinfo {author} {\bibfnamefont {A.}~\bibnamefont {Luna}}, \bibinfo
  {author} {\bibfnamefont {R.}~\bibnamefont {Roiban}}, \bibinfo {author}
  {\bibfnamefont {C.-H.}\ \bibnamefont {Shen}}, \ and\ \bibinfo {author}
  {\bibfnamefont {M.}~\bibnamefont {Zeng}},\ }\href {\doibase
  10.1103/PhysRevD.104.065014} {\bibfield  {journal} {\bibinfo  {journal}
  {Phys. Rev. D}\ }\textbf {\bibinfo {volume} {104}},\ \bibinfo {pages}
  {065014} (\bibinfo {year} {2021})},\ \Eprint
  {http://arxiv.org/abs/2005.03071} {arXiv:2005.03071 [hep-th]} \BibitemShut
  {NoStop}%
\bibitem [{\citenamefont {Kosmopoulos}\ and\ \citenamefont
  {Luna}(2021)}]{KL2021}%
  \BibitemOpen
  \bibfield  {author} {\bibinfo {author} {\bibfnamefont {D.}~\bibnamefont
  {Kosmopoulos}}\ and\ \bibinfo {author} {\bibfnamefont {A.}~\bibnamefont
  {Luna}},\ }\href {\doibase 10.1007/JHEP07(2021)037} {\bibfield  {journal}
  {\bibinfo  {journal} {JHEP}\ }\textbf {\bibinfo {volume} {07}},\ \bibinfo
  {pages} {037} (\bibinfo {year} {2021})},\ \Eprint
  {http://arxiv.org/abs/2102.10137} {arXiv:2102.10137 [hep-th]} \BibitemShut
  {NoStop}%
\bibitem [{\citenamefont {Bern}\ \emph
  {et~al.}(2022{\natexlab{c}})\citenamefont {Bern}, \citenamefont
  {Kosmopoulos}, \citenamefont {Luna}, \citenamefont {Roiban},\ and\
  \citenamefont {Teng}}]{BKLRT2022}%
  \BibitemOpen
  \bibfield  {author} {\bibinfo {author} {\bibfnamefont {Z.}~\bibnamefont
  {Bern}}, \bibinfo {author} {\bibfnamefont {D.}~\bibnamefont {Kosmopoulos}},
  \bibinfo {author} {\bibfnamefont {A.}~\bibnamefont {Luna}}, \bibinfo {author}
  {\bibfnamefont {R.}~\bibnamefont {Roiban}}, \ and\ \bibinfo {author}
  {\bibfnamefont {F.}~\bibnamefont {Teng}},\ }\href@noop {} {\  (\bibinfo
  {year} {2022}{\natexlab{c}})},\ \Eprint {http://arxiv.org/abs/2203.06202}
  {arXiv:2203.06202 [hep-th]} \BibitemShut {NoStop}%
\bibitem [{\citenamefont {de~la Cruz}(2022)}]{Cruz2022_kinthe}%
  \BibitemOpen
  \bibfield  {author} {\bibinfo {author} {\bibfnamefont {L.}~\bibnamefont
  {de~la Cruz}},\ }\href {\doibase 10.1103/PhysRevD.106.094041} {\bibfield
  {journal} {\bibinfo  {journal} {Phys. Rev. D}\ }\textbf {\bibinfo {volume}
  {106}},\ \bibinfo {pages} {094041} (\bibinfo {year} {2022})},\ \Eprint
  {http://arxiv.org/abs/2207.03452} {arXiv:2207.03452 [hep-th]} \BibitemShut
  {NoStop}%
\bibitem [{\citenamefont {Alessio}\ and\ \citenamefont
  {Di~Vecchia}(2022)}]{AV2022}%
  \BibitemOpen
  \bibfield  {author} {\bibinfo {author} {\bibfnamefont {F.}~\bibnamefont
  {Alessio}}\ and\ \bibinfo {author} {\bibfnamefont {P.}~\bibnamefont
  {Di~Vecchia}},\ }\href {\doibase 10.1016/j.physletb.2022.137258} {\bibfield
  {journal} {\bibinfo  {journal} {Phys. Lett. B}\ }\textbf {\bibinfo {volume}
  {832}},\ \bibinfo {pages} {137258} (\bibinfo {year} {2022})},\ \Eprint
  {http://arxiv.org/abs/2203.13272} {arXiv:2203.13272 [hep-th]} \BibitemShut
  {NoStop}%
\bibitem [{\citenamefont {Adamo}\ \emph {et~al.}(2022)\citenamefont {Adamo},
  \citenamefont {Cristofoli},\ and\ \citenamefont {Tourkine}}]{AAT2022}%
  \BibitemOpen
  \bibfield  {author} {\bibinfo {author} {\bibfnamefont {T.}~\bibnamefont
  {Adamo}}, \bibinfo {author} {\bibfnamefont {A.}~\bibnamefont {Cristofoli}}, \
  and\ \bibinfo {author} {\bibfnamefont {P.}~\bibnamefont {Tourkine}},\ }\href
  {\doibase 10.21468/SciPostPhys.13.2.032} {\bibfield  {journal} {\bibinfo
  {journal} {SciPost Phys.}\ }\textbf {\bibinfo {volume} {13}},\ \bibinfo
  {pages} {032} (\bibinfo {year} {2022})},\ \Eprint
  {http://arxiv.org/abs/2112.09113} {arXiv:2112.09113 [hep-th]} \BibitemShut
  {NoStop}%
\bibitem [{\citenamefont {Menezes}\ and\ \citenamefont
  {Sergola}(2022)}]{MS2022}%
  \BibitemOpen
  \bibfield  {author} {\bibinfo {author} {\bibfnamefont {G.}~\bibnamefont
  {Menezes}}\ and\ \bibinfo {author} {\bibfnamefont {M.}~\bibnamefont
  {Sergola}},\ }\href {\doibase 10.1007/JHEP10(2022)105} {\bibfield  {journal}
  {\bibinfo  {journal} {JHEP}\ }\textbf {\bibinfo {volume} {10}},\ \bibinfo
  {pages} {105} (\bibinfo {year} {2022})},\ \Eprint
  {http://arxiv.org/abs/2205.11701} {arXiv:2205.11701 [hep-th]} \BibitemShut
  {NoStop}%
\bibitem [{\citenamefont {Riva}\ \emph {et~al.}(2022)\citenamefont {Riva},
  \citenamefont {Vernizzi},\ and\ \citenamefont {Wong}}]{RVW2022}%
  \BibitemOpen
  \bibfield  {author} {\bibinfo {author} {\bibfnamefont {M.~M.}\ \bibnamefont
  {Riva}}, \bibinfo {author} {\bibfnamefont {F.}~\bibnamefont {Vernizzi}}, \
  and\ \bibinfo {author} {\bibfnamefont {L.~K.}\ \bibnamefont {Wong}},\ }\href
  {\doibase 10.1103/PhysRevD.106.044013} {\bibfield  {journal} {\bibinfo
  {journal} {Phys. Rev. D}\ }\textbf {\bibinfo {volume} {106}},\ \bibinfo
  {pages} {044013} (\bibinfo {year} {2022})},\ \Eprint
  {http://arxiv.org/abs/2205.15295} {arXiv:2205.15295 [hep-th]} \BibitemShut
  {NoStop}%
\bibitem [{\citenamefont {Chiodaroli}\ \emph {et~al.}(2022)\citenamefont
  {Chiodaroli}, \citenamefont {Johansson},\ and\ \citenamefont
  {Pichini}}]{CJP2022}%
  \BibitemOpen
  \bibfield  {author} {\bibinfo {author} {\bibfnamefont {M.}~\bibnamefont
  {Chiodaroli}}, \bibinfo {author} {\bibfnamefont {H.}~\bibnamefont
  {Johansson}}, \ and\ \bibinfo {author} {\bibfnamefont {P.}~\bibnamefont
  {Pichini}},\ }\href {\doibase 10.1007/JHEP02(2022)156} {\bibfield  {journal}
  {\bibinfo  {journal} {JHEP}\ }\textbf {\bibinfo {volume} {02}},\ \bibinfo
  {pages} {156} (\bibinfo {year} {2022})},\ \Eprint
  {http://arxiv.org/abs/2107.14779} {arXiv:2107.14779 [hep-th]} \BibitemShut
  {NoStop}%
\bibitem [{\citenamefont {Cangemi}\ \emph {et~al.}(2022)\citenamefont
  {Cangemi}, \citenamefont {Chiodaroli}, \citenamefont {Johansson},
  \citenamefont {Ochirov}, \citenamefont {Pichini},\ and\ \citenamefont
  {Skvortsov}}]{CCJOPS2022}%
  \BibitemOpen
  \bibfield  {author} {\bibinfo {author} {\bibfnamefont {L.}~\bibnamefont
  {Cangemi}}, \bibinfo {author} {\bibfnamefont {M.}~\bibnamefont {Chiodaroli}},
  \bibinfo {author} {\bibfnamefont {H.}~\bibnamefont {Johansson}}, \bibinfo
  {author} {\bibfnamefont {A.}~\bibnamefont {Ochirov}}, \bibinfo {author}
  {\bibfnamefont {P.}~\bibnamefont {Pichini}}, \ and\ \bibinfo {author}
  {\bibfnamefont {E.}~\bibnamefont {Skvortsov}},\ }\href@noop {} {\  (\bibinfo
  {year} {2022})},\ \Eprint {http://arxiv.org/abs/2212.06120} {arXiv:2212.06120
  [hep-th]} \BibitemShut {NoStop}%
\bibitem [{\citenamefont {Adamo}\ \emph {et~al.}(2023)\citenamefont {Adamo},
  \citenamefont {Cristofoli},\ and\ \citenamefont {Tourkine}}]{ACT2023}%
  \BibitemOpen
  \bibfield  {author} {\bibinfo {author} {\bibfnamefont {T.}~\bibnamefont
  {Adamo}}, \bibinfo {author} {\bibfnamefont {A.}~\bibnamefont {Cristofoli}}, \
  and\ \bibinfo {author} {\bibfnamefont {P.}~\bibnamefont {Tourkine}},\ }\href
  {\doibase 10.1007/JHEP02(2023)107} {\bibfield  {journal} {\bibinfo  {journal}
  {JHEP}\ }\textbf {\bibinfo {volume} {02}},\ \bibinfo {pages} {107} (\bibinfo
  {year} {2023})},\ \Eprint {http://arxiv.org/abs/2209.05730} {arXiv:2209.05730
  [hep-th]} \BibitemShut {NoStop}%
\bibitem [{\citenamefont {Bjerrum-Bohr}\ \emph {et~al.}(2023)\citenamefont
  {Bjerrum-Bohr}, \citenamefont {Chen},\ and\ \citenamefont
  {Skowronek}}]{BCS2023}%
  \BibitemOpen
  \bibfield  {author} {\bibinfo {author} {\bibfnamefont {N.~E.~J.}\
  \bibnamefont {Bjerrum-Bohr}}, \bibinfo {author} {\bibfnamefont
  {G.}~\bibnamefont {Chen}}, \ and\ \bibinfo {author} {\bibfnamefont
  {M.}~\bibnamefont {Skowronek}},\ }\href@noop {} {\  (\bibinfo {year}
  {2023})},\ \Eprint {http://arxiv.org/abs/2302.00498} {arXiv:2302.00498
  [hep-th]} \BibitemShut {NoStop}%
\bibitem [{\citenamefont {Kim}\ and\ \citenamefont {Steinhoff}(2023)}]{KS2023}%
  \BibitemOpen
  \bibfield  {author} {\bibinfo {author} {\bibfnamefont {J.-W.}\ \bibnamefont
  {Kim}}\ and\ \bibinfo {author} {\bibfnamefont {J.}~\bibnamefont
  {Steinhoff}},\ }\href@noop {} {\  (\bibinfo {year} {2023})},\ \Eprint
  {http://arxiv.org/abs/2302.01944} {arXiv:2302.01944 [hep-th]} \BibitemShut
  {NoStop}%
\bibitem [{\citenamefont {Cachazo}\ and\ \citenamefont
  {Guevara}(2020)}]{CG2017}%
  \BibitemOpen
  \bibfield  {author} {\bibinfo {author} {\bibfnamefont {F.}~\bibnamefont
  {Cachazo}}\ and\ \bibinfo {author} {\bibfnamefont {A.}~\bibnamefont
  {Guevara}},\ }\href {\doibase 10.1007/JHEP02(2020)181} {\bibfield  {journal}
  {\bibinfo  {journal} {JHEP}\ }\textbf {\bibinfo {volume} {02}},\ \bibinfo
  {pages} {181} (\bibinfo {year} {2020})},\ \Eprint
  {http://arxiv.org/abs/1705.10262} {arXiv:1705.10262 [hep-th]} \BibitemShut
  {NoStop}%
\bibitem [{\citenamefont {Bjerrum-Bohr}\ \emph {et~al.}(2018)\citenamefont
  {Bjerrum-Bohr}, \citenamefont {Damgaard}, \citenamefont {Festuccia},
  \citenamefont {Plant\'e},\ and\ \citenamefont {Vanhove}}]{BDFPV2018}%
  \BibitemOpen
  \bibfield  {author} {\bibinfo {author} {\bibfnamefont {N.~E.~J.}\
  \bibnamefont {Bjerrum-Bohr}}, \bibinfo {author} {\bibfnamefont {P.~H.}\
  \bibnamefont {Damgaard}}, \bibinfo {author} {\bibfnamefont {G.}~\bibnamefont
  {Festuccia}}, \bibinfo {author} {\bibfnamefont {L.}~\bibnamefont {Plant\'e}},
  \ and\ \bibinfo {author} {\bibfnamefont {P.}~\bibnamefont {Vanhove}},\ }\href
  {\doibase 10.1103/PhysRevLett.121.171601} {\bibfield  {journal} {\bibinfo
  {journal} {Phys. Rev. Lett.}\ }\textbf {\bibinfo {volume} {121}},\ \bibinfo
  {pages} {171601} (\bibinfo {year} {2018})},\ \Eprint
  {http://arxiv.org/abs/1806.04920} {arXiv:1806.04920 [hep-th]} \BibitemShut
  {NoStop}%
\bibitem [{\citenamefont {Cheung}\ \emph {et~al.}(2018)\citenamefont {Cheung},
  \citenamefont {Rothstein},\ and\ \citenamefont {Solon}}]{CRS2018}%
  \BibitemOpen
  \bibfield  {author} {\bibinfo {author} {\bibfnamefont {C.}~\bibnamefont
  {Cheung}}, \bibinfo {author} {\bibfnamefont {I.~Z.}\ \bibnamefont
  {Rothstein}}, \ and\ \bibinfo {author} {\bibfnamefont {M.~P.}\ \bibnamefont
  {Solon}},\ }\href {\doibase 10.1103/PhysRevLett.121.251101} {\bibfield
  {journal} {\bibinfo  {journal} {Phys. Rev. Lett.}\ }\textbf {\bibinfo
  {volume} {121}},\ \bibinfo {pages} {251101} (\bibinfo {year} {2018})},\
  \Eprint {http://arxiv.org/abs/1808.02489} {arXiv:1808.02489 [hep-th]}
  \BibitemShut {NoStop}%
\bibitem [{\citenamefont {Cristofoli}\ \emph {et~al.}(2019)\citenamefont
  {Cristofoli}, \citenamefont {Bjerrum-Bohr}, \citenamefont {Damgaard},\ and\
  \citenamefont {Vanhove}}]{CBDV2019}%
  \BibitemOpen
  \bibfield  {author} {\bibinfo {author} {\bibfnamefont {A.}~\bibnamefont
  {Cristofoli}}, \bibinfo {author} {\bibfnamefont {N.~E.~J.}\ \bibnamefont
  {Bjerrum-Bohr}}, \bibinfo {author} {\bibfnamefont {P.~H.}\ \bibnamefont
  {Damgaard}}, \ and\ \bibinfo {author} {\bibfnamefont {P.}~\bibnamefont
  {Vanhove}},\ }\href {\doibase 10.1103/PhysRevD.100.084040} {\bibfield
  {journal} {\bibinfo  {journal} {Phys. Rev. D}\ }\textbf {\bibinfo {volume}
  {100}},\ \bibinfo {pages} {084040} (\bibinfo {year} {2019})},\ \Eprint
  {http://arxiv.org/abs/1906.01579} {arXiv:1906.01579 [hep-th]} \BibitemShut
  {NoStop}%
\bibitem [{\citenamefont {Buonanno}\ and\ \citenamefont
  {Damour}(1999)}]{BD1999}%
  \BibitemOpen
  \bibfield  {author} {\bibinfo {author} {\bibfnamefont {A.}~\bibnamefont
  {Buonanno}}\ and\ \bibinfo {author} {\bibfnamefont {T.}~\bibnamefont
  {Damour}},\ }\href {\doibase 10.1103/PhysRevD.59.084006} {\bibfield
  {journal} {\bibinfo  {journal} {Phys. Rev. D}\ }\textbf {\bibinfo {volume}
  {59}},\ \bibinfo {pages} {084006} (\bibinfo {year} {1999})},\ \Eprint
  {http://arxiv.org/abs/gr-qc/9811091} {arXiv:gr-qc/9811091} \BibitemShut
  {NoStop}%
\bibitem [{\citenamefont {Buonanno}\ and\ \citenamefont
  {Damour}(2000)}]{BD2000}%
  \BibitemOpen
  \bibfield  {author} {\bibinfo {author} {\bibfnamefont {A.}~\bibnamefont
  {Buonanno}}\ and\ \bibinfo {author} {\bibfnamefont {T.}~\bibnamefont
  {Damour}},\ }\href {\doibase 10.1103/PhysRevD.62.064015} {\bibfield
  {journal} {\bibinfo  {journal} {Phys. Rev. D}\ }\textbf {\bibinfo {volume}
  {62}},\ \bibinfo {pages} {064015} (\bibinfo {year} {2000})},\ \Eprint
  {http://arxiv.org/abs/gr-qc/0001013} {arXiv:gr-qc/0001013} \BibitemShut
  {NoStop}%
\bibitem [{\citenamefont {Damour}\ \emph {et~al.}(2000)\citenamefont {Damour},
  \citenamefont {Jaranowski},\ and\ \citenamefont {Schaefer}}]{DJS2000}%
  \BibitemOpen
  \bibfield  {author} {\bibinfo {author} {\bibfnamefont {T.}~\bibnamefont
  {Damour}}, \bibinfo {author} {\bibfnamefont {P.}~\bibnamefont {Jaranowski}},
  \ and\ \bibinfo {author} {\bibfnamefont {G.}~\bibnamefont {Schaefer}},\
  }\href {\doibase 10.1103/PhysRevD.62.084011} {\bibfield  {journal} {\bibinfo
  {journal} {Phys. Rev. D}\ }\textbf {\bibinfo {volume} {62}},\ \bibinfo
  {pages} {084011} (\bibinfo {year} {2000})},\ \Eprint
  {http://arxiv.org/abs/gr-qc/0005034} {arXiv:gr-qc/0005034} \BibitemShut
  {NoStop}%
\bibitem [{\citenamefont {Damour}(2001)}]{Damour2001}%
  \BibitemOpen
  \bibfield  {author} {\bibinfo {author} {\bibfnamefont {T.}~\bibnamefont
  {Damour}},\ }\href {\doibase 10.1103/PhysRevD.64.124013} {\bibfield
  {journal} {\bibinfo  {journal} {Phys. Rev. D}\ }\textbf {\bibinfo {volume}
  {64}},\ \bibinfo {pages} {124013} (\bibinfo {year} {2001})},\ \Eprint
  {http://arxiv.org/abs/gr-qc/0103018} {arXiv:gr-qc/0103018} \BibitemShut
  {NoStop}%
\bibitem [{\citenamefont {Bini}\ and\ \citenamefont {Damour}(2012)}]{BD2012}%
  \BibitemOpen
  \bibfield  {author} {\bibinfo {author} {\bibfnamefont {D.}~\bibnamefont
  {Bini}}\ and\ \bibinfo {author} {\bibfnamefont {T.}~\bibnamefont {Damour}},\
  }\href {\doibase 10.1103/PhysRevD.86.124012} {\bibfield  {journal} {\bibinfo
  {journal} {Phys. Rev. D}\ }\textbf {\bibinfo {volume} {86}},\ \bibinfo
  {pages} {124012} (\bibinfo {year} {2012})},\ \Eprint
  {http://arxiv.org/abs/1210.2834} {arXiv:1210.2834 [gr-qc]} \BibitemShut
  {NoStop}%
\bibitem [{\citenamefont {Damour}(2008)}]{Damour2008}%
  \BibitemOpen
  \bibfield  {author} {\bibinfo {author} {\bibfnamefont {T.}~\bibnamefont
  {Damour}},\ }\href {\doibase 10.1142/S0217751X08039992} {\bibfield  {journal}
  {\bibinfo  {journal} {Int. J. Mod. Phys. A}\ }\textbf {\bibinfo {volume}
  {23}},\ \bibinfo {pages} {1130} (\bibinfo {year} {2008})},\ \Eprint
  {http://arxiv.org/abs/0802.4047} {arXiv:0802.4047 [gr-qc]} \BibitemShut
  {NoStop}%
\bibitem [{\citenamefont {Damgaard}\ and\ \citenamefont
  {Vanhove}(2021)}]{DV2021}%
  \BibitemOpen
  \bibfield  {author} {\bibinfo {author} {\bibfnamefont {P.~H.}\ \bibnamefont
  {Damgaard}}\ and\ \bibinfo {author} {\bibfnamefont {P.}~\bibnamefont
  {Vanhove}},\ }\href {\doibase 10.1103/PhysRevD.104.104029} {\bibfield
  {journal} {\bibinfo  {journal} {Phys. Rev. D}\ }\textbf {\bibinfo {volume}
  {104}},\ \bibinfo {pages} {104029} (\bibinfo {year} {2021})},\ \Eprint
  {http://arxiv.org/abs/2108.11248} {arXiv:2108.11248 [hep-th]} \BibitemShut
  {NoStop}%
\bibitem [{\citenamefont {Poisson}\ \emph {et~al.}(2011)\citenamefont
  {Poisson}, \citenamefont {Pound},\ and\ \citenamefont {Vega}}]{PPV2011}%
  \BibitemOpen
  \bibfield  {author} {\bibinfo {author} {\bibfnamefont {E.}~\bibnamefont
  {Poisson}}, \bibinfo {author} {\bibfnamefont {A.}~\bibnamefont {Pound}}, \
  and\ \bibinfo {author} {\bibfnamefont {I.}~\bibnamefont {Vega}},\ }\href
  {\doibase 10.12942/lrr-2011-7} {\bibfield  {journal} {\bibinfo  {journal}
  {Living Rev. Rel.}\ }\textbf {\bibinfo {volume} {14}},\ \bibinfo {pages} {7}
  (\bibinfo {year} {2011})},\ \Eprint {http://arxiv.org/abs/1102.0529}
  {arXiv:1102.0529 [gr-qc]} \BibitemShut {NoStop}%
\bibitem [{\citenamefont {Young}(1976)}]{Young1976_cappar}%
  \BibitemOpen
  \bibfield  {author} {\bibinfo {author} {\bibfnamefont {P.~J.}\ \bibnamefont
  {Young}},\ }\href {\doibase 10.1103/PhysRevD.14.3281} {\bibfield  {journal}
  {\bibinfo  {journal} {Phys. Rev. D}\ }\textbf {\bibinfo {volume} {14}},\
  \bibinfo {pages} {3281} (\bibinfo {year} {1976})}\BibitemShut {NoStop}%
\bibitem [{\citenamefont {Johnston}\ and\ \citenamefont
  {Ruffini}(1974)}]{JR1974}%
  \BibitemOpen
  \bibfield  {author} {\bibinfo {author} {\bibfnamefont {M.}~\bibnamefont
  {Johnston}}\ and\ \bibinfo {author} {\bibfnamefont {R.}~\bibnamefont
  {Ruffini}},\ }\href {\doibase 10.1103/PhysRevD.10.2324} {\bibfield  {journal}
  {\bibinfo  {journal} {Phys. Rev. D}\ }\textbf {\bibinfo {volume} {10}},\
  \bibinfo {pages} {2324} (\bibinfo {year} {1974})}\BibitemShut {NoStop}%
\bibitem [{\citenamefont {Barack}\ and\ \citenamefont {Pound}(2019)}]{BP2019}%
  \BibitemOpen
  \bibfield  {author} {\bibinfo {author} {\bibfnamefont {L.}~\bibnamefont
  {Barack}}\ and\ \bibinfo {author} {\bibfnamefont {A.}~\bibnamefont {Pound}},\
  }\href {\doibase 10.1088/1361-6633/aae552} {\bibfield  {journal} {\bibinfo
  {journal} {Rept. Prog. Phys.}\ }\textbf {\bibinfo {volume} {82}},\ \bibinfo
  {pages} {016904} (\bibinfo {year} {2019})},\ \Eprint
  {http://arxiv.org/abs/1805.10385} {arXiv:1805.10385 [gr-qc]} \BibitemShut
  {NoStop}%
\bibitem [{\citenamefont {Griffiths}\ and\ \citenamefont
  {Podolsky}(2009)}]{GP2009}%
  \BibitemOpen
  \bibfield  {author} {\bibinfo {author} {\bibfnamefont {J.~B.}\ \bibnamefont
  {Griffiths}}\ and\ \bibinfo {author} {\bibfnamefont {J.}~\bibnamefont
  {Podolsky}},\ }\enquote {\bibinfo {title} {{Exact Space-Times in Einstein's
  General Relativity}},}\ \ (\bibinfo  {publisher} {Cambridge University
  Press},\ \bibinfo {address} {Cambridge},\ \bibinfo {year} {2009})\ pp.\
  \bibinfo {pages} {208--209}\BibitemShut {NoStop}%
\bibitem [{Note1()}]{Note1}%
  \BibitemOpen
  \bibinfo {note} {Note a subtlety with applying eq. \protect \textup {\hbox
  {\mathsurround \z@ \protect \normalfont (\ignorespaces \ref
  {eq:chiscalar}\unskip \@@italiccorr )}}. If $\kappa $ depends on $b$, $h(r)$
  is no longer given by the usual scalar structure of eq. \protect \textup
  {\hbox {\mathsurround \z@ \protect \normalfont (\ignorespaces \ref
  {eq:hrscalar}\unskip \@@italiccorr )}}. However, as explained in ref. \cite
  {DHLV2022}, the scattering angle integral takes the form $d\phi /dr =
  dp_r/dL$ regardless. $L$ is angular momentum and $p_r$ is canonical radial
  momentum. Consequently eq. \protect \textup {\hbox {\mathsurround \z@
  \protect \normalfont (\ignorespaces \ref {eq:chiscalar}\unskip \@@italiccorr
  )}} still applies}\BibitemShut {NoStop}%
\bibitem [{\citenamefont {Newman}\ and\ \citenamefont {Janis}(1965)}]{NJ1965}%
  \BibitemOpen
  \bibfield  {author} {\bibinfo {author} {\bibfnamefont {E.~T.}\ \bibnamefont
  {Newman}}\ and\ \bibinfo {author} {\bibfnamefont {A.~I.}\ \bibnamefont
  {Janis}},\ }\href {\doibase 10.1063/1.1704350} {\bibfield  {journal}
  {\bibinfo  {journal} {J. Math. Phys.}\ }\textbf {\bibinfo {volume} {6}},\
  \bibinfo {pages} {915} (\bibinfo {year} {1965})}\BibitemShut {NoStop}%
\bibitem [{\citenamefont {Drake}\ and\ \citenamefont
  {Szekeres}(2000)}]{DS2000}%
  \BibitemOpen
  \bibfield  {author} {\bibinfo {author} {\bibfnamefont {S.~P.}\ \bibnamefont
  {Drake}}\ and\ \bibinfo {author} {\bibfnamefont {P.}~\bibnamefont
  {Szekeres}},\ }\href {\doibase 10.1023/A:1001920232180} {\bibfield  {journal}
  {\bibinfo  {journal} {Gen. Rel. Grav.}\ }\textbf {\bibinfo {volume} {32}},\
  \bibinfo {pages} {445} (\bibinfo {year} {2000})},\ \Eprint
  {http://arxiv.org/abs/gr-qc/9807001} {arXiv:gr-qc/9807001} \BibitemShut
  {NoStop}%
\end{thebibliography}%
\end{document}